\documentclass[12pt,oneside,letterpaper]{report}

\newcommand{\draftfinal}[2]{\ifdefined\draftversion#1\else#2\fi}

\newcommand{\finalonly}[1]{\draftfinal{}{#1}}

\newcommand{\thesistitle}{Emergent flows, irreversibility and unsteady effects in asymmetric and looped geometries}
\newcommand{\thesisauthor}{Quynh Minh Nguyen}
\newcommand{\thesisadvisor}{Dr.  \href{https://math.nyu.edu/~ristroph/}{\color{black}{Leif Ristroph}}}
\newcommand{\secondadvisor}{Dr.  \href{https://physics.nyu.edu/jz11/}{\color{black}{Jun Zhang}}}

\newcommand{\thesisdept}{Physics}
\newcommand{\gradmonth}{January}
\newcommand{\gradyear}{2021}
\newcommand{\thesisdedication}{To my mother Pham Thi Huyen Diu  who always supports my intellectual pursuits at any cost, and Vibeke Libby who always believes in me.}

\RequirePackage[margin=1in, includefoot, letterpaper]{geometry}

\RequirePackage[prologue]{xcolor}
\definecolor[named]{ThesisBlue}{cmyk}{1,0.1,0,0.1}
\definecolor[named]{ThesisYellow}{cmyk}{0,0.16,1,0}
\definecolor[named]{ThesisOrange}{cmyk}{0,0.42,1,0.01}
\definecolor[named]{ThesisRed}{cmyk}{0,0.90,0.86,0}
\definecolor[named]{ThesisLightBlue}{cmyk}{0.49,0.01,0,0}
\definecolor[named]{ThesisGreen}{cmyk}{0.20,0,1,0.19}
\definecolor[named]{ThesisPurple}{cmyk}{0.55,1,0,0.15}
\definecolor[named]{ThesisDarkBlue}{cmyk}{1,0.58,0,0.21}

\definecolor{SchoolColor}{rgb}{0.3412, 0.0235, 0.5490} 
\definecolor{chaptergrey}{rgb}{0.2600, 0.0200, 0.4600} 
\definecolor{midgrey}{rgb}{0.4, 0.4, 0.4}

\usepackage{chemformula}
\usepackage{array}

\RequirePackage[labelfont={bf,sf,small,singlespacing},
                textfont={sf,small,singlespacing},
                margin=0pt,
                figurewithin=chapter,
                tablewithin=chapter]{caption}

\usepackage{fix-cm}
\RequirePackage[raggedright,sc]{titlesec}
\definecolor{gray75}{gray}{0.75}
\newcommand{\hsp}{\hspace{20pt}}

\titleformat{\chapter}[hang]
{\Huge\sc}
{\textcolor{SchoolColor}{\thechapter}\hsp\textcolor{gray75}{|}\hsp}
{0pt}{\Huge\sc\raggedright}

\usepackage{setspace}

\finalonly{
  \doublespacing 
}

\usepackage{lipsum}

\input{defs}

\usepackage{comment}

\usepackage{hyperref}
\hypersetup{colorlinks,
  linkcolor=ThesisDarkBlue,
  citecolor=ThesisPurple,
  urlcolor=ThesisDarkBlue,
  filecolor=ThesisDarkBlue}
 
\usepackage[
    backend=biber,
    style=numeric-comp,
    natbib=true,
    url=false, 
    doi=true,
    eprint=false
]{biblatex}
 \usepackage{lscape}
 
\begin{document}

\pagenumbering{roman}
%
\thispagestyle{empty}
%

\vspace*{25pt}
\begin{center}

  {\Large
    \begin{doublespace}
      {\textcolor{SchoolColor}{\textsc{\thesistitle}}}
    \end{doublespace}
  }
  \vspace{.7in}

  by
  \vspace{.7in}

  \thesisauthor
  \vfill

  \begin{doublespace}
    \textsc{
    A dissertation submitted in partial fulfillment\\
    of the requirements for the degree of\\
    Doctor of Philosophy\\
    Department of \thesisdept\\
    New York University\\
    \gradmonth, \gradyear}
  \end{doublespace}
\end{center}
\vfill

\noindent\makebox[\textwidth]{\hfill\makebox[2.5in]{\hrulefill}}\\
\makebox[\textwidth]{\hfill\makebox[2.5in]{\hfill\thesisadvisor}}
\noindent\makebox[\textwidth]{\hfill\makebox[2.5in]{\hrulefill}}\\
\makebox[\textwidth]{\hfill\makebox[2.5in]{\hfill\secondadvisor}}

\newpage

\thispagestyle{empty}
\vspace*{25pt}
\begin{center}
  \scshape \noindent \small \copyright \  \small  \thesisauthor \\
  All rights reserved, \gradyear
\end{center}
\vspace*{0in}
\newpage

\cleardoublepage
\phantomsection
\chapter*{Dedication}
\addcontentsline{toc}{chapter}{Dedication}
  \thesisdedication
\vfill
\newpage

\chapter*{Acknowledgements}
\addcontentsline{toc}{chapter}{Acknowledgments}

 I am grateful to the many people who helped me during the doctoral program at New York University which culminated in this dissertation. 

 My advisor Leif Ristroph who has always been an inspiring, patient and encouraging mentor. He has helped me become a better thinker, doer and speaker, all of which are valuable beyond graduate school.

My co-advisor Jun Zhang who welcomed me to the  \href{https://math.nyu.edu/aml/}{Applied Math Lab}  and who helped me see through my shortcomings.

My dissertation committee, professor Paul M. Chaikin who generously spent time reading the early draft of my thesis, and professor Alexandra Zidovska and professor Alexander Y. Grosberg who both taught me valuable lessons. 

Professors Michael J. Shelley, Stephen Childress, and Charles Peskin whose wisdom and advices helped me in difficult times.

My collaborators Dr. Anand U. Oza and Dr. Christina Frederick at New Jersey Institute of Technology; undergraduate lab members Joanna Abouezzi, and Dean Huang; and high school students Evan Zauderer, Genevieve Romanelli and Charlotte L. Meyer. All of their works are important to my thesis.

 My friends at the Applied Math Lab and the Physics Department, among them Jinzi Mac Huang, Joel W. Newbolt, Digvijay Wadekar, Joshua N. Tong, Marco S. Muzio, Cristina Mondino, Pejman Sanaei, Quentin Brosseau, Kaizhe Wang and especially Michael Wang who proofread this thesis. Other friends, Ngan Nguyen and Duc Hoang who hosted me in their apartment in Massachusetts through out the coronavirus pandemic and the last months of graduate school, and Hoa Le and my partner Lan Do.

Our graduate program administrator Evette Ma who helped me navigate the program over the years.


\chapter*{Abstract}
\addcontentsline{toc}{chapter}{Abstract}

Fluid transport networks are important in many natural settings and engineering applications, from animal cardiovascular and respiratory systems to plant vasculature to plumbing networks and chemical plant,  \cite{nielsen1964animal,conway1971guide,lal1975hydraulic,waite2007applied,fung2013biomechanics}.  Understanding how network topology, connectivity, internal boundaries and other geometrical aspects affect the global flow state is a challenging problem that depends on complex fluid properties characterized by different length and time scales  \cite{tritton2012physical}. The study of flow  in micro-scale networks including  microvascular network of small animals, plant vasculature \cite{jensen2016sap} and artificial microfluidics \cite{whitesides2006origins} focuses on low Reynolds numbers where small volumes of fluids move at slow speeds. The flow physics at these scales is dominated by pressures overcoming viscous impedance, and the governing Stokes equation is linear \cite{squires2005microfluidics}. This linearity property allows for relatively simple theoretical and computational solutions that greatly aid in the understanding, modeling and designing of micro-scale networks.

At larger scales and faster flow rates,  macrofluidic networks such as in the  cardiovascular and respiratory systems of larger animals and numerous engineering applications are also important but the flow physics is quite different. The underlying Navier-Stokes equation is nonlinear, theoretical results are few, simulations are challenging, and the mapping between geometry and desired flow objectives are all much more complex \cite{tritton2012physical}. The  phenomenology for such high-Reynolds-number or inertially-dominated flows is well documented and well studied: Flows are retarded in thin boundary layers near solid surfaces; such flows are sensitive to geometry and tend to separate from surfaces; and vortices, wakes, jets and unsteadiness abound \cite{schlichting2016boundary}. The counter-intuitive nature of inertial flows is exemplified by the breakdown of reversibility: Running a given system in reverse, say by inverting pressures, does not necessarily cause the fluid to move in reverse but can instead trigger altogether different flow patterns \cite{tritton2012physical}. This dissertation explores two general ways of how rectified flows emerge in macrofluidic networks as a consequence of irreversibility and unsteady effects: When branches or channels of a network have asymmetric internal geometry and the second when a network contains loops. For the former, we focus on Tesla's valvular conduit or Tesla valve \cite{tesla1920valvular} and for the later, the bird respiratory system. Emergent circulation allows for valveless pumping that are of interest in both engineering and biological contexts.  

More than 100 years ago, the famous inventor Nikola Tesla was doing experiments in lower Manhattan, not far from the Applied Math Lab. Tesla was better known as an  ``electric wizard'' of the time, but he dabbled in fluid mechanics as well. In 1920, he patented the earliest fluidic device, the Tesla valve which is an asymmetric channel that intended to have strong asymmetric resistance. Since its invention, the device has spurred a lot of studies on its usage as a microfluidic device. However, fluid mechanical characterization of the device and related physics have been overlooked, especially at higher Reynolds number. We faithfully reproduce the device, and design an experimental system that can apply and control steady pressures on the device while measuring the resulting flow rate. This allows measuring resistances under steady condition extensively over a wide range of Reynolds numbers. We discover that the  Tesla valve works by inducing early turbulence. Flows in Tesla valve transition to turbulence at an expected low Reynolds number of 200. Our flow visualization at this transitional Reynolds number reveals the destabilization mechanism in the reverse direction that's a hallmark of turbulence. 

Another area that is also overlooked by existing literature is the behavior of the device under unsteady conditions. Nikola Tesla himself conjectured that the diodic behavior is significantly improved with pulsatile flows. To address that, we designed a fluidic circuit that acts as an AC-to-DC converter - converting oscillatory flows to directed flows or a valveless pump, and measured its output under different unsteady inputs using well-calibrated Particle Image Velocimetry (PIV). The response is found to be more than linear with both amplitude and frequency. Irreversible diodic behavior, expressed in terms of effectiveness of the circuit increases with both amplitude and frequency. In addition, our steady characterization of the device allows us to predict the effectiveness based on a steady assumption, which is then compared with the real effectiveness. Our findings confirm Tesla's conjecture, the diodic behavior is boosted by unsteady effect. In another unexpected and counter-intuitive result, we find that the output DC flows are smoothened as the driving amplitude increases. 

The scientific investigation of avian respiratory system  has a long history, and dates back to V. Coitier in 1573. Among other aspects of the lung, the airflow dynamics remained controversial and unresolved despite much scientific scrutiny. In mammal lungs, the air flow is inhaled into a tree-like structure ending in small sacs called aveoli, and simply reverse direction when exhaled. In contrast,the bird lung, also known as the most  efficient gas exchanger among living vertebrates \cite{maina2017pivotal}, has a complex and unique structure: it contains closed loops in which uni-directional flow is sustained during both inhalation and exhalation. Early researchers speculated that the bird lung must possess valves that open and close at the right time, much like a circulatory system. In fact, extensive physiological studies have shown that the lung is rigid and has no valves, and the mechanism responsible for directed flow is deemed 'aerodynamic valving'. Subsequently, many aerodynamic mechanisms have been suggested, and each has been ruled out as either non-existent, or not essential because it doesn't exhibit in all species, and the minimum ingredients required for directed flows remain a mystery for the last 100 years.  

We simplified the airway network into simple loopy networks of tubes. The tubes are filled with water, and oscillatory flow (AC) is forced in one of the branches, and the resulting flows are measured in the ``free" branch using PIV. Persistent DC flows emerge in the free branch for varying driving amplitude and frequency. Our finding demonstrates the minimum ingredients for DC flows are loops, which necessitate junctions (and for simplicity, we use T-junction), and the asymmetric connectivity of the junctions. Direct numerical simulation reproduced the qualitative phenomenon, and reveals that the valving mechanism is due to irreversibility. Flow separation and vortex shedding act to block a side branch of the T-junction only in half of the oscillation cycle, and with appropriate connectivity between partnered junctions, DC flows emerge in loops.  

Chapter \ref{ch1} reviews the physics relevant to macrofluidic networks, and flow rectification using static geometries. In Chapter \ref{ch2}, we take a pedagogical approach and draw on the electronic-hydraulic analogy to study Tesla's valve and other asymmetric channels under steady conditions. Chapter \ref{ch3} studies Tesla's valve across dynamical regimes and under unsteady conditions. Chapter \ref{ch4} presents our experiments and simulations on unidirectional flows in loopy network models of bird lungs. In Chapter \ref{ch5}, we propose a generalized modeling approach for flows in networks, and apply to an idealized network of a bird lung. Finally, Chapter \ref{ch6} summarizes the findings and discusses further studies.

\newpage

\tableofcontents

\cleardoublepage
\phantomsection
\addcontentsline{toc}{chapter}{List of Figures}
\listoffigures
\newpage


\pagenumbering{arabic} 




\chapter{Introduction: Macrofluidic networks and flow rectification} \label{ch1}
\section{Macrofluidic network}
 Networks appear everywhere, from biology to physics \cite{strogatz2001exploring, singer2013natural}. 
 Fluid transport networks are prevalent, both in nature and in man-made systems. Examples of such networks include the mammalian circulatory systems which transports nutrients throughout the body through blood circulation \cite{fung2013biomechanics-circulation} and channel networks associated with river drainage basins \cite{rodriguez2001fractal}. 
 Fluid transport networks or \emph{fluidics} generally consist of branches connected at nodes (or junctions), forming tree and loop topology. In addition, there are flow regulating elements such as pressure source and valves. The flow of fluid in these networks provides critical functions: Transport and distribution of important quantities such as liquid material, heat, nutrients, mechanical energy, etc. Thus the flow patterns within the network are required to be robust and effective. 
 
Microfluidics is the science and technology of fluid transport network that process or manipulate small amounts of fluids, using channels with dimensions of less than a milimeter \cite{whitesides2006origins,stone2004engineering}. \emph{ Macrofluidics} is similar, but the scales involved are macroscopic. A clearer distinction can be made based on not just volume or length scale but the relative magnitude of various effects as characterized traditionally using dimensionless parameters. The most important dimensionless parameter in fluid mechanics is the \emph{ Reynolds number}, $  \mathrm{ Re} = \rho U L / \mu $ where $\rho$ and $\mu$ are the density and viscosity of the fluid, respectively, and $U$ and $L$ are characteristic speed and length. $\mathrm{Re}$ indicates the relative importance of representative magnitude of inertial forces to that of viscous forces. Generally, at microscopic scales, $\mathrm{Re} \ll 1 $, viscous forces dominate, and the governing equation of motion is linear, flows are laminar and the flow physics is well-understood as microfluidics has been the main focus of recent research on fluidic networks. On the other hand, at macroscopic scales, $\mathrm{Re} \gg 1$, inertial forces dominate, and the governing equation of motion is non-linear, and the physics of the flows is more complicated with many aspects not well understood. The next section will review and analyze the governing equations relevant to macrofluidics.

\section{Governing equations and dynamical similarity} \label{ch2:section:governing eqs}
The incompressible Navier-Stokes equations for a Newtonian fluid are written as 
\begin{equation} \label{eq:navier-stokes}
   \rho \left[ \frac{\partial \boldsymbol{u}}{\partial t} + (\boldsymbol{u} \cdot \nabla) \boldsymbol{u} \right] =  -\nabla p + \mu \nabla^2 \boldsymbol{u} + \boldsymbol{F} \;, \;\; \nabla \cdot \boldsymbol{u} = 0
\end{equation}
where $\boldsymbol{u}$, $p$, $\boldsymbol{F}$, $\rho$ and $\mu$ are velocity field, pressure, body force, density and dynamic viscosity of the fluid, respectively \cite{tritton2012physical}. The first equation in \ref{eq:navier-stokes} is also referred to as the momentum equation or the equation of motion, while the second is the continuity equation for incompressible fluid. In many practical situations, the only body force is gravity which is negligible for uniform density, and we can assume   $\boldsymbol{F}=0$ and the main source of motion is due to pressure gradients or relative movement of the boundaries. This will be true of the problems studied here.

We define dimensionless variables as follows:
\begin{equation}
    \boldsymbol{u}^* = \boldsymbol{u}/U, \boldsymbol{x}^* = \boldsymbol{x}/L, \mathrm{and} \; p^* = p /(\mu U/L)  \; \mathrm{or}  \;  p/\rho U^2,
\end{equation}
where $U, L$ are characteristic velocity and length scales, respectively. The choice of which scale is used to nondimensionalize the pressure $p$ is a matter of convenience, depending on the limit being considered. Using the first scaling, the non-dimensional version of the dynamical equation is then
\begin{equation}
        \frac{\partial \boldsymbol{u}^*}{\partial t^*} + (\boldsymbol{u}^* \cdot \nabla^*) \boldsymbol{u}^*  = \frac{1}{\mathrm{Re}} \left[ -\nabla^* p^* + \nabla^{*2} \boldsymbol{u}^*  \right],
\end{equation}
where all star variables are of order unity (possibly excepting $\nabla p^*$).
The only parameter in the equation Re = $\rho UL/\mu$ is the non-dimensional Reynolds number. For a given problem, Re will fully determine the solutions (except some cases where well-formulated boundary-value problems for the Navier-Stoke equations don't have unique solutions \cite{landau1959fluid}). Re is the ratio of inertial force to viscous force, which indicates the relative importance of the viscous term $ \Delta \boldsymbol{u}^*$ to the inertial term $ (\boldsymbol{u}^* \cdot \nabla^*) \boldsymbol{u}^*$, Re $\sim $ inertial forces$/$viscous forces. The typical scales of inertial and viscous terms can be expressed in dimensional forms as $ |\rho(\boldsymbol{u} \cdot \nabla) \boldsymbol{u}| \sim \rho U^2/L $, and  $ |\mu \nabla^2 \boldsymbol{u}| \sim \mu U/L^2  $. 

When Re $\ll$ 1, the approximate form of the equation of motion is obtained by dropping terms on the LHS, which are of order unity. The pressure term, regardless of its magnitude, must be retained to match the number of variables and equations in Eq. \ref{eq:navier-stokes}.
\begin{equation*}
   0 = \frac{1}{\mathrm{Re}} \left[ -\nabla^* p^* + \nabla^{*2} \boldsymbol{u}^*  \right].
\end{equation*}
Combining with the continuity equation, we have what are referred together as the Stokes equations for incompressible flow:
\begin{equation} \label{eq:stokes}
    \nabla p - \mu\nabla^2 \boldsymbol{u} = 0\;,\;\; \nabla \cdot \boldsymbol{u} = 0. 
\end{equation}
Stokes equations describe fluid flow at Re $\ll 1$. Given a boundary condition, the solution of Stokes equations is unique \cite{acheson1991elementary}. In addition, because of the linearity in $\boldsymbol{u}$ and $p$, if such boundary condition is reversed, the reverse of the solution ( $-\boldsymbol{u}$ and $-p+c(t)$ where $c$ is a function of time only) is also a solution that uniquely solves the reversed boundary condition. This behavior is called \emph{ kinematic reversibility} \cite{taylor1967film,stone2017fundamentals}. Flow in microscale networks typically fall into the low Re dynamical regime \cite{squires2005microfluidics,whitesides2006origins}. 

In macroscale networks, Re $ \gg 1$ is more relevant. Rescaling the non-dimensional pressure as $ p^* = p/\rho U^2 $, the non-dimensional version of the momentum equation is then
\begin{equation}
        \frac{\partial \boldsymbol{u}^*}{\partial t^*} + (\boldsymbol{u}^* \cdot \nabla^*) \boldsymbol{u}^*  =  -\nabla^* p^* + \frac{1}{\mathrm{Re}} \nabla^{*2} \boldsymbol{u}^*  
\end{equation}

 At high Reynolds number, the viscous force term $ (1/\mathrm{Re}) \nabla^{*2} \boldsymbol{u}^*$ is small compared to other terms containing $\boldsymbol{u}^*$. Thus we can neglect  $(1/\mathrm{Re}) \nabla^{*2} \boldsymbol{u}^*$ and the pressure term must be retained as before. Putting dimensions back, we obtain the Euler's equations of inviscid, incompressible flow:
 \begin{equation} \label{eq:euler}
        \frac{\partial \boldsymbol{u}}{\partial t} + (\boldsymbol{u} \cdot \nabla) \boldsymbol{u}  =  -\nabla p \;,\;\; \nabla \cdot \boldsymbol{u} = 0. 
\end{equation}
 Euler's equations are an approximation to the full Navier-Stokes equations at Re $\gg 1$. The approximation reduces the differential equation to first order in space and time. The number of boundary conditions must also be reduced accordingly. For flow near a solid boundary, there are two boundary conditions: the no-penetration condition $ \boldsymbol{u} \cdot \boldsymbol{n} =0 $ (no flow normal to the boundary, where  $\boldsymbol{n}$ is the normal vector) and the no-slip condition $ \boldsymbol{u} \cdot \boldsymbol{t} =0 $ (no flow tangent to the boundary, where $\boldsymbol{t}$ is the tangent vector). Since Eq. \ref{eq:euler} is first order, one condition must be dropped to avoid over-constraint. Since the no-slip condition is due to viscosity, it can be dropped to be consistent with the disappearance of viscosity in Eq. \ref{eq:euler}. However, the no-slip boundary condition has been found to be empirically ever present and can not be discarded \cite{tritton2012physical}. Even though Euler's equation becomes a better approximation in the undisturbed bulk of the fluid as the Reynolds number increases, viscosity cannot be neglected near a boundary and acts to satisfy the no-slip boundary condition no matter how high the Reynolds number. Therefore, the viscous term $ \mu \nabla^2 \boldsymbol{u}$ is always important, and the full governing Eq.  \ref{eq:navier-stokes} must be considered.
 
 Flow in macrofluidic networks typically occur at Re $ \gg 1$, and while inertial effects dominate, the role of viscosity is always significant in the presence of boundaries. The equation of motion Eq. \ref{eq:navier-stokes} is non-linear in $\boldsymbol{u}$ and is kinematically irreversible (more on this later). This leads to many phenomena relevant to macrofluidic networks.

\section{Physics of flow in macrofluidic network}
The kinematic viscosity define as $\nu = \rho / \mu$, thus $\mathrm{Re}  = UL/\nu $. For water and air at common pressure and temperature are respectively $8.9 \times 10^{-3}\; \mathrm{cm}^2 \cdot \mathrm{s}^{-1}$ and $1.5 \times 10^{-1}\; \mathrm{cm}^2 \cdot \mathrm{s}^{-1}$ \cite{tritton2012physical}. In either fluid, the characteristic length $L$ and velocity $U$ need to be just a few cm and $\mathrm{cm}/$s to reach moderately high Reynolds number. 
Flows in macrofluidic networks are typically described by high Reynolds number flows interacting with the network geometry. A much wider range of phenomena occur at high than at low Reynolds number. In the following, the relevant physics of flow in macrofluidics will be reviewed.

\subsection{Inviscid flow approximations}

As previously reviewed, in regions far away from and undisturbed by what happens at the boundary, inviscid flow theory governed by Euler's equations \ref{eq:euler} accurately describes the flow. The same can be said about Bernoulli's equation which is a result of inviscid flow theory. The inertial force term can be re-expressed using the vector calculus  identity $ (\boldsymbol{u} \cdot \nabla ) \boldsymbol{u} = (\nabla \times \boldsymbol{u}) \times \boldsymbol{u} + \nabla (\boldsymbol{u}^2/2) $ \cite{batchelor2000introduction} and Euler's equations are recast as 
  \begin{equation} \label{eq:ch2:euler split}
     \rho \left[ \frac{\partial \boldsymbol{u} }{\partial t } +  (\nabla \times \boldsymbol{u}) \times \boldsymbol{u} \right]   =  -\nabla (p + \frac{1}{2} \rho \boldsymbol{u}^2) \;, \; \nabla \cdot \boldsymbol{u} =0.
\end{equation}
In steady flow or when the time derivative of $\boldsymbol{u}$ is negligible, Euler's equation becomes
 \begin{equation}\label{eq:ch2:rotational euler}
          \rho  (\nabla \times \boldsymbol{u}) \times \boldsymbol{u}   =  -\nabla H, 
\end{equation}
where $H = p + \frac{1}{2} \rho \boldsymbol{u}^2 $ is the internal energy density of the fluid. Taking the dot product with $\boldsymbol{u}$ on both sides of the equation, we obtain
\begin{equation}
    (\boldsymbol{u} \cdot \nabla ) H = 0.
\end{equation}
 The above equation is known as Bernoulli's principle which states that, for steady inviscid flow, $H$ is a constant for a fluid parcel following a streamline. In other words, energy is conserved when viscous dissipation is not present, and there is no work done. Thus for two points along a streamline, $ p_1 + \frac{1}{2} \rho \boldsymbol{u}_1^2 = p_2 + \frac{1}{2} \rho \boldsymbol{u}_2^2$. If the flow is irrotational, i.e. $\nabla \times \boldsymbol{u} =0$ everywhere, the LHS of Eq. (\ref{eq:ch2:rotational euler}) vanishes, and  $H$ is a constant across streamlines \cite{batchelor2000introduction}.  
 
 Another important result of inviscid flow theory is the Kelvin circulation theorem, which states that the circulation around a loop $C$ consisting of the same fluid particles is conserved. 
 \begin{equation}
     \frac{\mathrm{D}}{\mathrm{D}t} \oint_C \boldsymbol{u} \cdot \mathrm{d}l = 0,
 \end{equation}
 where $ \mathrm{D}/\mathrm{D}t = \partial/ \partial t + \boldsymbol{u} \cdot  \nabla $ is the derivative following the fluid motion (or material derivative). If fluid is set into motion from rest, the circulation will remain zero for all time \cite{batchelor2000introduction}. 


\subsection{Kinematic Irreversibility}
We briefly discussed in Section \ref{ch2:section:governing eqs} that flows at low Reynolds number or in microfluidics, which governed by linear equations, are kinematically reversible. As we shall see below, that is not the case in macrofluidics.

 In 3D Cartesian coordinates with  $\boldsymbol{u} = (u, v, w) $, the governing equation  \ref{eq:navier-stokes} can be expressed explicitly in 3 spatial components $x$, $y$, and $z$ as  
\begin{equation} \label{eq:NS x component}
   \rho \left[ \frac{\partial u}{\partial t} + u \frac{\partial u}{ \partial x} + v \frac{\partial u}{ \partial y} + w \frac{\partial u}{ \partial z} \right] =  - \frac{\partial p}{ \partial x} + \mu  \left[  \frac{\partial^2 u}{ \partial x^2} +  \frac{\partial^2 u}{ \partial y^2} + \frac{\partial^2 u}{ \partial z^2}  \right], \;\; \frac{\partial u}{ \partial x} +\frac{\partial v}{ \partial y} +\frac{\partial w}{ \partial z} =0,
\end{equation}
together with similar equation for $v$ and $w$. Let there be fluid in some region $V$ which is bounded by a closed surface S. Let $\boldsymbol{u}$ be given as $\boldsymbol{u} = \boldsymbol{f(x,t)}$ on the boundary $S$. Suppose unique solutions to Eq. \ref{eq:NS x component} satisfying the boundary condition  $  \Tilde{ \boldsymbol{u}}$ and $\Tilde{ p}$ exist. The reverse of the solutions are $- \Tilde{ \boldsymbol{u}}$ satisfying $-\Tilde{ \boldsymbol{u}} = -\boldsymbol{f(x,t)}$ on $S$ and $-\Tilde{ p} +c(t)$ where $c(t)$ is any function of time since pressure only appears as a spatial gradient. Plugging these in Eq. \ref{eq:NS x component},
\begin{equation} \label{eq:NS x component reverse}
   \rho \left[ -\frac{\partial \Tilde{u}}{\partial t} + \Tilde{u} \frac{\partial \Tilde{u}}{ \partial x} + \Tilde{v} \frac{\partial \Tilde{u}}{ \partial y} + \Tilde{w} \frac{\partial \Tilde{u}}{ \partial z} \right] =  + \frac{\partial \Tilde{p}}{ \partial x} - \mu  \left[  \frac{\partial^2 \Tilde{u}}{ \partial x^2} +  \frac{\partial^2 \Tilde{u}}{ \partial y^2} + \frac{\partial^2 \Tilde{u}}{ \partial z^2}  \right] \;, \;\; \frac{\partial \Tilde{u}}{ \partial x} +\frac{\partial \Tilde{v}}{ \partial y} +\frac{\partial \Tilde{w}}{ \partial z} =0,
\end{equation}
together with similar equations for $\Tilde{v}$ and $\Tilde{w}$. These are formally different from Eq. \ref{eq:NS x component} because of the presence of both linear terms ($\partial \boldsymbol{u} / \partial t $, $\mu \nabla^2 \boldsymbol{u}$) and non-linear terms ($ \rho \boldsymbol{u} \cdot \nabla \boldsymbol{ u}$). In general, because fundamental properties of the general solutions of the Navier-Stokes equations are not well understood, it can't be predetermined whether $- \Tilde{ \boldsymbol{u}}$ and $-\Tilde{ p}+c(t)$ would uniquely solve the equations. High Reynolds number flows  governed by the full Navier-Stokes equations are in general irreversible. Only in cases where there is a special symmetry, such as steady flow in a pipe, the LHS vanishes, and it can be said with certainty that $- \Tilde{ \boldsymbol{u}}$ and $-\Tilde{ p}+c(t)$ also solve Eq. \ref{eq:NS x component}. In those cases, kinematic reversibility is granted. 

By the same token, Euler equations \ref{eq:euler}, which have both linear terms $\partial \boldsymbol{u} / \partial t $ and non-linear terms $ \rho \boldsymbol{u} \cdot \nabla \boldsymbol{ u}$ on the LHS of the momentum equation, are also irreversible under the transformation ($\boldsymbol{u},p) \mapsto (-\boldsymbol{u}, -p+c(t))$. A special case of Euler equations is when the flow of inviscid fluid is irrotational $ \nabla \times \boldsymbol{u} =0$, and Euler's equations \ref{eq:ch2:euler split} become:
 \begin{equation}
        \rho \frac{\partial \boldsymbol{u} }{\partial t }    =  -\nabla (p + \frac{1}{2} \rho \boldsymbol{u}^2) \;,\;   \nabla^2 \phi =0,
 \end{equation}
where $\phi$ is the velocity potential and $ \nabla \phi = \boldsymbol{u}$, hence the name potential flow. The velocity obeys Laplace's equation. Thus if $\boldsymbol{u}$ is a solution, $ -\boldsymbol{u}$ is also a solution, however the pressure still depends non-linearly on $\boldsymbol{u}$.

Note that kinematic irreversibility is a result of mathematical properties of the governing equations, which assume fluid is a continuum. It's different from  thermodynamical irreversibility (and time-irreversibility) which comes  from coarse-graining of microscopic processes such as Brownian motion of fluid molecules. In what follows, ``irreversibility" refers to kinematic irreversibility unless otherwise indicated.
The governing equations, and their properties (in 3 dimension) are summarized in Table \ref{table:all-eqs}. 

\begin{landscape}

\begin{table}
    \centering
   \begin{tabular}{ |c|c|>{\centering\arraybackslash}m{5cm}|>{\centering\arraybackslash}m{3cm}| } 
 \hline
 Equations & Initial-boundary value problem &  Applicability &  Under ($\boldsymbol{u},p) \mapsto (-\boldsymbol{u}, -p+c(t))$\\ 
 \hline
 Stokes & \begin{tabular}{@{}c@{}}$    \nabla p - \mu\nabla^2 \boldsymbol{u} = 0$ \\  $\nabla \cdot \boldsymbol{u} = 0 $, \\ with boundary conditions \end{tabular} &  Re $ \ll 1$. The solution is exact near rigid boundaries \cite{childress2009introduction}.   & Reversible  \\ 
 \hline
  Navier-Stokes  & \begin{tabular}{@{}c@{}} $ \rho \left[ \partial \boldsymbol{u}/\partial t + (\boldsymbol{u} \cdot \nabla) \boldsymbol{u} \right] =  -\nabla p + \mu \nabla^2 \boldsymbol{u}$  \\  $\nabla \cdot \boldsymbol{u} = 0$, \\ with initial and boundary conditions \end{tabular} &  Re $> 1$ & Irreversible  \\ 
 \hline
 Euler's &   \begin{tabular}{@{}c@{}} $\partial \boldsymbol{u}/\partial t + (\boldsymbol{u} \cdot \nabla) \boldsymbol{u}  =  -\nabla p$ \\ $\nabla \cdot \boldsymbol{u} = 0$, \\ with initial and boundary conditions \end{tabular}  &  Re$\gg 1$, in free flow regions \cite{batchelor2000introduction}. & Irreversible   \\ 
 \hline
\end{tabular}
    \caption{The governing equations of fluid flow at different dynamical regimes and kinematic (ir)reversibility}
    \label{table:all-eqs}
\end{table}

\end{landscape}

\subsection{Boundary layer and related phenomena}

As reviewed in Section \ref{ch2:section:governing eqs}, as Re increases, the viscous term $\mu \nabla^2 \boldsymbol{u}$ becomes negligible and we obtain Euler's equation. However because of the no-slip boundary condition we must retain the viscous term in the vicinity of the boundary. 
The region where viscosity is significant is known as the boundary layer, first introduced by Ludwig Prandtl in 1904. The flow develops an internal length scale, the boundary layer thickness $\delta$ that is much smaller than the imposed length scale $L$. In this region, the inertia term $ |\rho(\boldsymbol{u} \cdot \nabla) \boldsymbol{u}| \sim \rho U^2/L $ is of the same order as the viscous term  $ |\mu \Delta \boldsymbol{u}| \sim \mu U/\delta^2  $, and thus $\delta / L \sim 1/\sqrt{\mathrm{Re}}$. As Re increases, the disparity  between the two length scale gets larger.

Far from the boundary layer and regions disturbed by it, inviscid flow theory such as Euler's equations,  Bernoulli's principle, and Kelvin Circulation theorem are valid and describe most of the flow. Near the boundary however, the dynamics are much more complex and lead to phenonmenon such as boundary layer separation, vortex shedding, jets, and turbulence \cite{schlichting2016boundary}.





\section{Flow rectification with wall geometry: Asymmetric conduits}

An important aspect of fluid transport networks is controlling the direction of flow. In many natural and artificial settings, flow direction is ensured by valves that open and close, such as heart valves in animals, or various types of mechanical valves in engineering. Valves, which have moving parts, are effective in allowing fluid flow only in one direction while blocking the other completely, but they are not without limitations. For example, the moving parts, which can be delicate, are subjected to malfunction, wear and tear and in the case of mechanical valves, are harder to manufacture and maintain. In addition, valves might be limited in operating frequency, and the presence of moving parts might be inappropriate for certain fluids, such as fluid laden with sensitive particles. 
 Simpler methods for directing fluid traffic take advantage of fixed and asymmetric geometries. At the nanoscale, Brownian motions of suspended particles in an asymmetric channel could be rectified into macroscopic drifts by a time periodic perturbation \cite{magnasco1993forced,marquet2002rectified}. These channels are referred to as Brownian ratchets. When there are free surfaces, surface tension force could be used to transport small droplets of fluid in so-called capillary ratchets. For example, shorebirds use their long beaks as asymmetric capillary ratchets to move water upward \cite{Prakash931}, and Texas horned lizards have asymmetric scales that break the wetting symmetry and direct dew on their skin towards the snout \cite{comanns2015directional}. This strategy, which has inspired biomimetic microfluidics designs \cite{li2017topological,buchberger2018fluidic}. There are more types fluid transport using asymmetric geometries \cite{stroock2003fluidic,lagubeau2011leidenfrost}, but we focus the discussion on general networks where the flow physics is governed by the Navier-Stokes equations and its variations given in Table \ref{table:all-eqs}. At the microscale when Re $\ll1$ and viscous forces dominate, asymmetric geometry doesn't lead to transport as the governing Stokes equations are reversible, unless the fluid behaves non-linearly  \cite{groisman2004microfluidic}. In constrast, when Re $\gg 1$ as is common in macrofluidics, the governing equations are irreversible and asymmetric geometry could be exploited to direct flow. In an asymmetric conduit when the flow is in one direction, the fluid-boundary interaction is different from that of the reverse direction. This leads to differences in flow state and in hydraulic resistance.

The irreversible behavior of fast fluid flow was apparent to the famous inventor Nikola Tesla who more than 100 years ago was doing experiments in lower Manhattan, not far from the Applied Math Lab. Tesla was more famous for his work on electricity and AC-to-DC transformer but he dabbled in fluid mechanics as well. In 1920, Tesla patented a macrofluidic device that makes use of an asymmetric geometry \cite{tesla1920valvular}. The conduit, according to Tesla's claim, can function as a fluidic diode. It has inspired many inventions and studies of channels with asymmetric geometry to be used as flow control devices \cite{paul1969fluid,bardell2000diodicity,hampton2018fluidic,dyson2015flow,reed1993fluidic,gilbert2019three,thompson2014numerical,truong2003simulation,zhang2007performance,mohammadzadeh2013numerical,nobakht2013numerical,gamboa2005improvements,morris2003low,forster2002parametric,forster2007design,anagnostopoulos2005numerical,ansari2018flow,de2017design,abdelwahed2019reconstruction,lin2015topology,deng2010optimization,thompson2013transitional}. They fall under a class of ``Tesla-type'' channels. Most of these designs are planar periodic patterns, and have a directionality built into the geometry. 
Planar design are easier to design, manufacture and simulate.

A much simpler type of asymmetric channels that is also commonly used is the nozzle/diffuser type. They consist of straight walls that are tapered and while they exhibit weaker irreversibility, they nonetheless can be used to direct flows \cite{stemme1993valveless,forster1995design}. Other types of asymmetric channels have also been explored to a limited extent \cite{fadl2007experimental,nejat2013study,hawa2000viscous}. 

Existing literature on asymmetric conduits are mostly focused on practical engineering applications at microscales, with Re $\sim 10^2$. We are interested in the irreversible and unsteady dynamics which manifest stronger at larger scales, and so we will focus on the Tesla's valve as a case study in Chapters \ref{ch3} and \ref{ch4}. 


\subsection{Fluid resistance and diodicity}

Hydraulic resistance characterizes the pressure drop required to force a certain flow rate through the channel. The most common definition is analogous to that of  electric resistance which is the ratio of voltage to current, $R= \Delta p /Q$, where $\Delta p$ and $Q$ are the pressure drop across the channel and the corresponding flow rate, respectively. This definition of resistance is convenient in the linear approximation when $R$ is constant, and the fluidic circuit can be solved similarly to an electric circuit using Kirchhoff's rules. Another definition used in hydraulic engineering is the non-dimensional Darcy friction factor \cite{moody1944friction} $f_\mathrm{D} = (\Delta p /L) / ( \rho U^2/2D)  $ where $L$ is the length of the channel, $D$ its hydraulic diameter \cite{white1999fluid}, $\rho$ the fluid density, and $U$ the average speed in the channel. The Darcy friction factor is convenient for turbulent flows when it's fairly constant with the flow rate, and increases with irregularity (roughness) of the internal surface as shown in Moody's chart \cite{moody1944friction}. Finally, a less common definition of the hydraulic resistance is the ratio of dimensionless pressure, known as the Hagen number $\textrm{Hg}  =  \left(  \Delta p/L \right) \left( D^3 \rho / \mu^2 \right)$ \cite{martin2002generalized,tritton2012physical}, to the dimensionless flow rate or Reynolds number, $R= \mathrm{Hg}/\mathrm{Re}$. This fully non-dimensional expression characterizes resistance across all parameters: driving force, geometry, and properties of working fluid. We will take advantage of all three definitions. 

In an asymmetric channel, resistance depends on the flow direction and the designated forward direction generally has lower resistance than the reverse direction. Diodicity (coming from the word diode) quantifies this anisotropy, and is defined as the ratio of resistances in two directions, Di $= R_\mathrm{R}/R_{\mathrm{F}}$ where R and F stand for reverse and forward. It can be shown that Di is equal to the ratio of pressure drops at fixed flow rate, and is independent of how resistance is defined. When Re $ \ll 1$, Di $ =1$ regardless of the spatial asymmetry. When Re $ \gg 1$, generally Di $ >1$ if the geometry is asymmetric. For directing flows, the larger Di the better. 




\section{Flow rectification with network topology: unidirectional flow in avian respiratory system}

The most studied macrofluidic networks in nature are the cardiovascular and respiratory networks of animals. The blood flow in cardiovascular networks is driven by the heart. Periodic contractions and expansions of the heart provide alternating pressures, while the heart valves regulate the one-directional circulation of blood flow \cite{fung2013biomechanics-circulation}. The respiratory system of mammals, in particular the airways of the lower respiratory tract such as the lung has a tree architecture \cite{weibel1963morphometry}. While the structure is more complex, the flow pattern is straight forward. As the lung expands in volume, fresh air is drawn into every branch and to the alveoli, sac-like dead ends where gas is exchanged with the blood. As the lung contracts, air flows out in the reverse direction. In every branch, the flow is oscillatory. 

Unlike the pure AC flow in mammalian lungs, flow takes a distinctively different character in  avian lungs. Surprisingly, the topology of airways involve a looped  architecture \cite{hazelhoff1951structure,biggs1957new,dunker1971lung,duncker1972structure,duncker1974structure} as shown schematically in Fig. \ref{fig:ch2:lungphysio}. 
The oscillations of air sacs act as bellows and provide oscillating pressure, and flow in the main airways is AC, but flow in the loops is one directional (DC). Mysteriously, there are no valves involved and yet AC-to-DC conversion is achieved. Flows are directed  without anatomical valves as in the case of the circulatory system. The mechanism behind  unidirectional circulation is not well understood. In the following, the physiology of avian respiratory systems and observations of unidirectional flows will be reviewed. 

\subsection{Physiology of the avian respiratory system: Complex networks with loopy topology}
\begin{figure*}
\centering
\includegraphics[width=8cm]{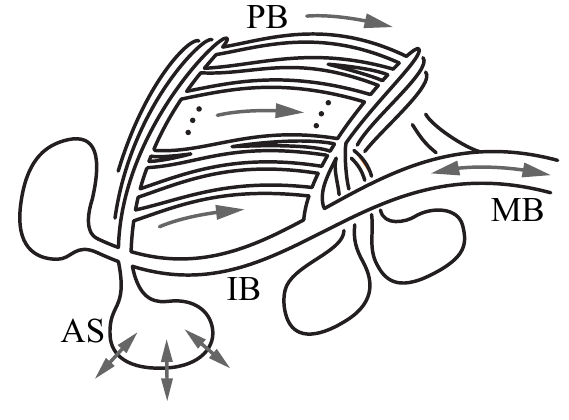}\vspace{-0.2cm}
\caption{Schematic of bird lung anatomy modified from \cite{scheid1972mechanisms}. Reciprocal expansion/contraction of air sacs (AS) drives oscillatory inhalation/exhalation flows in the main bronchus (MB). Directed flow is observed in the parabronchi (PB) that span the intrapulmonary bronchus (IB). }
\label{fig:ch2:lungphysio}\vspace{-0.4cm}
\end{figure*}

Among air-breathing vertebrates, the avian respiratory system 
is geometrically the most complex and functionally the most efficient gas exchanger \cite{maina2017pivotal}. A schematic is provided in Fig. \ref{fig:ch2:lungphysio}. The avian respiratory system is an open system that starts from the trachea and bifurcates into two main airways called primary or main bronchus (MB) before reaching each lung. Each primary bronchus branches off to about 10--15 smaller airways (secondary bronchi). The secondary bronchi can be considered in two roughly equal subgroups based on their location on the primary bronchi: one subgroup of 3--6 \cite{maina2009inspiratory} originating from the front portion of the primary bronchi (dorsobronchi), and the other subgroup consists of 6--10 originating from the back (ventrobronchi). The secondary bronchi reconnects front-to-back through hundreds of the third generation airway called parabronchi (PB). As the airways divide from first to third generation, they increase in number and decrease in size. Typical diameters of the airways are $\sim$1 cm for both the primary and secondary and $\sim$ 0.05--0.2 mm for the parabronchi \cite{king1966structural}. There are exceptions  to the description above, where a small portion of  parabronchi connect directly to the primary bronchus or the air sacs \cite{scheid1972mechanisms,casteleyn2018anatomy}  and the parabronchi can cross-connect within themselves. 

The primary and secondary bronchi simply transport air while the parabronchi performs all gas exchange. From parabronchi arise numerous air capillaries which divide repeatedly and reconnect with each other inside the wall of parabronchi. The system of veins and arteries enmeshes with the parabronchi, and gas exchange happens in the wall of parabronchi. It has been shown that the dominant mode of gas exchange is cross-current \cite{maina2017pivotal}. 
This is thought to contribute to the increased gas exchange efficiency of birds as  compared to mammals. 


Attached to the primary and secondary bronchi are the air sacs, of which there are 9 in most species. The air sacs comprise most of the volume of the respiratory system. During respiration, as the chest cavity contracts and expands, air sacs change volume synchronously \cite{brackenbury1973respiratory,king1975outlines}, providing oscillating pressure ( typical pressure of $\pm$ 1 cm \ch{H2O} \cite{powell2015respiration} and typical frequency of 1 Hz \cite{king1964normal}) that drive the system. Air sacs act purely as bellows and don't participate in gas exchange \cite{hazelhoff1951structure}. The air sacs are the compliant part of the lung while the rest of the lung is rigid. Between inhaling and exhaling cycles, the volume of the lung only changes by 1.4 \% \cite{jones1985lung}.

\subsection{Air flow in the avian respiratory system}
\subsubsection{Unidirectionality}

Air flow in the avian respiratory system has been measured in many studies. In all species, while the air flow in the primary and secondary bronchi is oscillatory \cite{brackenbury1971airflow,brackenbury1972physical,brackenbury1972lung,scheid1979mechanisms,fedde1980structure,maina2017pivotal}, it is unidirectional in the parabronchi \cite{bretz1970patterns}. The travel path of air is in the following order: the atmosphere, primary bronchi, secondary bronchi and air sacs in the back, parabronchi, front air sacs, front secondary bronchi, primary bronchi and back to the atmosphere. It takes 2 cycles of breathing for air to circulate through the system, and the flow in the parabronchi is back to front during both inhalation and exhalation \cite{brackenbury1971airflow,scheid1979mechanisms,fedde1980structure,powell2015respiration,kuethe1988fluid}.
To be more precise, the air flow is unidirectional in the paleopulmonary parabronchi. Some species have neopulmonary parabronchi where the air flow is oscillatory. This is not of great importance as they form a small part of the lung (up to 25\%, \cite{king1975outlines}), and not all species have these parabronchi. From here on, parabronchi refers to paleopulmonary parabronchi unless otherwise indicated.

\textit{Experimental evidence}. Directed flow exists in all birds, despite variances in anatomy and experimental conditions. Because of the  structural complexity of the lung, there have been many approaches and techniques to measuring the air flows inside. Flow patterns in the avian
lung have been inferred by deposition of aerosols, by measurement of tracer gas or respiratory gas concentration, by visualization of liquid flow, and by various flow meters \cite{butler1988inspiratory}. Such studies have been carried out in awake, anesthetized, and dead birds, and even in fixed lungs removed from the body. In all cases, unidirectional flow was observed. The earliest indication of unidirectional flow came from experiments with aerosols \cite{maina2017pivotal}. 
The deposition of aerosols (such as soots or charcoal powder) was used to infer flow directions. More direct measurements with flow meters implanted in the primary and secondary bronchi \cite{brackenbury1971airflow,bretz1971bird,scheid1971direct}
allows observation of the flow at  different times during the respiratory cycle. Less intrusive measurements of flow patterns make use of tracer gas (\ch{CO2} and \ch{O2}) concentration \cite{cohn1968respiration,banzett1987inspiratory,powell1981airflow}.
The flow patterns from these experiments agree well with flow meter measurements. The flow direction is further confirmed by three-dimensional reconstructions of the avian respiratory system and computer simulations \cite{casteleyn2018anatomy,maina2006development,maina2009inspiratory,moyes2005animal}.


\subsubsection{Fluid dynamical flow control}

\textit{Aerodynamic valving}. Early researchers 
hypothesized that, much like in the circulatory system, anatomical valves are responsible for controlling air flow \cite{maina2017pivotal} by opening and closing air ways at appropriate times. Despite the observed rectification phenomenon, there is no evidence for the existence of anatomical valves  \cite{king1966structural,maina2006development}.
 Valving action exhibits even in fixed lung \cite{scheid1972mechanisms} and  
 static model \cite{wang1988bird}. Cineradiography of airways during spontaneous breathing also shows no narrowing or closure of airways \cite{jones1981control}. They recognized that the mechanism must due to 
 inertial flows which is consistent with the typical Reynolds numbers in the lung Re  $\sim 10^2$ -- $10^3 \gg 1$. The mechanism, also called aerodynamic valving, has been shown to be more effective at higher flow rate and density of gas \cite{butler1988inspiratory,Brown1995the,wang1988bird,banzett1987inspiratory,kuethe1988fluid}
 both of which contribute to an increase in inertial forces $\sim  \rho U^2/L$ where $\rho$ is the density of gas, and $U$ is the  typical velocity and $D$ the typical airway diameter. It was further demonstrated that when air is blown into the trachea, the correct direction is only achieved at higher flow rate \cite{BRACKENBURY1979corrections}. 
 Following the consensus on aerodynamic valving, many fluid dynamical mechanisms have been suggested, from fixed to time-dependent anatomical geometries. 


\textit{Guiding dam}. It was initially posited that a vane-like structure (guiding dam) locates near the primary-secondary junction guides the flow. However, there seems to be no anatomical evidence for the existence of such a structure \cite{butler1988inspiratory}.

\textit{Constriction}. It was also hypothesized that the  constriction of the airway plays a role in the valving. In some species, a constriction exists on the primary bronchi before the first branching point \cite{wang1992aerodynamic,maina2000inspiratory}. The authors speculated that the constriction accelerates the flow of inspired air and thrusts the air past the secondary branches. Wang  \cite{wang1988bird} and Kuethe \cite{kuethe1988fluid} showed that the diameter of the primary bronchus before the first secondary bronchus was wider during exercise and narrower during rest, confirming that aerodynamic valving is more efficient at high velocities.  However, the constriction was not found in some species of birds.
Thus the presence of such constriction is not essential for aerodynamic valving.  


\textit{Air sacs}.  
The air sacs are not responsible directly for the valving mechanism (nor gas-exchange).
Rather, they are the compliant part of the lung and change volume as the body cavity expands and contracts and provide oscillatory pressure necessary to inhale and exhale air. Furthermore, the discovery of unidirectional airflow in some species of reptiles \cite{cieri2016unidirectional,farmer2015evolution} without air sacs rules out their role.

Other anatomical geometries such as diverging diameter of the primary bronchi \cite{brackenbury1972lung} and branching angles between the primary and secondary bronchi \cite{butler1988inspiratory} were also considered.

Mathematical modelling has mainly focused on the gas exchange in the parabronchi \cite{PIIPER1975gas,SCOTT2006flying}. 
Computational fluid dynamics of airflow in simplified geometries have limited success in producing unidirectional flow \cite{urushikubo2013effects,maina2009inspiratory}. Other mathematical models of the air flow, while successful in producing directed flow, are not fluid dynamical in nature nor based on the interaction of flow with the network geometry \cite{harvey2016robust}.

In summary, each suggested mechanism has been ruled out as either non-existent or not a prerequisite for flow rectification. There are no valves, sphincters, or guiding dams. Air sacs are not directly involved in valving. Constriction and specific bifurcation angle might enhance valving but are not essential. In addition, inspiratory valving and expiratory valving are studied as two different phenomena with their relative importance unexplored, and with much more focus on the former than the latter. It is unclear what are the essential ingredients for rectification.


\chapter{Tesla's 
fluidic diode and the electronic-hydraulic analogy} \label{ch2}



This chapter is adapted from the preprint version of Q. M. Nguyen, D. Huang, E. Zauderer, G. Romanelli, C. L. Meyer, and L. Ristroph, \href{https://doi.org/10.1119/10.0003395}{"Tesla's fuidic diode and the electronic-hydraulic analogy"}, submitted to American Journal of Physics (Jun 2020)\cite{nguyen2021tesla}. In this chapter, we take a pedagogical approach to study steady flows in the Tesla's valve. 

\section*{abstract}
Reasoning by analogy is powerful in physics for students and researchers alike, a case in point being electronics and hydraulics as analogous studies of electric currents and fluid flows. Around 100 years ago, Nikola Tesla proposed a flow control device intended to operate similarly to an electronic diode, allowing fluid to pass easily in one direction but providing high resistance in reverse. Here we use experimental tests of Tesla’s diode to illustrate principles of the electronic-hydraulic analogy. We design and construct a differential pressure chamber (akin to a battery) that is used to measure flow rate (current) and thus resistance of a given pipe or channel (circuit element). Our results prove the validity of Tesla’s device, whose anisotropic resistance derives from its asymmetric internal geometry interacting with high-inertia flows, as quantified by the Reynolds number (here, $\textrm{Re} \sim 10^3$). Through the design and testing of new fluidic diodes, we explore the limitations of the analogy and the challenges of shape optimization in fluid mechanics. We also provide materials  that may  be  incorporated  into  lesson plans  for  fluid  dynamics  courses, laboratory modules and further research projects.

\section{Introduction}

\begin{figure}
\centering
\includegraphics[width=16.5cm]{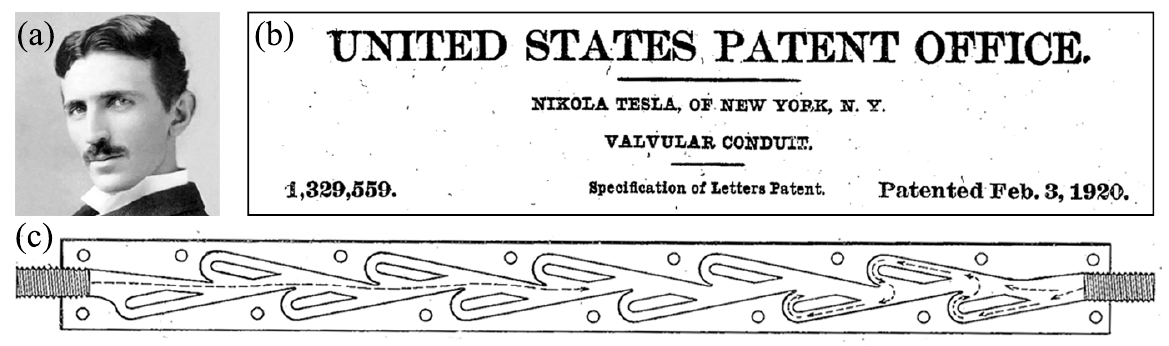}\vspace{-0.2cm}
\caption{Nikola Tesla's valvular conduit. (a) The inventor Nikola Tesla (1856-1943). (b) Title to Tesla's 1920 patent for the valvular conduit \cite{tesla1920valvular}. (c) Schematic showing the conduit's internal geometry. The channel is intended to allow fluid to flow from left to right (forward direction) with minimal resistance while providing large resistance to reverse flow. }
\label{fig:patent}
\end{figure}

Nikola Tesla is celebrated for his creativity and ingenuity in electricity and magnetism. Perhaps part of his genius lies in connecting ideas and concepts that do not at first glance appear related, and Tesla's writings and record of inventions suggest he reasoned by analogy quite fluidly. While he is best known for inventing the AC motor -- which transforms oscillating electric current into one-way mechanical motion -- Tesla also invented a lesser known device intended to convert oscillating fluid flows into one-way flows. Just around 100 years ago, and while living in New York City not far from our Applied Math Lab, Tesla patented what he termed a \textit{valvular conduit} \cite{tesla1920valvular}, as shown in Fig. \ref{fig:patent}. The heart of the device is a channel through which a fluid such as water or air can pass, and whose intricate and asymmetric internal geometry is intended to present strongly different resistances to flow in one direction versus the opposite direction. From his writing in the patent, Tesla has clear purposes for the device: To transform oscillations or pulsations, driven perhaps by a vibrating piston, to one-way motion either of the fluid itself (the whole system thus acting as a pump or compressor) or of a rotating mechanical component (\textit{i.e.} a rotary motor).

Whether called pumping, valving, rectification or AC-to-DC conversion, this operation is one example of the analogy between electrodynamics and hydrodynamics. And the electronic-hydraulic analogy is but one of many such parallels that show up in physics and across all fields of science and engineering. As emphasized by eminent physicists such as Maxwell and Feynman, reasoning by analogy is one of the powerful tools that allow scientists, having understood one system, to quickly make progress in understanding others \cite{feynmanvol2,maxwell,pask2003mathematics}. It is also a valuable tool in teaching challenging scientific concepts \cite{duit1991role}. Here, we explore this style of reasoning in the context of Tesla's invention, whose operation is analogous to what we now call an electronic \textit{diode}. In hydraulic terms, this device plays the role of a \textit{check valve}, which typically involves an internal moving element such as a ball to block the conduit against flow in the reverse direction. The practical appeal of Tesla's diode, in addition to its pedagogical value \cite{stith2019tesla}, is that it involves no moving parts and thus no components that wear, fail or need replacement \cite{tesla1920valvular}. 

In exploring the electronic-hydraulic analogy through Tesla's diode, we also provide pedagogical materials that may be incorporated into lesson plans for fluid dynamics courses and especially laboratory courses. For students of physics, electricity and electronics are often motivated by hydraulics \cite{smith1974electrical,bauman1980hydraulic,greenslade2003hydraulic,pfister2004illustrating,pfister2014sponge}, in which voltage is akin to pressure, current to flow rate, electrical resistance in a wire to fluidic resistance of a pipe, etc. However, in the modern physics curriculum, electrodynamics quickly outpaces hydrodynamics, and most students have better intuition for voltage than pressure! Lesson plans and laboratory modules based on this work would have the analogy work in reverse, \textit{i.e.} employ ideas from electronics to guide reasoning about Tesla's device, and thereby learn the basics of \textit{fluidics} or the control of fluid flows. As such, we present an experimental protocol for testing Tesla's diode practically, efficiently and inexpensively while also emphasizing accuracy of measurements and reproducibility of results. The feasibility of our protocol is vetted through the direct participation of undergraduate (DH and EZ) and high school (GR and CM) students in this research. We also suggest and explore avenues for further research, such as designing and testing new types of fluidic diodes.

\section{Tesla's device, proposed mechanism, efficacy and utility}

Tesla's patent is an engaging account of the motivations behind the device, its design, proposed mechanism and potential uses. In this section, we briefly summarize the patent and highlight some key points, quoting Tesla's words wherever possible \cite{tesla1920valvular}. The general application is towards a broad class of machinery in which ``fluid impulses are made to pass, more or less freely, through suitable channels or conduits in one direction while their return is effectively checked or entirely prevented.'' Conventional forms of such valves rely on ``carefully fitted members the precise relative movements of which are essential'' and any mechanical wear undermines their effectiveness. They also fail ``when the impulses are extremely sudden or rapid in succession and the fluid is highly heated or corrosive.'' Tesla aims to overcome these shortcomings through a device that carries out ``valvular action... without the use of moving parts.'' The key is an intricate but static internal geometry consisting of ``enlargements, recesses, projections, baffles or buckets which, while offering virtually no resistance to the passage of the fluid in one direction, other than surface friction, constitute an almost impassable barrier to its flow in the opposite sense." Figure \ref{fig:patent}(c) is a view of the channel internal geometry, where the fluid occupies the central corridor and the eleven ``buckets'' around the staggered array of solid partitions.

Without making any concrete claims to having investigated its mechanism, Tesla relates the function of the device to the character of the flows generated within the channel, as indicated by dashed arrows in Fig. \ref{fig:patent}(c). In the left-to-right or forward direction, the flow path is ``nearly straight''. However, if driven in reverse, the flow will ``not be smooth and continuous, but intermittent, the fluid being quickly deflected and reversed in direction, set in whirling motion, brought to rest and again accelerated.'' The high resistance is ascribed to these ``violent surges and eddies'' and especially the ``deviation through an angle of $180^\circ$'' of the flow around each ``bucket'' and ``another change of $180^\circ$... in each of the spaces between two adjacent buckets,'' as indicated by the arrows on the right of Fig. \ref{fig:patent}(c).

The effectiveness of the device can be quantified as ``the ratio of the two resistances offered to disturbed and undisturbed flow.'' Without directly stating that he constructed and tested the device, Tesla repeats a claim about its efficacy: ``The theoretical value of this ratio may be 200 or more;'' ``a coefficient approximating 200 can be obtained;'' and ``the resistance in reverse may be 200 times that in the normal direction.'' The experiments described below will directly assess this effectiveness or \textit{diodicity} \cite{forster1995design}.

Much of the remaining portions of Tesla's patent are devoted to example uses. The first is towards the ``application of the device to a fluid propelling machine, such as, a reciprocating pump or compressor.'' The second is intended to drive ``a fluid propelled rotary engine or turbine.'' The explanations and accompanying diagrams are quite involved, and at the end of this paper we offer our own applications that we think capture Tesla's intent in simpler contexts.

\section{Experimental method to test Tesla's diode}

\begin{figure}
\centering
\includegraphics[width=16.5cm]{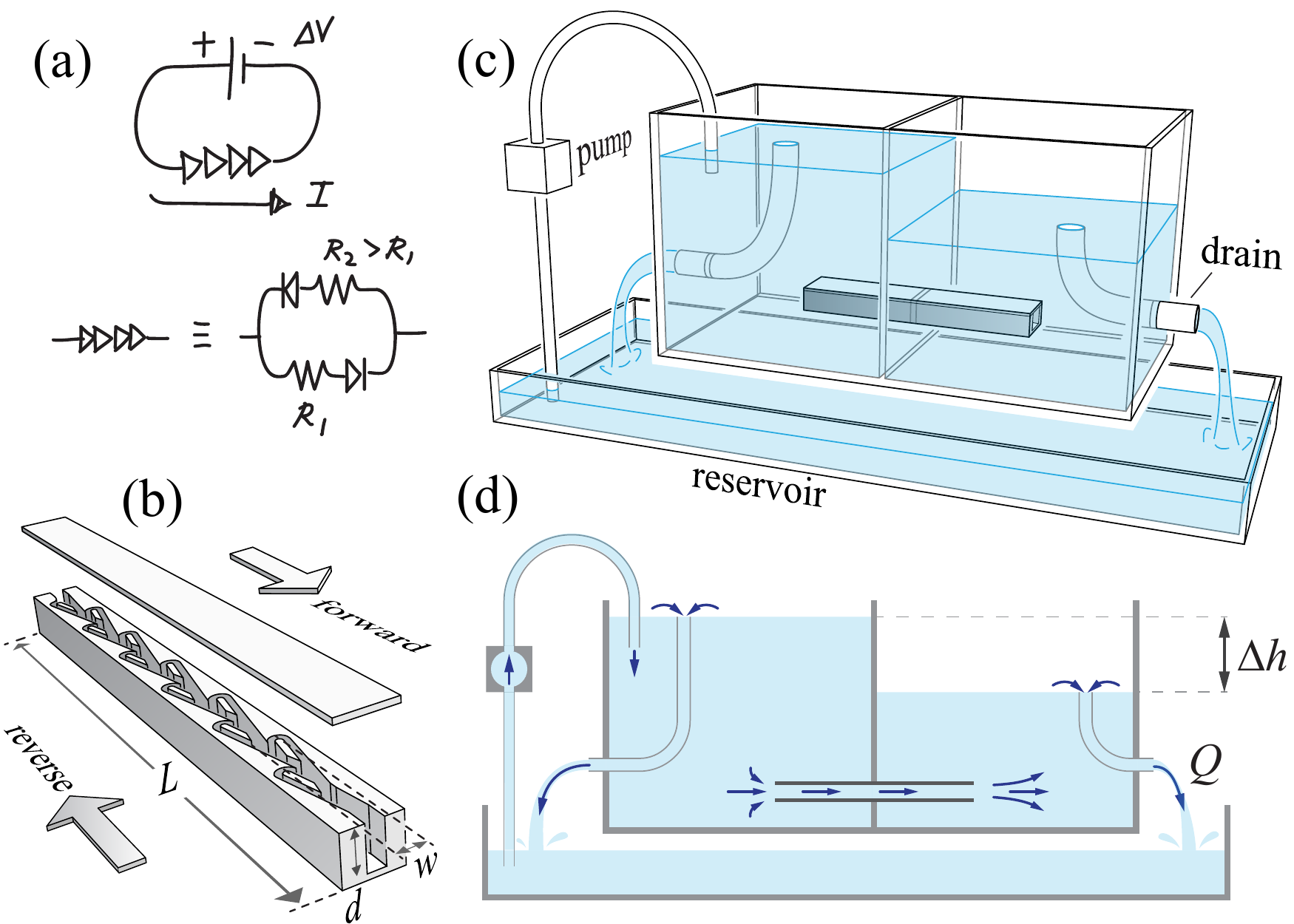}\vspace{-0.2cm}
\caption{Experimental apparatus for testing Tesla's diode. (a) Circuit analogs. The resistance of an unknown component is characterized by imposing a voltage difference and measuring current. A `leaky' diode may be represented in terms of resistors and ideal diodes. (b) Rendering of Tesla's conduit that can be 3D printed or laser cut. Relevant dimensions include total length $L$, depth or height $d$, and average wetted width $w$ (c) Perspective view of experimental apparatus. Two upper chambers are connected only via the conduit to be tested, and both overflow through internal drains to a lower reservoir. The water level of the higher is maintained by an overspill mechanism with a pump. Flow through the channel into the lower chamber induces overflow into the reservoir. (d) Schematic of pressure source chamber. The height difference $\Delta h$ is set by variable-height drains, and the resulting volume flow rate $Q$ is measured. }
\label{fig:setup}
\end{figure}

It is unclear whether Tesla ever constructed and tested a prototype, and this fact is obscured by the vague language used in the patent \cite{tesla1920valvular}. In any case, no data are provided. In the 100 years since, there has been much research into modified versions of Tesla's conduit
\cite{paul1969fluid,bardell2000diodicity,hampton2018fluidic,dyson2015flow,reed1993fluidic,gilbert2019three,thompson2014numerical,truong2003simulation,zhang2007performance,mohammadzadeh2013numerical,nobakht2013numerical,gamboa2005improvements,morris2003low,forster2002parametric,forster2007design,anagnostopoulos2005numerical,ansari2018flow,de2017design,abdelwahed2019reconstruction,lin2015topology,deng2010optimization,thompson2013transitional} and asymmetric channels generally for micro- and macro-fluidic applications \cite{forster1995design,stemme1993valveless,izzo2007modeling,groisman2004microfluidic,thiria2015ratcheting,lin2018passive,nguyen2008improvement,sousa2010efficient,fadl2007experimental,chen2010high}. However, to our knowledge, there are no studies on conduits faithful to Tesla's original geometry. Here we use modern rapid prototyping techniques to manufacture such a channel, and we outline an experimental characterization of its hydraulic resistance that uses everyday instruments like rulers, beakers and stopwatches to yield high-precision measurements.

We start with motivation from the electronic-hydraulic analogy. Suppose we have a circuit element of unknown and possibly anisotropic resistance, which we give the symbol of four arrowheads in Fig. \ref{fig:setup}(a). To characterize the element, we wish to impose a voltage difference $\Delta V$ using a battery or voltage source and measure the resulting current $I$, perhaps using an ammeter (not shown). The resistance is then given by Ohm's law $R=\Delta V / I$. Following the usual analogy, we wish to impose a pressure difference $\Delta p$ across the conduit, measure the resulting volume flow rate $Q$ and thus infer the resistance $R=\Delta p / Q$.

The conduit plays the role of the unknown element and is readily manufactured thanks to the modern convenience of rapid prototyping. We first digitize the channel geometry directly from the patent to arrive at a vector graphics file, and we have tested both 3D-printed and laser-cut realizations. Having achieved highly reproducible results on the latter, here we report on a design cut from clear acrylic sheet, a rendering of which is shown in Fig. \ref{fig:setup}(b). The channel tested has height or depth $d = 1.9$ cm, length $L = 30$ cm and average wetted width $w=0.9$ cm, and its planform geometry faithfully reflects Tesla's design. Gluing a top using acrylic solvent ensures a waterproof seal. If the desired channel is deeper than the maximum thickness permitted for a given laser-cutter, several copies may be cut and bonded together in a stack. Channels may also be 3D-printed, in which case waterproofing can be achieved by painting the interior with acrylic solvent to seal gaps between printed layers.

What serves the function of a hydraulic battery? We desire a means for producing a pressure difference across the channel, thereby driving a flow, and it is natural to employ columns of water as sources of hydrostatic pressure. If the channel bridges two columns of different heights, water flows through the channel from the higher to the lower. The challenge is to achieve an \textit{ideal pressure source} (akin to an ideal voltage source) that maintains the heights and thus pressures even as flows drain from one and into the other. This is accomplished using overflow mechanisms, as detailed in the experimental apparatus of Fig. \ref{fig:setup}(c). Two water-filled chambers are connected only by the channel being tested. Each chamber has an internal drain that can be precisely positioned vertically via a translation stage. The heights of the drains set the water levels and thus the flow direction, which can be reversed by switching which drain is higher. The draining of the high side through the channel is compensated by a pump that draws from a reservoir; the pump is always run sufficiently fast so as to just overflow the chamber and thus maintain its level. The lower side is fed only by the flow from the channel, and hence the flow through its drain and out the side to the reservoir represents the flow through the channel itself. The whole system is closed and can be run indefinitely.

A sectional view of the apparatus is shown in Fig. \ref{fig:setup}(d). A desired difference in water heights $\Delta h$ can be obtained by adjusting the drain heights. Considering hydrostatic pressure, we argue that the pressure difference across the channel is $\Delta p = \rho g \Delta h$, where $\rho = 1.0 ~\textrm{g}/\textrm{cm}^3$ is the density of water and $g = 980 ~\textrm{cm}/\textrm{s}^2$ is gravitational acceleration. (We use the centimeter-gram-second or CGS system of units throughout this work, as it proves convenient given the experimental scales.) A more thorough analysis of the pressures is detailed in Section VI. The resulting volume rate $Q$ of water flowing from high to low through the channel can then be measured by timing with a stopwatch the filling of a beaker of known volume that intercepts the flow exiting from the lower chamber. Large vessels and consequently long measurement times ensure highly accurate results, and measurements may be repeated to ensure reproducibility. Resistance can then be calculated as $R=\Delta p / Q$. Importantly, the experimental scales and working fluid are chosen to achieve strongly inertial flows, as quantified by the Reynolds number $\textrm{Re}$ introduced in Section V.

\section{Resistance measurements and the leaky diode}

\begin{figure}
\centering
\includegraphics[width=16.5cm]{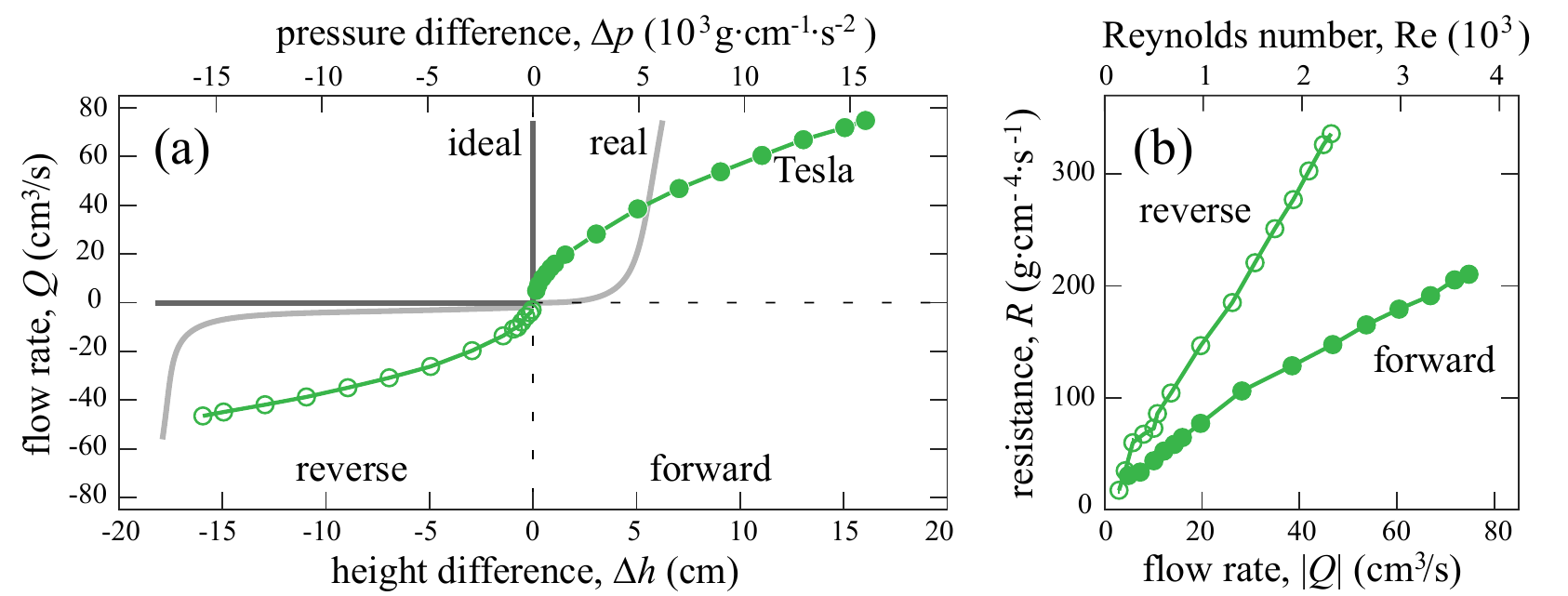}\vspace{-0.2cm}
\caption{Experimental tests of Tesla's conduit. (a) Green markers and curves indicate measured flow rate $Q$ across varying differences in water column heights $\Delta h$ and thus differential pressure $\Delta p$. The forward direction (filled symbols) permits greater flow rates for the same $\Delta h \propto \Delta p$. Curves are also shown for the behavior of an ideal diode (dark gray) and a typical real electronic diode (light gray). (b) Inferred resistance $R=\Delta p / Q$ versus $|Q|$ for flow in both directions. The conduit acts as a leaky diode with higher but finite resistance in the reverse direction. Error bars are smaller than the symbol size.}
\label{fig:teslaplot}
\end{figure}

In Fig. \ref{fig:teslaplot}(a), we present as the green markers and curves measurements of flow rate $Q$ for varying height differential $\Delta h$. Here $\Delta h > 0$ corresponds to $Q>0$ or flow in the forward or `easy' direction (filled markers), while the reverse or `hard' direction corresponds to $\Delta h < 0$ and $Q<0$ (open markers). As might be expected, increasing the height differential yields higher magnitudes of flow rate in both cases. The absolute flow rate $|Q|$ increases monotonically but nonlinearly with $|\Delta h|$ for flow in both directions. More important but more subtle is that, for the same $|\Delta h|$, the values of $|Q|$ differ for forward versus reverse, the former being greater than the latter across all values $|\Delta h|$. This anisotropy is more clearly seen in Fig. \ref{fig:teslaplot}(b), where the resistance $R=\Delta p / Q$ is plotted versus $|Q|$ for the forward and reverse cases. Across all values of $|Q|$ explored here, the resistance in the reverse direction is higher than that of the forward direction.

A note about errors: The flow rate $Q$ is determined by triggering a stopwatch when a given volume of liquid is collected, and so errors are set by the human visual reaction time ($\sim 0.3$ s). Relative errors under 1\% are achieved simply with long collection times ($>30$ s). The differential height $\Delta h$ is determined by visually reading the water column heights on vertical rulers in each chamber, and so errors are set by the height of the meniscus ($\sim 1$ mm). For a typical $\Delta h$ of 10 cm, the errors are about 1\%. We suppress error bars in Fig. \ref{fig:teslaplot} and elsewhere when they are smaller than the symbol size.

The data of Fig. \ref{fig:teslaplot} provide direct experimental validation for Tesla's main claim of anisotropic resistance. But, at least for the conditions studied here, the ratio of hard to easy resistances is far less than that reported in Tesla's patent, being closer to 2 times rather than 200 times. This factor is quantified and compared against other channel designs in Section VIII, and we discuss possible reasons for this discrepancy in Section IX.

Returning to the electronic-hydraulic analogy, our results suggest that the conduit acts as a \textit{leaky diode}. To appreciate this term, it is useful to compare our results against the performance of ideal and real electronic diodes, as shown in Fig. \ref{fig:teslaplot}(a). An ideal diode offers no resistance in the forward direction and thus $Q$ is infinite for all $\Delta h > 0$. It has infinite resistance in reverse and thus $Q=0$ for all $\Delta h < 0$. A real electronic diode typically requires a finite voltage to `turn on' in the forward direction, has some small leakage current in reverse, and breaks down for very large reverse voltages. Our measurements indicate that Tesla's conduit deviates in all such features, but a common trait is the leakage in reverse, which is quite substantial for the conditions studied here.

These results can be summarized by the representation of the leaky diode as shown in Fig. \ref{fig:setup}(a). Symbolized as four arrowheads, it is equivalent to a parallel pair of resistors and ideal diodes, themselves arranged in series within each pair. Positive voltages drive current through a forward resistance $R_1$, and negative voltages drive lower current through a higher reverse resistance $R_2 > R_1$. The analogy is made more exact if the resistance values are functions of current.

\section{Irreversibility of high Reynolds number flows}

The fundamental characteristic of Tesla's device borne out by the above measurements is that, when the applied pressures are reversed, the flows do not simply follow suit by reversing as well. Rather, substantially different flow rates result, and presumably all details of the forward versus reverse flows through the conduit differ as well. This is a manifestation of \textit{irreversibility}, a property that arises more generally in many physical contexts \cite{gollub2006microscopic,hollinger2012nature}. Flow irreversibility was anticipated by Tesla, whose drawing reproduced in Fig. \ref{fig:patent}(c) shows a rather straight trajectory (dashed line and arrow) down the central corridor for the forward direction and a more circuitous route around the islands for the reverse direction. This outcome could be viewed as unsurprising; after all, the channel is clearly asymmetric or directional. For those new to fluid mechanics, it may be counter-intuitive that there exist conditions for which flows are exactly reversible even for asymmetric geometries, implying equality of the forward and reverse resistance values. (Tesla may not have been aware of this.) Fluid dynamical (ir)reversibility can be derived from the governing Navier-Stokes equation of fluid dynamics, an analysis taken up elsewhere \cite{stone2004engineering,stone2017fundamentals}. Its central importance to the function of Tesla's device warrants a brief overview of known results.

While other forces may participate in various situations, three effects are intrinsic to fluid motion: pressure, inertia and viscosity. It is useful to think of flows as being generated by pressure differences overcoming the inertia of the dense medium and its viscous resistance. In well known results for laminar flows that can be found in fluid mechanics textbooks \cite{tritton2012physical}, the inertial pressure scales as $\rho U^2$ and the viscous stress as $\mu U/L$, where $\rho$ is the fluid density, $\mu$ its viscosity, $U$ a typical velocity, and $L$ a relevant length scale. The relative importance of inertia to viscosity can be assessed by their ratio, which is the dimensionless Reynolds number $\textrm{Re} = \rho U^2 / (\mu U / L) = \rho U L /\mu$. In the low Reynolds number regime of $\textrm{Re} \ll 1$, inertial effects can be ignored, and the resulting linear Stokes equation is reversible \cite{tritton2012physical,stone2017fundamentals}. Qualitatively, viscosity causes flows to stick to solid boundaries and conform to identical paths for forward and reverse directions. This general property of viscous flows has many important consequences and is beautifully demonstrated by stirring a viscous fluid and then `unstirring' with precisely reversed motions, causing a dispersed dye to recollect into its original form \cite{taylor1967film,fonda2017unmixing}. When Re is not small, inertial effects participate and flows are governed by the full Navier-Stokes equation, whose nonlinearity leads to irreversibility \cite{tritton2012physical}. Qualitatively, inertia allows flows to depart or separate from solid surfaces, this tendency being sensitive to geometry and thus directionality. Among many other phenomena, this relates to the observation that one can blow out but not suck out a candle, the flows being markedly different under the reversal of pressures.  

For computing Reynolds numbers for pipe flows, it is customary to set $U$ as the average flow speed and to use the diameter $D$ (or a corresponding dimension for non-circular conduits) as the length scale. Saving a deeper discussion of these quantities for Section VII, the parameters in our experiments yield $\textrm{Re} = \rho U D /\mu \sim 10^2-10^4$, as reported on the upper axis of Fig. \ref{fig:teslaplot}(b). Such high values of Re indicate that the flows are strongly inertial and thus irreversible, which is consistent with the different forward versus reverse resistance values reported here. The directional dependence of high-Re flows has been observed previously in computational fluid dynamics simulations for modified forms of Tesla's channel \cite{bardell2000diodicity,nobakht2013numerical,anagnostopoulos2005numerical,zhang2007performance,thompson2013transitional}. For low Re, flow reversibility has been confirmed by experimental visualization \cite{stone2017fundamentals}, and future work should verify the symmetric resistance expected in this regime.

\section{Analysis of pressures in two-chamber system}

In the above analysis of the experimental data, we have assumed that our two-chamber apparatus imposes $\Delta p = \rho g \Delta h$ across the channel. This formula strictly represents the gravitational \textit{hydrostatic} pressure difference between liquid columns, whereas the fluid is in motion throughout our system, so what justifies its application here? The short answer is that the flows in the chambers outside of the channel are ``slow enough'' to safely ignore velocity-dependent pressures. The long answer is provided in this section, where we analyze the contributing pressures in the system via Bernoulli's law. This analysis also sets up the following section by showing that Bernoulli's law is violated in the channel itself, requiring a characterization of friction or dissipation. 



Bernoulli's law is a statement of conservation of energy for steady flows of an inviscid (zero-viscosity) fluid \cite{tritton2012physical}. Of course, real fluids like water have finite viscosity, and later we discuss the validity of this approximation for the flows in the chambers. Following streamlines of the flow, the pressure, the gravitational potential energy density (\textit{i.e.} energy per unit volume), and the kinetic energy density must each change in a way such that their sum (total energy density) is unchanged: $H = p + \rho g z + \frac{1}{2} \rho U^2$ is constant. Here $p$ is the pressure, which can be thought of as a measure of internal energy density, $z$ is the vertical coordinate, $U$ is the speed at any location along a streamline, and $H$ is the total energy density. 

\begin{figure}
\centering
\includegraphics[width=8cm]{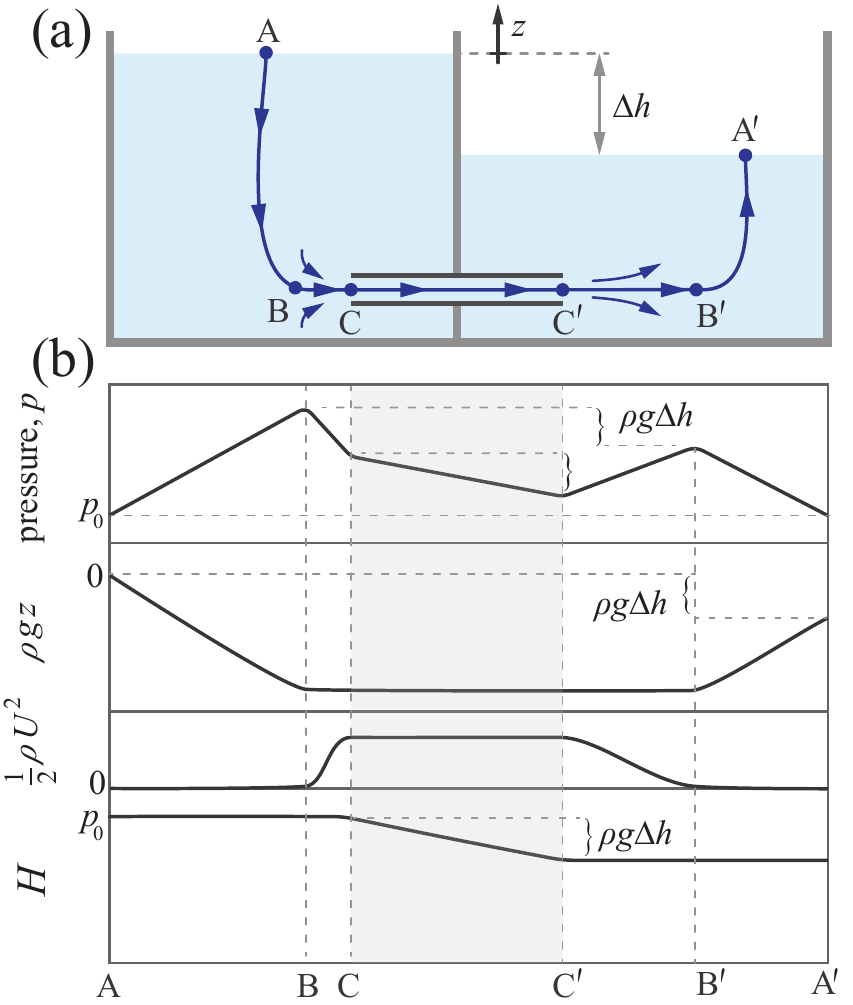}\vspace{-0.2cm}
\caption{An analysis of the two-chamber system guided by Bernoulli's law. (a) Chambers of high and low water levels are connected only by a channel at some depth. A flow streamline ABCC$'$B$'$A$'$ starts at the surface of the higher column, enters the channel, and emerges in the lower column where it meets the surface. (b) Hypothetical data showing how pressure $p$, gravitational potential energy density $\rho g z$, and flow kinetic energy density $\frac{1}{2} \rho U^2$ and total energy density $H= p + \rho g z + \frac{1}{2} \rho U^2$ vary along the streamline.}
\label{fig:bernoulli}
\end{figure}

In Fig. \ref{fig:bernoulli}(a) we examine a hypothetical streamline assumed to originate at the surface of the high-level chamber, descend to the channel opening, transit through the channel and out into the low-level chamber, and finally up to the surface. In its descent from the point marked A at the free surface to the point B somewhat upstream of the channel inlet, it is reasonable to assume that flow speed remains always small $U\approx 0$ and so too does the kinetic energy (more on this approximation later). In this case, the primary energy exchange involves a drop in the gravitational term $\rho g z$ and a consequent rise in $p$, as shown in the segment A-B in Fig. \ref{fig:bernoulli}(b), which tracks the terms of Bernoulli's law with hypothetical data.

A similar exchange occurs for the end segment B$'$-A$'$. Because the free surface points A and A$'$ are both at atmospheric pressure $p_0$, we conclude that pressure difference between points B and B$'$ is $\rho g \Delta h$. The regions B-to-C just before the inlet and C$'$-to-B$'$ just after the outlet are approximated as horizontal and so involve exchanges of pressure with kinetic energy only. The flow becomes faster from B-to-C as it is constricted and becomes slower C$'$-to-B$'$ as it spreads out, and these changes in speed $U$ must come with changes in pressure. However, the increase in speed from B-to-C would seem to be matched by the decrease from C$'$-to-B$'$, and so the pressure drop B-to-C is matched by the rise C$'$-to-B$'$. If true, then indeed the pressure drop across the channel C-to-C$'$ is $\Delta p = \rho g \Delta h$.

This conclusion rests on the assumptions that the fluid has negligibly small viscosity (so that Bernoulli's law may be used) and that the flow speeds outside of the channel are negligibly slow (so motion-dependent pressures may be ignored). Such approximations are statements about the \textit{relative} strengths of effects, which can be quantified by dimensionless numbers representing ratios of participating forces \cite{tritton2012physical}. For the flows in the chambers of our device, we have not only the intrinsic effects of fluid inertia or kinetic energy and viscous stresses or dissipation but also gravitational pressure or potential energy. Following up on the force scales introduced in the preceding section, the associated stresses or energy densities are $\rho U^2$, $\mu U/L$ and $\rho g L$, respectively, where $\rho$ is the fluid density, $\mu$ its viscosity, $U$ a typical velocity, and $L$ a relevant length scale \cite{tritton2012physical}. Concerned only with orders of magnitude, $g \sim 10^3~ \textrm{cm}/\textrm{s}^2$, $\rho \sim 1~\textrm{g}/\textrm{cm}^3$ and $\mu = 10^{-2}~\textrm{dyn} \cdot \textrm{s} / \textrm{cm}^{2}$ for water, $L \sim 10$ cm is the chamber size, and $U = Q/L^2 \sim 0.1$ cm/s for the typical flow rates explored here. The gravitational-to-viscous ratio is then $\rho g L / (\mu U / L) = \rho g L^2 / \mu U \sim 10^8 \gg 1$, which certainly justifies the neglect of viscosity. (For those familiar with dimensionless numbers in fluid mechanics, this ratio can be expressed in terms of the Galileo and Reynolds numbers.) The gravitational-to-kinetic ratio is $\rho g L / \rho U^2 = g L / U^2 \sim 10^6 \gg 1$, which justifies the neglect of kinetic effects. (This ratio is related to the Froude number.) In essence, the pressure is very nearly balanced by the gravitational hydrostatic pressure throughout the chambers, with all other effects being many orders of magnitude smaller.

\section{Characterizing fluid friction in Tesla's conduit}

The above reasoning about pressures in the system indicates that Bernoulli's law is violated in the channel itself: The total energy density $H = p + \rho g z + \frac{1}{2} \rho U^2$ shown in the lowest panel of Fig. \ref{fig:bernoulli}(b) is not constant over the gray region C to C$'$. The gravitational energy density $\rho g z$ is unchanged over the horizontal length, and the kinetic energy density $\frac{1}{2} \rho U^2$ is unchanged due to mass conversation and thus uniformity of flow speed, so pressure varies without any variation in potential or kinetic energy. This is expected since the flow inside the channel is resisted due to viscosity, which dissipates energy and may trigger turbulence or unsteady flows, effects which are not accounted for in Bernoulli's law. The associated hydraulic resistance or friction is, of course, well studied in the engineering literature due to its practical importance, and in this section we apply established characterizations to our measurements on Tesla's channel. We also compare our findings to previous results on rough-walled pipes, which may serve as a crude (rough?) way to view Tesla's channel.

\begin{figure}
\centering
\includegraphics[width=7cm]{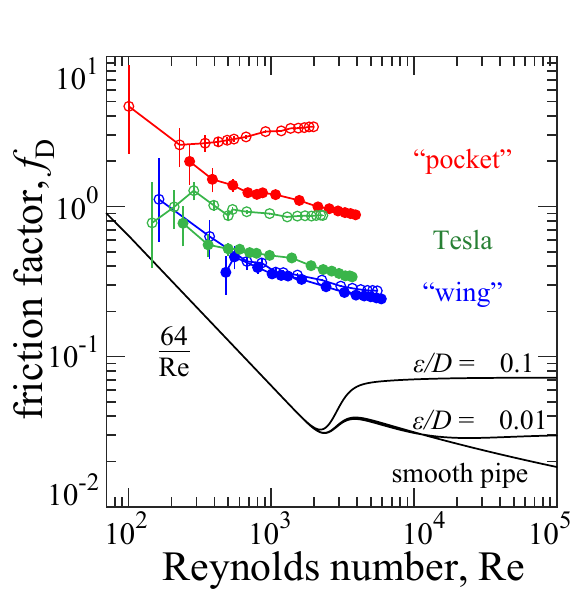}\vspace{-0.2cm}
\caption{Fluid friction factor versus Reynolds number (Moody diagram) for Tesla's channel and compared to smooth and rough pipes. Friction factors are plotted for forward and reverse flow through Tesla's channel (green) as well as two additional diode designs (red and blue) and circular pipes of different relative roughness $\varepsilon/D$ (black).}
\label{fig:moody}
\end{figure}


Hydraulic resistance depends on the conduit geometry as well as the Reynolds number, which assesses the relative importance of fluid inertia to viscosity. Following the discussions in previous sections, internal flows are characterized by $\textrm{Re} = \rho U D /\mu$, where $\rho$ and $\mu$ are the fluid density and viscosity, and $U$ is the average flow speed, and $D$ is the pipe diameter or a corresponding dimension in the case of non-circular cross-sections. For our realization of Tesla's channel, we use the so-called hydraulic diameter $D = 4V/S = 0.8$ cm, where $V$ is the total wetted volume of the conduit and $S$ is its total wetted surface area. (This is a generalization of the conventional form $D = 4A/P$ for a conduit whose cross-section shape is uniform and of wetted area $A$ and perimeter length $P$ \cite{white1999fluid}.) The average flow speed is $U = Q/A = 1 - 100 $ cm/s, where $A=wd$ is the average wetted area of the cross-section. These parameters yield $\textrm{Re} \sim 10^2-10^4$, as mentioned in Section V and reported on the upper axis of Fig. \ref{fig:teslaplot}(b).

A dimensionless measure of hydraulic resistance used often in engineering is the Darcy friction factor \cite{weisbach1845lehrbuch} $f_{\mathrm{D}} = (\Delta p/L)/(\frac{1}{2} \rho U^2/D  )$, where $L$ is the conduit length and $\Delta p$ is the pressure loss along the conduit or, equivalently for our set-up, the applied pressure difference. This choice of nondimensionalization has been shown to yield values of $f_{\mathrm{D}}$ that vary only weakly with Re for turbulent flow through long pipes \cite{moody1944friction,bellos2018friction}. The green markers and curves of Fig. \ref{fig:moody} represent the measured friction factors versus Reynolds number for forward and reverse flow through Tesla's channel, and data are included for two additional diode designs that will be introduced in Section VIII. For comparison, we include the so-called Moody diagram \cite{moody1944friction,bellos2018friction}, which is a log-log plot summarizing measurements of $f_{\mathrm{D}}(\textrm{Re})$ for circular pipes of varying degrees of wall roughness. Here the relative roughness $\varepsilon/D$ represents the ratio of typical surface deviations $\varepsilon$ to the mean diameter $D$. Smooth and rough pipes alike follow a well-known form of $f_{\mathrm{D}} = 64/\textrm{Re}$ in the laminar flow regime of $\textrm{Re} \lesssim 2 \times 10^3$. (This form can be derived from the Hagen-Poiseuille law for developed, laminar flow in cylindrical pipes \cite{tritton2012physical}.) At higher $\textrm{Re} \gtrsim 4 \times 10^3$, the flow tends to be turbulent, and $f_{\mathrm{D}}$ varies weakly with Re but increases with wall roughness.

Interestingly, the friction factors for flow through Tesla's conduit are far higher than those reported for smooth and rough pipes at comparable Re. This likely reflects the extreme degree of roughness ($\varepsilon/D \sim 1$, if such a quantity is at all meaningful) of the channel and consequent disturbances to the flow presented by its baffles and islands. It is clear that our results do not follow the form $f_{\mathrm{D}} = 64/\textrm{Re}$ for the range $\textrm{Re} \approx 200-2000$ explored here, nor is there any clear feature in the curves that would indicate a laminar-to-turbulent transition. Making sense of these observations, and better understanding the hydraulics of very rough channels generally \cite{gloss2010wall,liu2019roughness}, would benefit from further work that varies Re over a wider range and includes flow visualization.

\section{Alternative diode designs and comparison of diodicity}

\begin{figure}
\centering
\includegraphics[width=16.5cm]{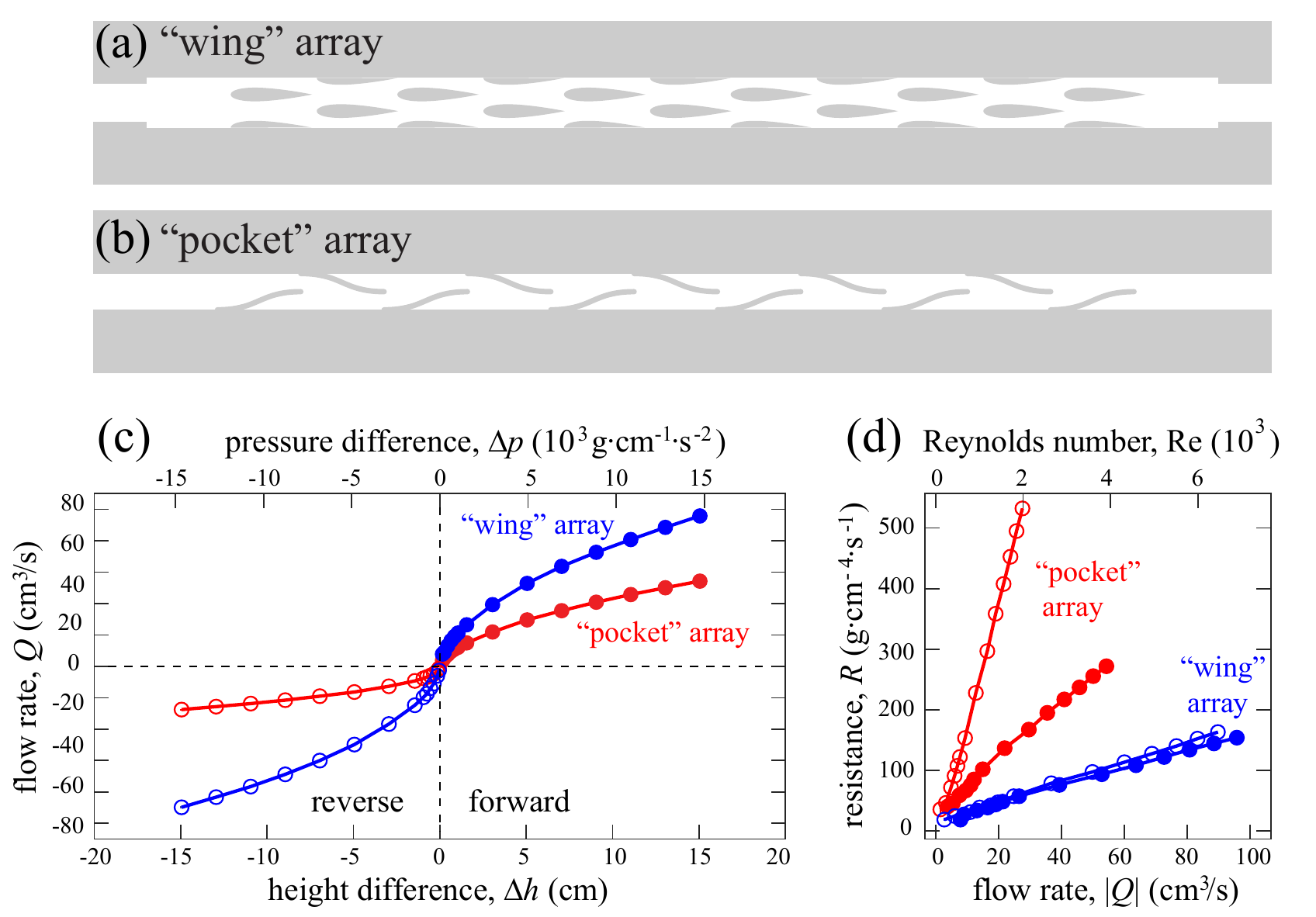}\vspace{-0.2cm}
\caption{Two alternative diode designs and measurements of their performance. (a) An array of wing or airfoil sections. The fluid is to occupy the white regions between the gray regions representing solid material. (b) An array of sigmoidal baffles forms dead-end `pockets' when viewed from the reverse direction. (c) Characterization of the designs through measurements of volume flow rate $Q$ versus water level height differential $\Delta h$ or differential pressure $\Delta p$. (d) Resistance $R = \Delta p / Q$ versus absolute flow rate $|Q|$ for forward and reverse operation of both designs.}
\label{fig:twodiodes}
\end{figure}

Equipped with the experimental methods and with a grasp of the basic fluid dynamics involved, we next pose as a challenge to design a fluidic diode that outperforms Tesla's valvular conduit. In principle, any channel with asymmetric internal geometry may have asymmetric resistance at appreciably high Re. However, designing a channel with high resistance ratio is a challenging shape optimization problem, and the electronic-hydraulic analogy is not informative when it comes to issues of detailed fluid-structure interactions. In the absence of any such well-informed strategy for ``intelligent design'', and without the patience for evolutionary algorithms that may iteratively and systematically improve the shape, we borrow Tesla's use of intuition and inspiration to arrive at the two alternative diodes shown in Figs. \ref{fig:twodiodes}(a) and (b), which we then construct and test.

To facilitate direct comparison against Tesla's design, we imposed the following criteria on our channels: They must be periodic with the same number (11) of repeating units and the same length (30 cm), depth (1.9 cm), and average width (0.9 cm). The first design of Fig. \ref{fig:twodiodes}(a) employs a staggered array of wings or airfoil shapes whose rounded leading edges face into the flow in the forward operation of the diode. The reasoning is that wings, at least when used individually in their typical application of forward flight, are intentionally streamlined for low resistance. Flow in reverse, however, is can trigger flow separation near the thickest portion of the airfoil section and thus a wide wake. Our second design shown in Fig. \ref{fig:twodiodes}(b) replaces Tesla's `buckets', which reroute flows in the reverse mode, with dead-end `pockets' formed by sigmoid-shaped baffles. 

Repeating the experimental procedures outlined above, we arrive at curves for $Q(\Delta h)$ for both designs operating in both forward and reverse, as shown by the plots of Fig. \ref{fig:twodiodes}(c). The corresponding resistance curves $R(Q)$ are shown in Fig. \ref{fig:twodiodes}(d). Surprisingly, the wing design is nearly isotropic with the forward and reverse resistances almost equal across all flow rates tested. Perhaps the similarity in resistance values could be explained by the suppression of flow separation in both directions due to the confined geometry of the channel. In any case, it is clear that not all asymmetric geometries lead to strongly asymmetric resistances even for high Re flows. The pocket design fares better, with a resistance in reverse that is substantially higher than the forward resistance.

\begin{figure}
\centering
\includegraphics[width=10cm]{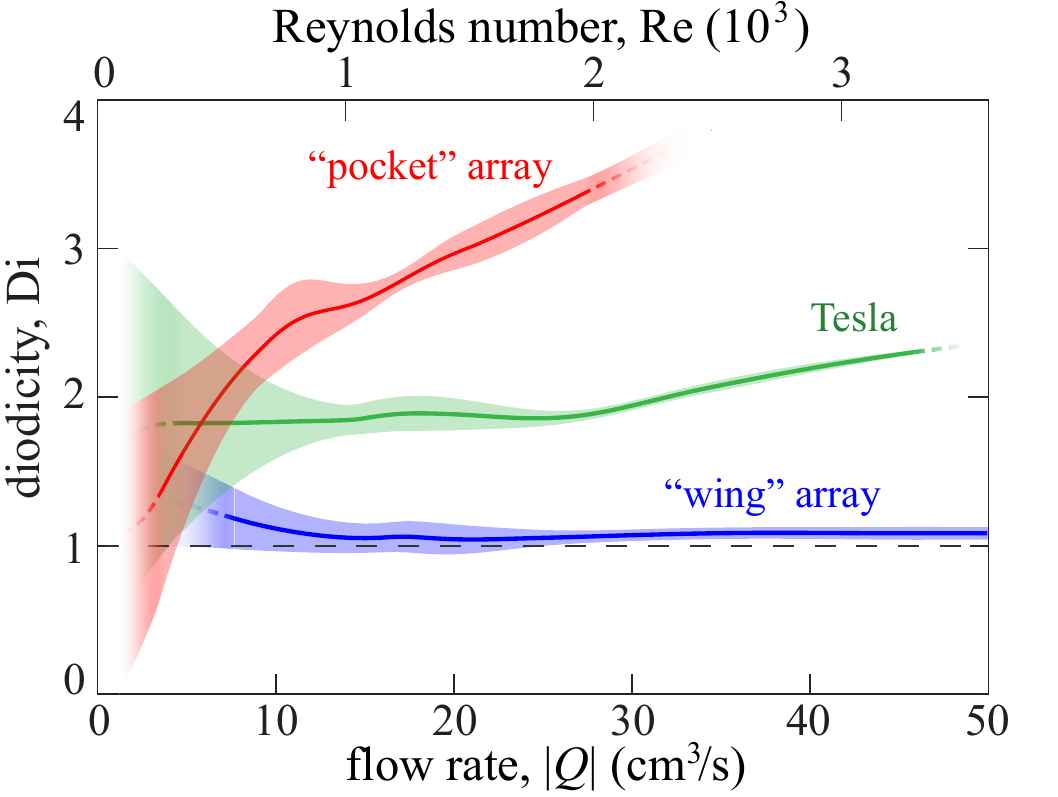}\vspace{-0.2cm}
\caption{Comparing the performance of three channel designs. The diodicity $\mathrm{Di}(|Q|) = R_2(|Q|)/R_1(|Q|)$ of a given channel is the ratio of reverse to forward resistances at the same magnitude of flow rate. The pocket-array design outperforms Tesla's conduit, while the wing-array is nearly symmetric in resistance.  Each $R(|Q|)$ is interpolated from $R$ versus $|Q| $ data. Shaded regions reflect standard error of the mean propagated from measurement errors of $Q$ and $\Delta h$.}
\label{fig:diodicity}
\end{figure}

To quantitatively compare the performance of all three diode designs, we define the ratio of reverse resistance $R_2$ to forward resistance $R_1$ as the \textit{diodicity}\cite{forster1995design}, $\mathrm{Di} = R_2/R_1$. Specifically, we may evaluate $\mathrm{Di}(|Q|) = R_2(|Q|)/R_1(|Q|) = \Delta p_2(|Q|) / \Delta p_1(|Q|) $, which is also equivalent to $\mathrm{Di}(\mathrm{Re}) = f_2(\textrm{Re})/f_1(\textrm{Re})$. Because $R_1$ and $R_2$ are not in general measured at the same $|Q|$, we fit curves to these data and compute their ratio, resulting in the plots shown for Tesla's design and our two diodes in Fig. \ref{fig:diodicity}. The shaded regions indicated errors propagated from the raw measurements. The diodicity of Tesla's conduit is a weakly increasing function of flow rate with a typical value of $\mathrm{Di} \approx 2$ for the conditions studied here. The wing-array design has weak diodicity near unity. Interestingly, the pocket-array design has a more strongly increasing $\mathrm{Di}(|Q|)$ curve than does Tesla's conduit, and it significantly outperforms Tesla's design over most of the range of $Q$ tested. If the trend continues to higher flow rates (higher Re), then even greater diodicity values can be expected.

\section{Discussion and conclusions}

Nikola Tesla's valvular conduit is an engaging context to introduce students of all levels to the role played by creativity in scientific research and specifically the power and limitations of reasoning by analogy. The information presented here can be used as a guide for lectures in an introductory fluid mechanics course, a module in a laboratory course, or as a springboard for a further research projects into fluidics and fluidic devices. As such, we have emphasized the practicalities of the experiments and their pedagogical value. The experimental apparatus can readily be made from household and standard laboratory items such as tanks, tubing and pumps, and the measurements may be accurately and efficiently carried out by students using rulers, beakers and stopwatches. The basic data analysis and plotting is rather straightforward, but at the same time there are ample opportunities for further analysis into errors and their propagation, curve fitting, dynamic resistance and differentiation of data, and so on. The fluid mechanics concepts of channel/pipe flow, hydraulic resistance/friction, Reynolds number, Bernoulli's law, low (ir)reversibility, etc. naturally arise from these investigations and our discussions are but brief introductions that may be followed up in depth. An additional project stems naturally from the challenge to design and test yet better diodes, in which case our pocket-array design might serve as the new standard to beat.

\begin{figure}
\centering
\includegraphics[width=13.5cm]{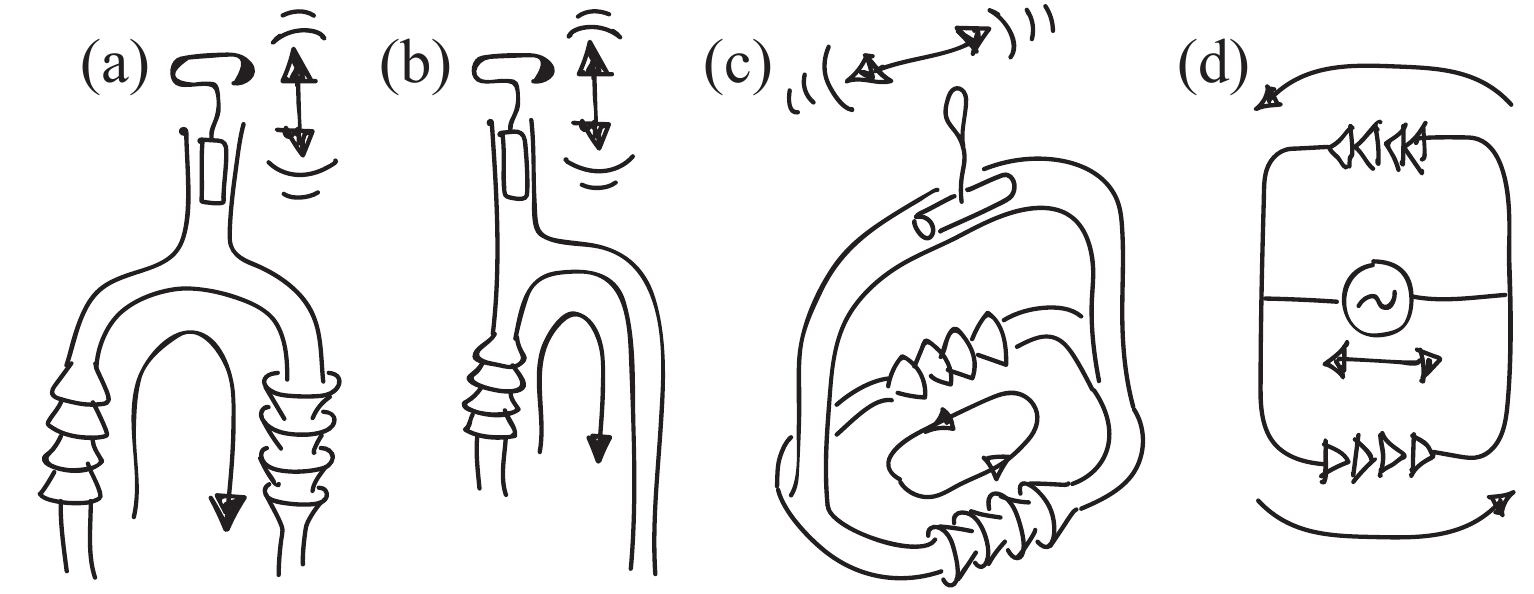}\vspace{-0.2cm}
\caption{Potential applications of fluidic diodes. (a) A hand pump in which oscillations of a piston are rectified by diodes to move fluid from one container to another. (b) A pump that uses a single diode to `prime' or fill a tube, which then drains a container by siphon action. (c) A closed system whereby oscillations drive one-way circulation. (d) An equivalent electronic circuit with AC source for the application shown in (c).}
\label{fig:applications}
\end{figure}

Another idea for an additional lesson plan or further direction for research involves the use of fluidic diodes in practical applications. In Fig. \ref{fig:applications}(a) we sketch a type of hand pump that might be used to transfer liquid from one container to another. The device is a bifurcating tube with a well-sealed piston on the upper branch and two diodes oppositely directed within the lower branches. Reciprocating motions of the piston may drive liquid up one branch and down the other. A related version with a single diode might exploit the rectification effect to `prime' or fill the tube, which then drains as a siphon if the outlet is held below a tank to be emptied [Fig. \ref{fig:applications}(b)]. An analogous but closed system is shown in Fig. \ref{fig:applications}(c). This represents a kind of circulatory system in which oscillations are rectified to drive flow around a loop. This could be used to pump coolant, fuel, lubricant or any of the other fluids that must be moved within machinery. The equivalent electronic circuit of Fig. \ref{fig:applications}(d) involves an alternating current (AC) source transformed into direct current (DC) by the diodes, the whole circuit acting as an AC-to-DC converter or rectifier.

It is also left for future work to explain the source of the large discrepancy between our measured diodicity of about 2 for Tesla's conduit and the `theoretical' and `approximate' value of 200 stated in his patent \cite{tesla1920valvular}. In any case, it is a worthwhile goal to enhance diodicity, which could involve modifying not only the conduit geometry but also the form of the imposed pressures and fluid properties beyond density and viscosity. Describing the reverse mode, Tesla conjectured that unsteady motions could be advantageous \cite{tesla1920valvular}: ``[T]he resistance offered to the passage of the medium will be considerable even if it be under constant pressure, but the impediments will be of full effect only when it is supplied in pulses and, more especially, when the same are extremely sudden and of high frequency.'' He may also have conceived of using air as the working fluid, for which compressibility effects would become important at extremely high speeds and pressures.

\chapter{Early turbulence and pulsatile flows enhance diodicity of Tesla’s macrofluidic valve}  \label{ch3}



This chapter is adapted from the preprint version of Q. M. Nguyen, J. Abouezzi, and L. Ristroph, "Early turbulence and pulsatile flows enhance diodicity of Tesla’s macrofluidic valve", submitted to Nature Communications (July 2020).

\section*{Abstract}
Microfluidics has enabled a revolution in the manipulation and transport of small volumes of fluids. Controlling flows at larger scales and faster rates, or \textit{macrofluidics}, also has applications spanning the sciences, engineering and industry but involves the unique complexities of high-Reynolds-number (high-Re) flows. Here, we show how such effects are exploited in a device proposed by Nikola Tesla that acts as a macrofluidic diode or valve whose asymmetric internal geometry leads to direction-dependent hydraulic resistance. For steady forcing, systematic tests reveal a resistance ratio or diodicity of about 2 that turns on abruptly at $\textrm{Re} \approx 200$ and which is accompanied by nonlinear scaling of pressure with flow rate and flow instabilities in the reverse mode. These results indicate a laminar-to-turbulent transition triggered at an order of magnitude lower Re than than that observed for pipe flow. To assess performance under unsteady forcing, we devise a macrofluidic circuit that functions as an AC-to-DC converter, rectifier or pump in which diodes transform imposed oscillations into directed flow. Mapping pump rate across varying driving conditions reveals boosts as high as 100\%, confirming Tesla's conjecture of enhanced unsteady performance. The connections between diodicity, early turbulence and pulsatility uncovered here can guide applications in fluidic mixing and pumping.

\section{Introduction}
From circulatory and respiratory systems to chemical and plumbing networks, controlling flows is important in many natural settings and engineering applications \cite{nielsen1964animal,conway1971guide,lal1975hydraulic,waite2007applied,fung2013biomechanics}. For flows within vessels, pipes, channels and networks of such conduits, the simplest means for directing fluidic traffic is through internal geometries. How geometry maps to flow patterns and distribution, however, is a challenging problem that depends on flow regime, as characterized by dimensionless quantities involving length- and time-scales and fluid material properties \cite{tritton2012physical}. The field of microfluidics focuses on low Reynolds numbers in which small volumes are conveyed at slow speeds, and recent progress stems from advances in microscale manufacturing and lab-on-a-chip applications \cite{whitesides2006origins}. The flow physics at these scales is dominated by pressures overcoming viscous impedance, and the linearity of the governing Stokes equation enables theoretical and computational approaches that greatly aid in the design of microfluidic devices \cite{squires2005microfluidics}. At larger scales and faster rates, the applications are as numerous and important \cite{conway1971guide,lal1975hydraulic} but the flow physics quite different. The underlying Navier-Stokes equation is nonlinear, theoretical results are fewer, simulations are challenging, and the mapping between geometry and desired fluidic objectives all the more complex \cite{tritton2012physical}. The phenomenology of high-Reynolds-number or inertia-dominated flows is well documented: Flows are slowed in boundary layers near surfaces and tend to separate in a manner sensitive to geometry to yield vortices, wakes, jets and turbulence \cite{tritton2012physical,schlichting2016boundary}. Such complexities are exemplified by the breakdown of reversibility: Running a given system in reverse, say by inverting pressures, does not in general cause the fluid to move in reverse but instead triggers altogether different flow patterns \cite{tritton2012physical}.
 
\begin{figure*}
\centering
\includegraphics[width=15cm]{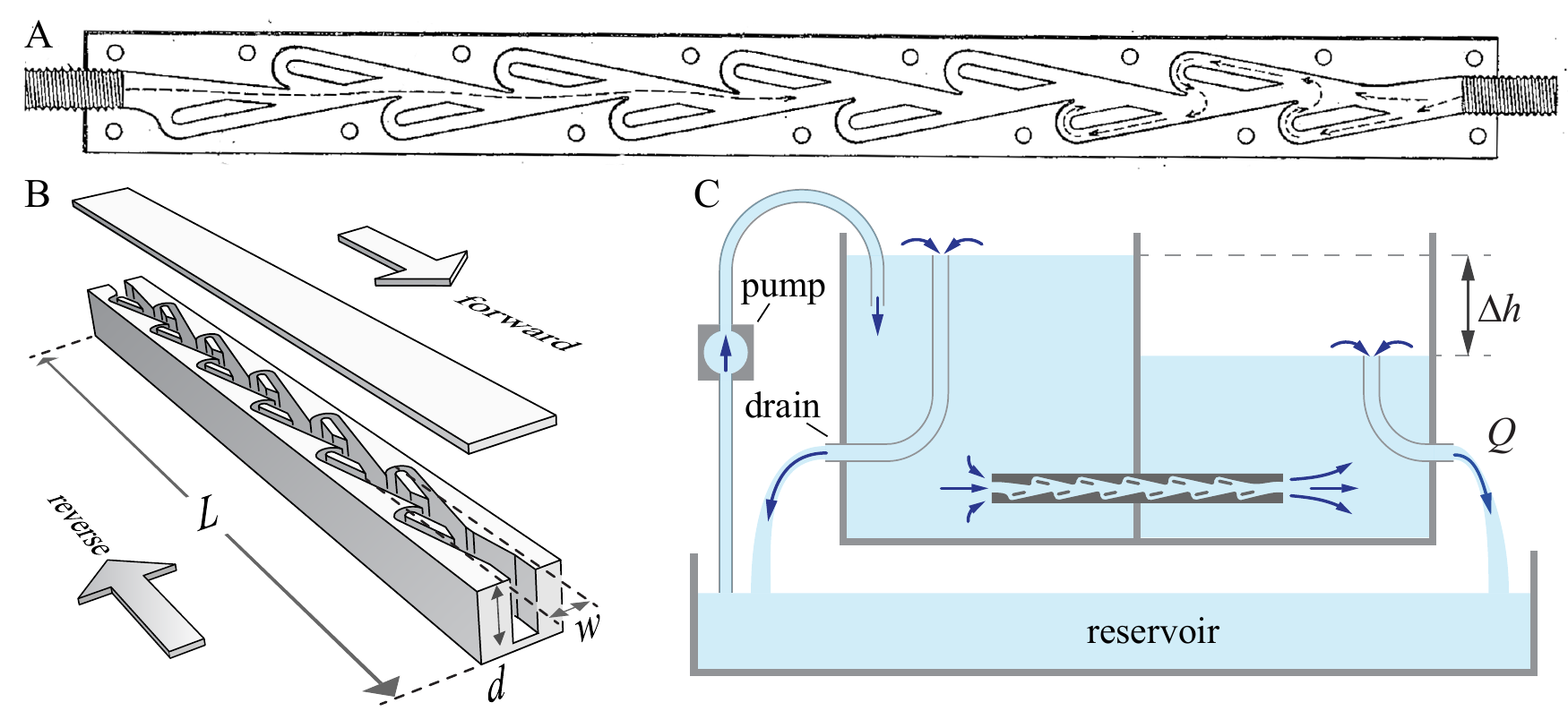}\vspace{-0.1cm} 
\caption{Experimental tests of Tesla's conduit under steady pressures. (\textit{A}) Schematic showing planform view of the `valvular conduit' from Tesla's patent \cite{tesla1920valvular}. (\textit{B}) Rendering of the conduit used in experiments. Upper and lower lids sandwich the internal geometry, which is digitized from Tesla's design, laser-cut and bonded. Relevant dimensions include total length $L$, average wetted width $w$ and depth $d$. (\textit{C}) Schematic of pressure chamber. Overflow mechanisms ensure fixed water levels that drive flow through the conduit, whose actual orientation is upright as shown in \textit{B}. The differential height $\Delta h$ is varied and volumetric flow rate $Q$ measured for both forward and reverse directions. }\vspace{-0.3cm}
\label{fig:DCsetup}
\end{figure*}

 
Here we explore how inertial-flow physics is exploited in an intriguing device proposed by the inventor Nikola Tesla \cite{tesla1920valvular}. Tesla intended this `valvular conduit' -- an image of which is reproduced from his 1920 patent in Fig. \ref{fig:DCsetup}A -- to allow fluid to pass easily in one direction while presenting substantially higher hydraulic resistance in the reverse direction. Such a fluidic diode or valve can be used as a fundamental component for directing flows. While it is unclear if Tesla ever fabricated and tested a prototype, its unique geometry of linked and looped lanes has attracted many studies into its operation and potential applications \cite{forster1995design,schluter2002micro,bardell2000diodicity,gamboa2005improvements,paul1969fluid,kord2012hybrid,zhang2007performance,anagnostopoulos2005numerical,forster2002parametric,bendib2001analytical,kim2015hydrodynamic,losapio2018,nobakht2013numerical,thompson2014numerical,truong2003simulation,mohammadzadeh2013numerical,ansari2018flow,arunachala2019stability,chandavar2019stability,hong2004novel,hossain2010analysis,yang2013development,thompson2013transitional,wang2014tesla,morris2003low,de2017design,forster2007design,deng2010optimization,lin2015topology,abdelwahed2019reconstruction,stone2017fundamentals,magoon2017,porwal2018}. Previous fluid mechanical tests have focused on steady conditions of constant flow rate, confirming the resistance asymmetry or diodicity at high Reynolds numbers (Re) for geometries modified from Tesla's original design \cite{paul1969fluid,forster1995design,bardell2000diodicity,schluter2002micro,gamboa2005improvements,kord2012hybrid}. While it is clear that valving action is absent at very low $\textrm{Re} \ll 1$ for which flows are reversible \cite{stone2017fundamentals} but permissible for high $\textrm{Re} \gg 1$ \cite{bardell2000diodicity,bendib2001analytical,forster2002parametric,gamboa2005improvements,anagnostopoulos2005numerical,zhang2007performance,nobakht2013numerical,kim2015hydrodynamic,losapio2018,paul1969fluid,kord2012hybrid}, there is lacking a characterization spanning flow regimes that would reveal how diodicity varies over a wide range of Re. Consequently, it is unclear if and how diodicity relates to phenomena such as the laminar-to-turbulent transition \cite{tritton2012physical,schlichting2016boundary}. Further, Tesla's patent strongly emphasizes \textit{unsteady} or pulsatile flow conditions \cite{tesla1920valvular}, which arise in pumping or rectification applications \cite{forster1995design,bardell2000diodicity,anagnostopoulos2005numerical,gamboa2005improvements,morris2003low,schluter2002micro,kord2012hybrid,mohammadzadeh2013numerical} but for which the potential for enhanced diodic performance remains to be determined.

In this work, we experimentally test the valvular conduit for a wide range of steady and unsteady conditions and thereby evaluate Tesla's conjecture \cite{tesla1920valvular} that diodic performance is strongest when flow ``is supplied in pulses and, more especially, when the same are extremely sudden and of high frequency.'' On a conduit whose planform shape is faithful to Tesla's original design, we first carry out systematic characterizations for fixed pressure differences, these measurements providing a steady-flow baseline. We then propose and implement a fluidic circuit or network in which unsteady forcing is imposed through oscillatory flows, which the diodes transform or rectify into one-way flows, the whole system serving as an AC-to-DC converter or pump. A quasi-steady model links these two studies by using the steady-forcing characteristics to predict the response for unsteady forcing. Comparing predicted and measured pumping performance permits an assessment of Tesla's conjecture and suggests optimal operating conditions for fluidic rectifiers more generally.

\section{Experimental tests under steady forcing}

\begin{figure*}
\centering
\includegraphics[width=13cm]{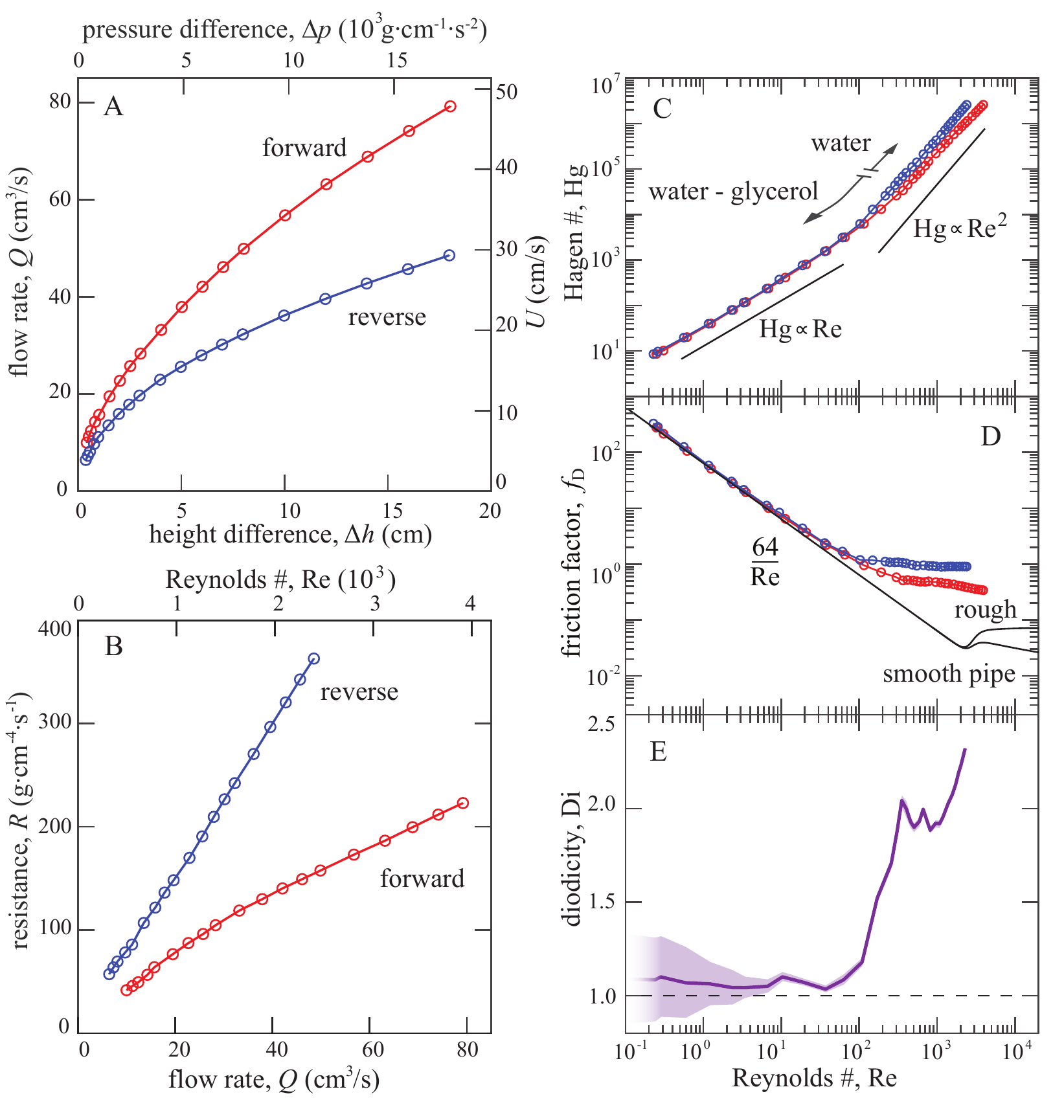}\vspace{-0.3cm}
\caption{Experimental characterization of Tesla's conduit under steady pressures. (\textit{A}) Flow rate $Q$ across varying pressure heads $\Delta h$ and pressure differences $\Delta p = \rho g \Delta h$ for the case of pure water as the working fluid. The forward (red) and reverse (blue) directions exhibit different $Q$ for the same $\Delta p$. (\textit{B}) Hydrodynamic resistance $R = \Delta p /Q  $ versus $Q$ for the forward and reverse directions. (\textit{C}) Dimensionless forms of pressure difference (Hagen number Hg) versus flow rate (Reynolds number Re). The plot combines data on pure water and water-glycerol solutions to cover a wide range of Re. (\textit{D}) Friction factors $f_{\mathrm{D}} = (\Delta p / L) / (\rho U^2 /2D)$, a dimensionless form of pressure drop appropriate for turbulent flow. Also shown are curves representing previous measurements for smooth and rough pipes. (\textit{E}) Diodicity Di or ratio of reverse to forward resistances versus Re. The band represents standard errors determined from repeated measurements. }\vspace{-0.3cm}
\label{fig:DCplots}
\end{figure*}

We first characterize experimentally the hydraulic resistance of Tesla's diode under conditions of fixed pressure differences, this quantity varied systematically to explore flow rates in both directions. We realize a conduit whose planform or overhead-view geometry is faithful to Tesla's original design \cite{tesla1920valvular}, and we pursue Reynolds numbers ranging over orders of magnitude, the latter being important for our later comparison of steady versus unsteady (oscillatory) forcing. We digitize the planform of Fig. \ref{fig:DCsetup}A and manufacture a macroscale version of the conduit via laser-cutting and bonding clear acrylic plastic, a 3D rendering of which is shown in Fig. \ref{fig:DCsetup}B. We select a scale that, together with the use of water and water-glyercol mixtures as the working fluids, allow for characterization of the channel across low to high Re. The overall length is $L = 30$ cm, average wetted width $w = 0.9$ cm and depth $d = 1.9$ cm.

To impose and controllably vary the pressure difference across the channel, we design and construct the apparatus whose sectional view is shown in Fig. \ref{fig:DCsetup}C. Two chambers of a tank are connected only via the conduit, and the liquid level in each can be set and stably maintained via overspill mechanisms. The level difference $\Delta h$ is set by two adjustable internal drains whose heights can be independently varied. The hydrostatic pressure difference across the channel is $\Delta p = \rho g \Delta h $, where $\rho$ is the density of the fluid and $g = 980 ~\textrm{cm}/\textrm{s}^2$ is gravitational acceleration \cite{tritton2012physical}. Liquid flows from the high side to the low side through the channel at a volumetric flow rate $Q$ and out to the reservoir at the same rate. A pump takes fluid from the reservoir, slightly overflowing the high side and thus maintaining its level. The system is closed and runs indefinitely. In this way, a pressure difference may be imposed and recorded by measuring the column heights with rulers, and the volumetric flow rate $Q$ is measured by intercepting the lower drain with a beaker of known volume and reading the fill-up time with a stopwatch. The flow direction is changed by simply changing which chamber has higher level.

The measured flow rate $Q$ versus $\Delta h $ for pure water is shown in Fig. \ref{fig:DCplots}A. As might be expected, increasing the height difference yields higher flow rates for both the forward and reverse directions, as defined in Fig. \ref{fig:DCsetup}B. The flow rate $Q$ increases monotonically but nonlinearly with $\Delta h$. Importantly, for the same $\Delta h$, $Q$ is greater for the forward direction than reverse, and this is true across all values of $\Delta h$. This anisotropy is more clearly seen in Fig. \ref{fig:DCplots}B, where the resistance $R = \Delta p /Q $ is plotted versus $ Q $ for the forward and reverse directions. Across all $Q$, the resistance in reverse is greater, and this disparity increases with $Q$.

\section{Resistance and diodicity across flow regimes}

Experiments carried out with pure water yield high Reynolds numbers $\mathrm{Re} = \rho U D / \mu \sim 10^3$, where $\mu$ is the fluid viscosity, $U = Q / wd$ is the section-averaged flow speed through the channel, and $D = 4V/S = 0.8$ cm is its hydraulic diameter calculated from the total wetted volume $V$ and surface area $S$. [The latter generalizes the conventional form $D = 4A/P$ for a conduit whose cross-section shape is uniform and of wetted area $A$ and perimeter length $P$ \cite{white1999fluid}.] To explore lower Reynolds numbers, we conduct further experiments with water-glycerol solutions of varying viscosity, and all results are combined in Fig. \ref{fig:DCplots}C. To account for the different fluid properties, we plot non-dimensional pressure or Hagen number $\textrm{Hg}  =  \left(  \Delta p/L \right) \left( D^3 \rho / \mu^2 \right)$ versus non-dimensional flow rate or Reynolds number \cite{tritton2012physical,martin2002generalized}. As expected, the disparity between the forward and reverse directions is significant only at sufficiently high Re. Also shown for comparison are reference lines indicating linear and quadratic scalings of pressure with flow rate. For low Re, it is seen that $\textrm{Hg} \sim \textrm{Re}$, which is expected for well-developed and laminar flow \cite{tritton2012physical}. For higher Re, the data follow a stronger dependence of approximately $\textrm{Hg} \sim \textrm{Re}^2$, which is characteristic of turbulent flow \cite{tritton2012physical}. Interestingly, the disparity in resistance occurs together with the nonlinearity of the Hg-Re scaling.

A conventional measure of hydraulic resistance is the Darcy friction factor $f_{\mathrm{D}} = (\Delta p / L) / (\rho U^2 /2D)$, which normalizes pressure drop based on inertial scales \cite{moody1944friction}. In Fig. \ref{fig:DCplots}D we plot our measurements of $f_{\mathrm{D}}(\textrm{Re})$ for forward and reverse flow through Tesla's conduit, and we include for comparison previous results on pipes \cite{bellos2018friction}. The two curves shown correspond to a smooth pipe and to one of high roughness in which wall variations measure 10\% of the mean diameter. The form $f_{\mathrm{D}} = 64/\textrm{Re}$ corresponds to the Hagen-Poiseuille law and applies well to both smooth and rough-walled pipes in the laminar flow regime of $\mathrm{Re} < 2000$ \cite{kandlikar2005characterization}. Following a transitional region, well-developed turbulence tends to be triggered at higher $\mathrm{Re} > 4000$, for which $f_{\mathrm{D}}$ is more constant with Re and increases with roughness. By comparison, Tesla's conduit transitions away from the laminar-flow scaling at significantly lower $\mathrm{Re} \approx 200$. The order-one friction values at higher Re are substantially higher than those for turbulent flow through smooth and rough pipes.

\begin{figure*}
\centering
\includegraphics[width=16.5cm]{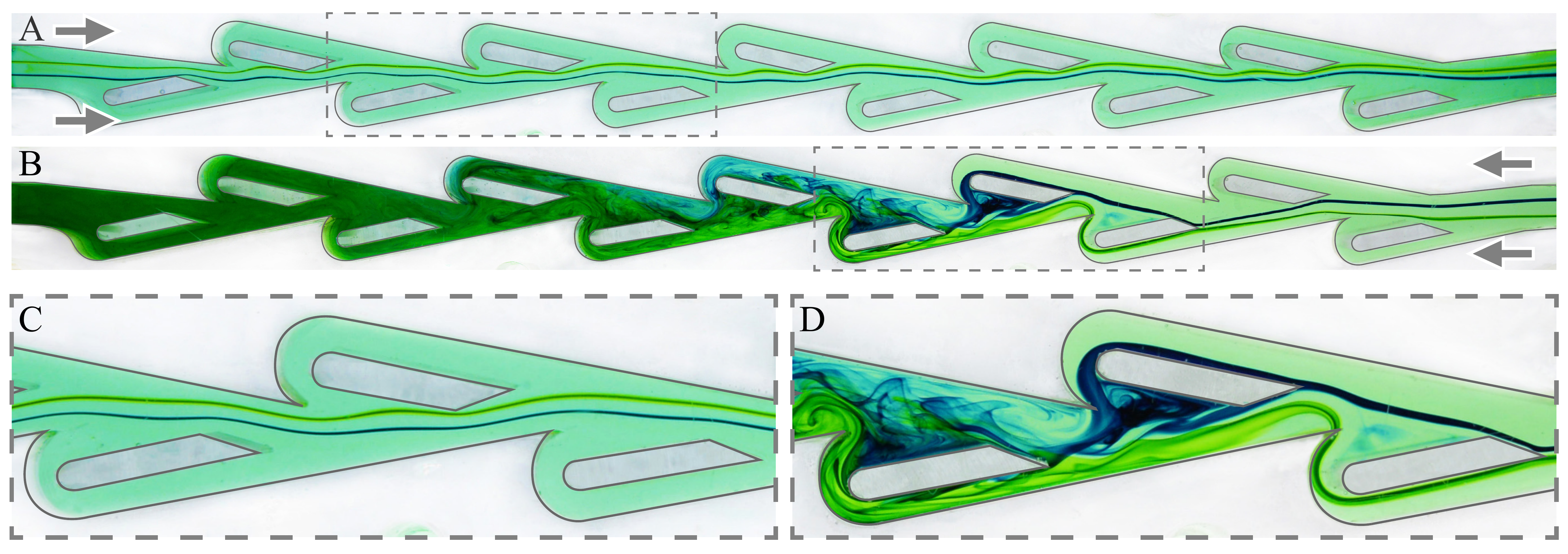}\vspace{-0.2cm}
\caption{Streakline flow visualization at $\textrm{Re}=200$ using dye injected upstream. (\textit{A}) and (\textit{C}) Forward direction. Two adjacent filaments remain in the central corridor of the conduit with only small lateral deflections. (\textit{B}) and (\textit{D}) Reverse direction. The filaments ricochet off the periodic structures, deflecting increasingly sharply before being rerouted around the `islands' and mixing.}
\label{fig:flowvis}\vspace{-0.1cm}
\end{figure*}

The performance of the channel as an asymmetric resistor can be quantified by its \textit{diodicity} or ratio of reverse to forward resistance values \cite{li2015nanofluidic}. Equivalently, we define this ratio using dimensionless forms of pressure drops at the same Re: $\mathrm{Di}(\mathrm{Re}) = \textrm{Hg}_\mathrm{R}(\mathrm{Re}) / \textrm{Hg}_\mathrm{F}(\mathrm{Re}) = f_{\mathrm{D, R}}(\mathrm{Re}) / f_{\mathrm{D, F}}(\mathrm{Re})$, where the subscripts indicate the reverse (R) and forward (F) directions. In Fig. \ref{fig:DCplots}E the curve indicates how Di varies with Re, with the band representing propagated errors based on repeated measurements. For low Re, Di is close to unity and remains so up until $\mathrm{Re} \approx 100$. Over a transitional range $\textrm{Re} = 100-300$, the diodic function of the channel `turns on' or  is activated, and for $\mathrm{Re} = 300-1500$ we find $\textrm{Di} \approx 2$. The trend suggests yet higher diodicity at higher $\mathrm{Re} > 2000$. 

Interestingly, the turn-on of diodicity apparent in Fig. \ref{fig:DCplots}E comes along with the nonlinear scaling of pressure drop with flow rate (Fig. \ref{fig:DCplots}C) and the departure from the laminar-flow friction law (Fig. \ref{fig:DCplots}D). These results suggest that diodic function is closely linked to a transition to turbulent flow that occurs significantly earlier (at lower Re) than that observed for smooth and rough pipes.

\section{Flow visualization and early turbulence}

To corroborate the link between diodicity and flow state, we next carry out visualizations in the transitional regime. We focus on $\textrm{Re}=200$, for which we inject neutrally buoyant dye upstream and record photographs and video using a camera positioned to view the planform. The conduit is clear and backlit, and the resulting images reveal flow streaklines \cite{tritton2012physical}. Two adjacent streaklines near the middle of the channel are color-coded using blue and green dyes. Figure \ref{fig:flowvis}A shows the case of flow in the forward direction. The streaklines remain in the central corridor along the entire length of the channel and are deflected only slightly as they pass the periodic structures. Details of the gently meandering path can be seen in the zoomed-in image of Fig. \ref{fig:flowvis}C. In contrast, the reverse direction involves amplified lateral deflections of the streams that eventually results in total mixing, as shown in Fig. \ref{fig:flowvis}B. The incoming filaments ricochet off the internal structures, with the redirections being only slight in passing the first `islands' or baffles but quickly growing downstream after repeated interactions. The flows are eventually rerouted into the recesses, and total mixing is observed by the end of the channel. Some of the steps that destabilize the initially-laminar flow can be seen in the zoomed-in view of Fig. \ref{fig:flowvis}D.

These results confirm the high-Re irreversibility reported in previous studies, which have emphasized the circuitous route taken by reverse flows \cite{bardell2000diodicity,thompson2014numerical,nobakht2013numerical,ansari2018flow}. Our visualizations reveal the nature of the reverse flow instability and also the extent of mixing, which we associate with increased dissipation and resistance. Unsteady flows and increased resistance are hallmarks of turbulent flow, which is triggered for Re in the thousands for pipe and channel flow \cite{tritton2012physical,pope2001turbulent}. Our visualizations of flow destabilization in Tesla's conduit at considerably lower $\textrm{Re} = 200$ offer further evidence for an early transition to turbulence.  

\section{Unsteady forcing of a fluidic AC-to-DC converter}

\begin{figure*}
\centering
\includegraphics[width=16.5cm]{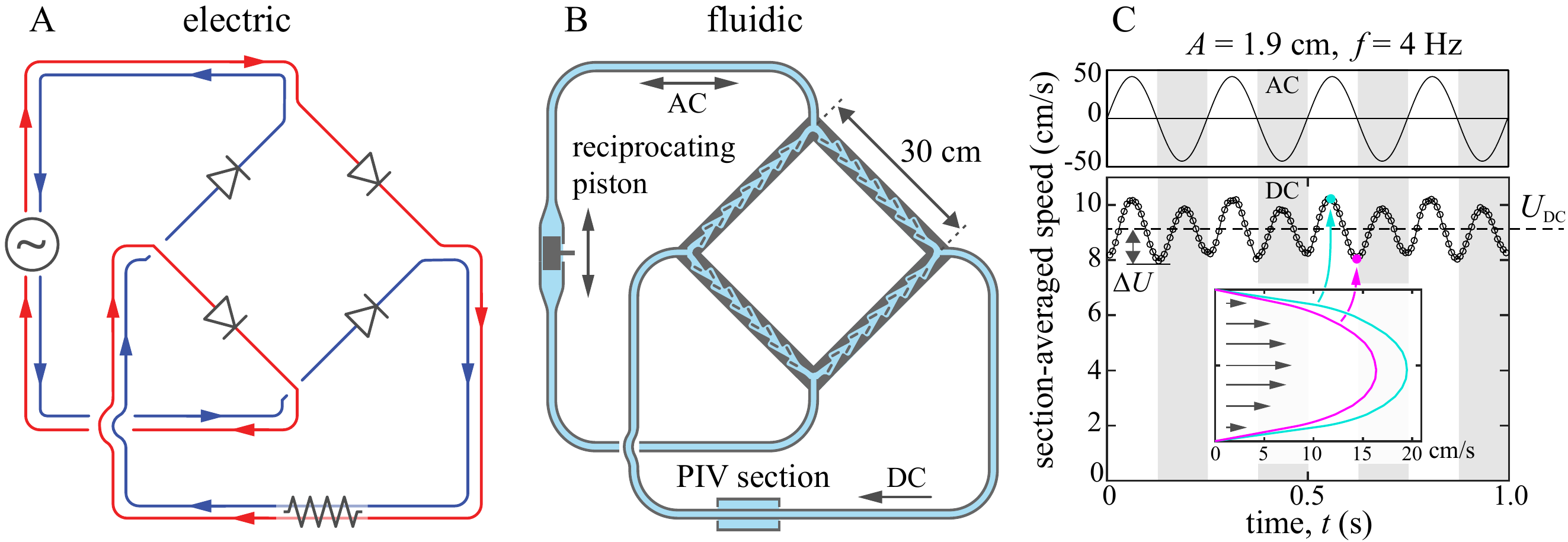}\vspace{-0.2cm}
\caption{Electronic AC-to-DC converter and an analogous fluidic pump. (\textit{A}) Electric circuit with four ideal diodes that converts alternating current source (AC, left branch) to direct current (DC, lower branch). Red and blue lines highlight the path and direction of current at different phases in the AC cycle. (\textit{B}) Sectional view of an analogous fluidic circuit with four Tesla diodes and a pulsatile flow source. The experimental device employs a reciprocating piston of amplitude $A$ and frequency $f$ as an AC source in one branch, and the DC flow is measured in a second branch. (\textit{C}) Section-averaged flow imposed in the AC branch (top) and measured in the DC branch (bottom) for $A=1.9$ cm and $f=4$ Hz. The mean flow rate $U_{\mathrm{DC}}>0$ indicates successful AC-DC conversion or pumping. Inset: sample flow velocity profiles measured by PIV.}\vspace{-0.3cm}
\label{fig:ACschematic}
\end{figure*}

Having characterized Tesla's conduit for steady pressure differences, we next consider unsteady forcing in which the internal flows are driven to oscillate. To assess Tesla's conjecture of enhanced performance for pulsatile flows \cite{tesla1920valvular}, we draw on the analogy between electric and fluidic circuits and consider a full-wave rectifier that uses four diodes arranged in a bridge configuration in order to convert an imposed alternating current (AC) in one branch into directed current (DC) in a second branch \cite{horowitz1989art}. The electric circuit is shown schematically in Fig. \ref{fig:ACschematic}A. An AC current source is on the left, and the directionality of each ideal diode is indicated by the arrowhead. These elements are linked by conducting wires, and the current direction in these wires are shown in red and blue for the two half-cycles of the AC source. When current is driven upwards through the source, only the two diodes under favorable bias conduct, and the current follows the red path. In the next half-cycle, the other pair of diodes conducts, and the current follows the blue path. Thus, while the input branch is purely AC or oscillatory, the output branch on the bottom exhibits a DC component or non-zero mean.

Figure \ref{fig:ACschematic}B shows the schematic of the fluidic analog that we design, construct and test. Laser-cut and bonded Tesla conduits serve as diodes, a reciprocating piston replaces the AC current source, and these elements are connected in bridge configuration through piping. The circuit is filled with water, and the position of the piston is driven sinusoidally in time with amplitude $A$ and frequency $f$ controllably varied via a high-torque stepper motor (Longs Motor) and Arduino controller. Because the piston completely seals the surrounding cylinder, the flow in the AC branch is purely oscillatory. The diodic behavior of the conduits then manifest as one-way or directed current (DC) in the lower branch. To assess this, we use Particle Image Velocimetry (PIV) to measure the flow velocity field along a segment of the transparent DC branch pipe. The 5-cm-long interrogation region is encased in a rectangular water jacket to minimize optical distortion \cite{eguchi2003development}. Following procedures from earlier studies \cite{becker2015hydrodynamic,ramananarivo2016flow,newbolt2019flow}, we seed the water with particles (hollow glass microspheres of approximate diameter 50 $\mu$m, 3M) whose near neutral buoyancy is ensured by selection from a fractionation column in water.  A laser sheet (1.25W CW green, CNI) of thickness 0.5 mm is shone across the mid-plane along the PIV section, and the resulting particle motions are recorded via high-speed camera (12 MP, 150 fps, Teledyne Dalsa Falcon2). Post-processing via an established PIV algorithm \cite{thielicke2014pivlab}, these data provide the flow velocity profile across the pipe, resolved in time within each oscillation cycle and over a total duration of at least 10 cycles.

Representative data are shown in Fig. \ref{fig:ACschematic}C for one set of $A$ and $f$. The upper panel presents the flow speed averaged across the cross-section in the AC branch, where the sinusoidal oscillations $2 \pi A f \sin (2 \pi f t)$ are slave to the piston motions. The lower panel represents the measured section-averaged flow speed in the DC branch, and the inset shows the flow velocity profile furnished by PIV at two points in the cycle. At each time $t$, the two halves of the profile (bisectioned by the tube's axis) are averaged and axisymmetry is assumed to arrive at the section-averaged speed. Strikingly, the flow has a dominant DC component $U_{\mathrm{DC}}$, and the flow profiles remain unidirectional throughout the oscillation cycle. Thus the circuit achieves the goal of AC-to-DC conversion or pumping. The output flow also shows a weak AC component of amplitude $\Delta U$. These ripple oscillations occur at twice the driving frequency $f$, as both half-strokes of the AC input contribute to the DC output.

\begin{figure*}
\centering
\includegraphics[width=17.5cm]{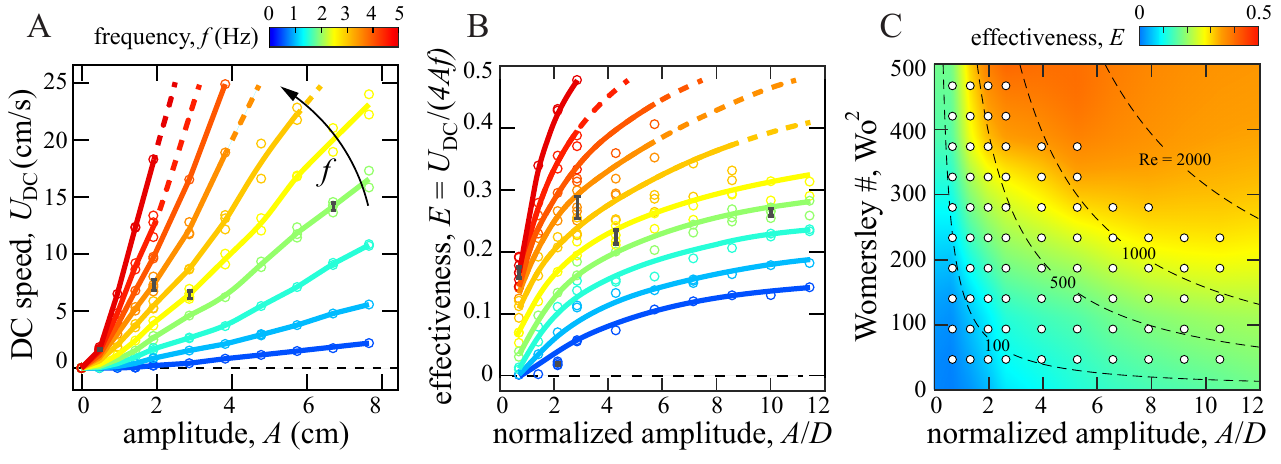}\vspace{-0.3cm}
\caption{Performance of the fluidic AC-to-DC converter or pump. (\textit{A}) Average DC flow speed $U_{\mathrm{DC}}$ versus driving amplitude $A$ and at different frequencies $f$. Representative error bars (black) show standard errors of the mean. (\textit{B}) Effectiveness of rectification versus amplitude normalized by the hydraulic diameter $D$. (\textit{C}) Experimentally-measured effectiveness. The axes correspond to normalized amplitude and non-dimensional frequency or Womersley number $\textrm{Wo}^2 \propto f$. White markers indicate data points, and the colormap is interpolated and extrapolated elsewhere. The dashed hyperbolic curves are contours of the driving or oscillatory Reynolds number. }
\label{fig:ACplots}\vspace{-0.3cm}
\end{figure*}

To assess the pumping performance of the circuit more generally, we systematically vary the AC input parameters $A$ and $f$ and measure the section-averaged DC flow speed $U_{\mathrm{DC}}$, which is equivalent to volumetric flux (volume per unit cross-sectional area and time). Figure \ref{fig:ACplots}A shows how $U_{\mathrm{DC}}$ varies with $A$ and $f$. Across all driving conditions, $U_{\mathrm{DC}}>0$ and the system achieves AC-to-DC conversion. As expected, the output $U_{\mathrm{DC}}$ increases with the inputs $A$ and $f$. Less apparent in Fig. \ref{fig:ACplots}A is that the response is nonlinear. To clarify this, we define the \textit{effectiveness} of the pump as $E = U_{\mathrm{DC}} / 4Af $. This normalization is chosen such that ideal or perfect diodes yield $E=1$: A volume of fluid proportional to the piston displacement $2A$ is injected into the DC branch in each half-stroke of duration $1/2f$. In Fig. \ref{fig:ACplots}B we plot effectiveness $E$ versus frequency $f$ and dimensionless amplitude $A/D$. The fact that $E<1$ for all conditions reflects the non-ideal nature or `leakiness' of the diode. Interestingly, it is seen that $E$ itself increases with both $A$ and $f$, quantifying the nonlinear response of $U_{\mathrm{DC}}$. That is, doubling either $A$ or $f$ leads to disproportionately higher $U_{\mathrm{DC}}$. For the conditions studied here, we achieve $E \approx 0.5$, and the trends suggest yet higher efficacy would be attained for stronger driving.

A fully dimensionless characterization is displayed in Fig. \ref{fig:ACplots}C, where $E$ is mapped across varying $A/D$ and the Womersley number $\textrm{Wo}^2=  (\pi/2) (\rho f D^2/\mu)$, which is proportional to frequency and characterizes the unsteadiness of pulsatile flow \cite{womersley1955method}. The variations in the map again highlight the nonlinearity of the pump, which is most effective in the red region of high $\textrm{Wo}^2 \propto f$. For reference, we include contours (dashed hyperbolic curves) of constant driving or oscillatory Reynolds number, defined as $\textrm{Re} = \rho A f D/ \mu = (2/\pi)(\textrm{Wo}^2)(A/D)$. Significant pumping occurs for $\textrm{Re}$ in the hundreds, when diodicity is observed to turn on for steady flow (Fig. \ref{fig:DCplots}E).

\section{A quasi-steady model of the AC-to-DC converter}

The rectifying circuit provides a clean context for assessing Tesla's conjecture of enhanced performance of the diode for pulsatile flows. Our strategy involves formulating a model that predicts the pump rate of the system based on its steady-flow characteristics, and then comparing this prediction to the actual performance measured experimentally. The quasi-steady model views the AC-DC rectifier as a network of nonlinear resistors whose resistance values vary with flow rate and direction as measured and characterized in Fig. \ref{fig:DCplots}. The network can then be analyzed by standard methods for electronic circuits, \textit{i.e.} by solving for unknown currents through all segments via equations for current/flow conservation at each node or junction and voltage/pressure-drops around closed loops \cite{horowitz1989art}.



\begin{figure}[hbt!]
\centering
\includegraphics[width=0.8\textwidth]{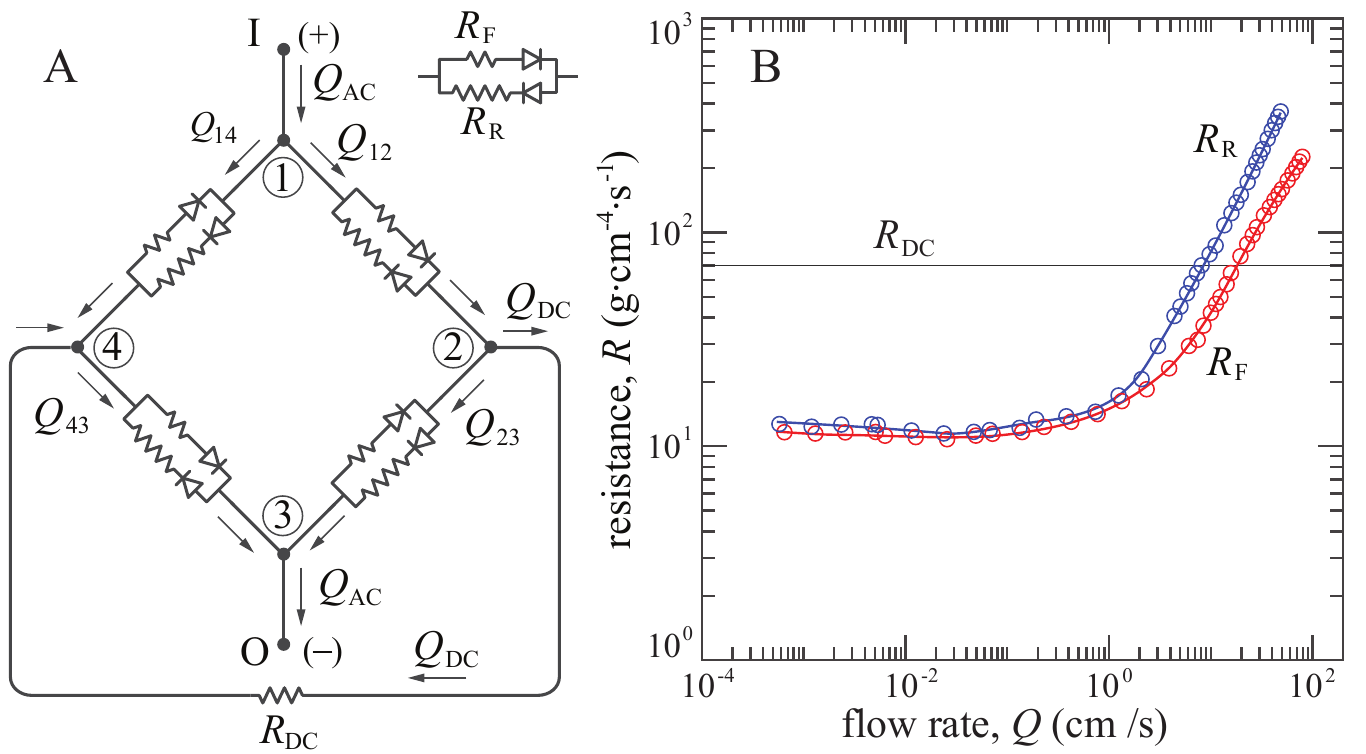}
\caption{(A) Equivalent electric circuit of the fluidic AC-to-DC converter, where the Tesla conduit is modeled as a `leaky diode' with asymmetric resistance values. (B) Resistances vs. flow rate for forward (red) and reverse (blue) flow through Tesla's conduit, as inferred from the data shown in Fig. 2C. The Dc branch resistance value is estimated via the Hagen-Poiseuille law.}
\label{circuit}
\end{figure}

We model the AC-to-DC converter as the equivalent electric circuit shown in Fig. \ref{circuit}A. The equivalent of each Tesla conduit is a combination of two ideal diodes and two resistances  $R_{\mathrm{F}}$ and  $R_{\mathrm{R}}$ with $R_{\mathrm{R}} > R_{\mathrm{F}} $ and both resistance values depending nonlinearly on flow rate. For terminals I and O connected to an oscillating source $Q_\mathrm{AC}(t)$, we seek to calculate the resulting time-averaged DC flow rate $\left< Q_\mathrm{DC}(t)\right>$ going from node 2 to node 4. We quantify the effectiveness of rectification by $E_{\mathrm{M}}$, defined as the ratio of time-averaged DC flow rate to the time-averaged absolute AC flow rate: 
\begin{equation} \label{effective}
    E_{\mathrm{M}} = \frac{\left< Q_{\mathrm{DC}}(t)\right>}{
    \left< |Q_{\mathrm{AC}}(t)| \right>
    } .
\end{equation}
Here, $R_{\mathrm{F}}(Q)$ and  $R_{\mathrm{R}}(Q)$ are positive functions of flow rate $Q$, $R_\mathrm{DC}$ is positive and fixed, and the rectified flow rate $\left< Q_{\mathrm{DC}}(t)\right>$ is always positive. To find the solution, we apply the equivalent of Kirchoff's laws for nodes and loops: Total incoming flows and total outgoing flows are equal at a node, and the sum of all pressure drops around a closed loop is equal to zero. The equivalent of Ohm's law $ \Delta p  = Q R $ describes the relationship of pressure $\Delta p $ and flow rate $Q$ for any given resistor of resistance $R$.

In general, given a time-dependent input flow rate $Q_\mathrm{AC}$, the resulting flow rates $Q_{12}$, $Q_{23}$, $Q_{43}$, $Q_{14}$ and $Q_{\mathrm{DC}}$ also depend on time and thus so do $R_{\mathrm{F}} $ and  $R_{\mathrm{R}}$. To simplify the notation, the time dependence of these quantities is suppressed in what follows. The number of unknown flow rates can be reduced by invoking symmetry. Consider when the source polarity is as indicated in Fig. \ref{circuit}A, that is, positive source flow rate $Q_\mathrm{AC}$ enters at I and exits at O. In the direction indicated by arrows, the sign of $Q_\mathrm{DC}$ is positive, while the signs of $Q_{12}$, $Q_{43}$, $Q_{23}$ and $Q_{14}$ are not known \textit{a priori}. By symmetry, it must be true that $Q_{12}= Q_{43}$ and $Q_{23} = Q_{14}$. In the labelled directions, $Q_{12}$ and $Q_{14}$ must be like-signed (both positive or both negative) so that the net pressure drop in the closed loop 1-2-4-1 is zero. Further, in order for the flow rates to conserve at node 1, the sign in question must be positive. It then follows that all flow rates are positive in the labelled directions. There are then two independent equations corresponding to Kirchoff's node law and one corresponding to the loop law:
\begin{align} \label{kirchoff}
\begin{split}
   & Q_{12} - Q_{23} - Q_{\mathrm{DC}} =0\\
    & Q_{12}+Q_{23}-Q_\mathrm{AC}=0\\
     Q_{12} R_{\mathrm{F}}(Q_{12}&)  - Q_{23} R_{\mathrm{R}}(Q_{23}) + Q_{\mathrm{DC}} R_{\mathrm{DC}} = 0.
\end{split}
\end{align}
Again by symmetry, the set of equations describing the circuit when polarity of the source is reversed (i.e. $Q_{\mathrm{AC} } \mapsto -Q_{\mathrm{AC} }$) is the same. So the system of equations is valid in general and describes the circuit with an oscillating source at all instances in time. The algebraic system of 3 equations is non-linear in the 3 unknowns $Q_{12}$, $Q_{23}$ and $Q_{\mathrm{DC}}$. Further, note that $Q_{\mathrm{AC}}(t)$ represents a time-dependent (harmonic) forcing, and all unknown flow rates are also time dependent. There is no explicit closed-form solution for $Q_{\mathrm{DC}}$ nor its time average, and thus the system must be solved numerically.  

For harmonic input flow rate $Q_\mathrm{AC}(t)= Q_{0} $sin$(2 \pi f t) $, the system is periodic in time, so it is sufficient to solve and average over a single period $T=1/f$. This together with the symmetry of the system under reversing  $Q_{\mathrm{AC}}$ as discussed above, we conclude that it is only necessary to solve and average over a half-period. We solve the system numerically using MATLAB by stepping time through a half-period $t \in [0, T/2]$, where the exact functions for $R_{\textrm{F}}(Q)$ and $R_{\textrm{R}}(Q)$ are specified by spline fits to the experimental measurements, as shown in Fig. \ref{circuit}B. Since the DC branch pipe has only a weak oscillatory component, we estimate its resistance $R_{\mathrm{DC}}$ using Hagen-Poiseuille theory for developed and laminar pipe flow \cite{tritton2012physical}. Accordingly, the resistance $R = 128 \mu L / \pi D^4$ of a pipe is constant in time and depends on the fluid (water at room temperature) viscosity $\mu=8.9\times10^{-3} \textrm{dyn} \cdot \textrm{s} / \textrm{cm}^2$ as well as the pipe length $L$ and diameter $D$. For experimental reasons, the DC branch consists of four different segments connected in series. Segments 1, 2 and 3 are pipes with different lengths (30, 150, 30 cm) and diameters (1, 2.4, 1 cm respectively), and segment 4 is a flow conditioner consisting of a bundle of 20 parallel straws, each 0.25 cm in diameter and 10 cm in length. Following the law of adding linear resistors, the resistance of segment 4 is $1/20$ the resistance of each straw. We then estimate $R_{\mathrm{DC}}$ as the sum of four resistances, $R_{\mathrm{DC}} = (128 \mu / \pi) (L_1 /  D_1^4 + L_2 / D_2^4 +  L_3 /  D_3^4 + L_4 / 20 D_4^4)  = 70 $ g $\cdot$ cm$^{-4}\cdot$ s$^{-1} $. At each time step, an instantaneous value of $Q_{\mathrm{DC}}(t)$ is obtained using the MATLAB root-finding function \textit{fsolve}. Averaging the solutions over the half-period $[0, T/2]$ yields the rectified flow rate $ \left< Q_{\mathrm{DC}}(t)\right> $. The effectiveness of rectification as defined in Eq. (\ref{effective}) is then
\begin{equation} \label{effective2}
   E_{\mathrm{M}} = \frac{\left< Q_{\mathrm{DC}}(t)\right>}{
    \frac{1}{T/2} \int_0^{T/2} Q_{0} \mathrm{sin} (2 \pi f t) \;\mathrm{d}t 
    } =\frac{\left< Q_{\mathrm{DC}}(t)\right>}{2Q_0/\pi} = \frac{U_{\mathrm{DC}}}{4Af},
\end{equation}
which is consistent with the definition of $E$ given in the text in the context of the experiments. Here $Q_0 = (2 \pi A f) wd $ and $\left< Q_{\mathrm{DC}}(t)\right> = \left< U_{\mathrm{DC}}(t)\right> wd = U_{\mathrm{DC}} wd $, where $wd$ is equal to the average cross-sectional area of Tesla's conduit.

Though not true for Tesla's conduit, the simplified case of $R_{\mathrm{F}}$ and $R_{\mathrm{R}}$ being constants independent of flow rate gives some intuition for $E_{\mathrm{M}}$. The system of equations (\ref{kirchoff}) are then a linear system in the 3 unknowns $Q_{12}, Q_{23}$ and $Q_{\mathrm{DC}}$. The output DC flow rate can be calculated analytically as
\begin{equation}
  Q_{\mathrm{DC}} =   \frac{R_{\mathrm{R}}-R_{\mathrm{F}}}{R_{\mathrm{R}}+R_{\mathrm{F}}+2 R_{\mathrm{DC}}} Q_{\mathrm{AC}},
\end{equation}{}
\noindent which is valid at all instances in time for time-dependent flow rates $Q_{\mathrm{DC}}(t) $ and $Q_{\mathrm{AC}}(t)$. As in the full problem with flow rate-dependent resistance values, given the input $Q_\mathrm{AC}(t)= Q_{0} $sin$(2 \pi f t) $ and averaging over a half-period $t\in [0,T/2]$ yields
\begin{equation}
   \left< Q_{\mathrm{DC}} (t)\right> =  \frac{1}{T/2}  \int_{0}^{T/2} \frac{R_{\mathrm{R}}-R_{\mathrm{F}}}{R_{\mathrm{R}}+R_{\mathrm{F}}+2 R_{\mathrm{DC}}}
   Q_{0} \mathrm{sin} (2 \pi f t) \;\mathrm{d} t  = \frac{2Q_0}{\pi}  \frac{R_{\mathrm{R}}-R_{\mathrm{F}}}{R_{\mathrm{R}}+R_{\mathrm{F}}+2R_{\mathrm{DC}}}.
\end{equation}{}
Following Eq. (\ref{effective2}), the effectiveness $E_{\mathrm{M}}$ is then given by
\begin{equation}
    E_{\mathrm{M}} = \frac{R_{\mathrm{R}}-R_{\mathrm{F}}}{R_{\mathrm{R}}+R_{\mathrm{F}}+2R_{\mathrm{DC}}}  = \frac{\mathrm{Di} -1}{ \mathrm{Di} +1+2 R_{\mathrm{DC}}/R_{\mathrm{F}}}, 
\end{equation}{}
\noindent where $\textrm{Di} = R_{\textrm{R}} / R_{\textrm{F}}$. Because $\textrm{Di} \geq 1$ and all resistance values are non-negative, $E_{\mathrm{M}} \leq 1$. Some special cases and limits:
\begin{itemize}

\item For ideal diodes with $R_{\mathrm{F}}=0$ and $R_{\mathrm{R}} = \infty$, $\mathrm{Di} = \infty$ and $E_{\mathrm{M}}=1$, indicating perfect rectification.

\item For isotropic or symmetric resistors with $R_{\mathrm{F}}=R_{\mathrm{R}}$, $\mathrm{Di} = 1$ and $E_{\mathrm{M}}=0$, indicating no rectification.

\item If $R_{\mathrm{DC}} \ll R_{\mathrm{F}}$, then
$E_{\mathrm{M}} = (\mathrm{Di}-1)/(\mathrm{Di}+1)$ depends on the resistance values $R_{\mathrm{R}}$ and $R_{\mathrm{F}}$ only through their ratio Di. 

\end{itemize}{}

\section{Comparing steady and unsteady performance}

\begin{figure*}
\centering
\includegraphics[width=10.5cm]{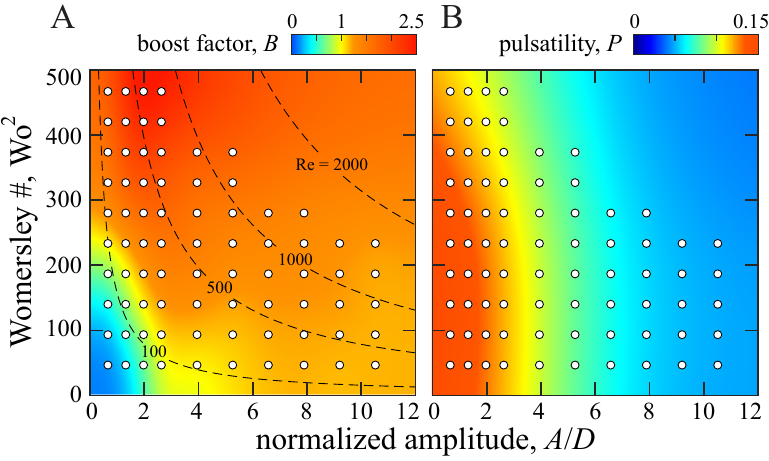}\vspace{-0.3cm}
\caption{Comparison of pump rate and pulsatility in experiments and a quasi-steady model. (\textit{A}) Boost factor $B$ quantifying enhancement of pumping effectiveness in experiments relative to model prediction. Markers indicate locations  of measurements, and the colormap is interpolated and extrapolated elsewhere. (\textit{B}) Pulsatility of the experimentally-measured rectified flow, defined as the ratio of ripple amplitude to mean flow rate in the DC branch.}
\label{fig:comparison}\vspace{-0.3cm}
\end{figure*}

The model furnishes predictions across varying inputs $A$ and $f$, and these results serve as a quasi-steady baseline against which the measured performance under unsteady conditions can be compared. In the colormap of Fig. \ref{fig:comparison}A, we plot the so-called \textit{boost} or relative enhancement of the experimentally measured effectiveness over the model prediction: $B = E / E_\mathrm{M}$. The axes are again dimensionless forms of amplitude $A/D$ and frequency $\textrm{Wo}^2$. Warmer colors with $B > 1$ indicate conditions for which the actual circuit outperforms quasi-steady expectations. It can be seen that the device performs better than expected for all but the lowest values of $A$ and $f$, providing validation of Tesla's conjecture of enhanced diodic performance for pulsatile flows \cite{tesla1920valvular}. Further, unsteady effects seem to be optimally exploited for low amplitude and high frequency oscillations (red region), for which we observe boosts as high as $B \approx 2.5$ and thus more than a doubling in pump rate over the quasi-steady baseline. Extrapolation of these data suggests yet greater enhancement for higher frequencies.

Another point of comparison between model and experiment involves the oscillations or ripples apparent in the DC branch signal, an example from experiments shown in the lower panel of Fig. \ref{fig:ACschematic}C. We define the \textit{pulsatility} $P = \Delta U / U_{\mathrm{DC}}$ as the ratio of the ripple amplitude to the mean pump rate, which can be assessed from the experimental measurements and from the model output. In both cases, we fit the form $U_{\mathrm{DC}} + \Delta U \sin (2 \pi f t + \phi)$ to the section-averaged flow speed to extract the necessary quantities. Smooth output flows and thus low $P$ are generally preferable in pumping applications. In the quasi-steady model, we observe uniformly high $P_{\mathrm{M}} \approx 1$ for all driving conditions (data not shown). This behavior is similar to an electronic diode-bridge rectifier, whose output current reaches a minimum of zero whenever the source current crosses zero, leading to oscillations as large as the mean. As shown in the map of Fig. \ref{fig:comparison}B, the actual fluidic circuit proves to be much smoother with $P \approx 0.1$ across the conditions explored here. Thus the fluctuations are an order of magnitude less than that predicted by the quasi-steady model. Surprisingly, the DC output in experiments is less pulsatile for stronger AC driving, and high $A$ in particular yields low $P$. We speculate that this effect and the general mitigation of pulsatility as compared to quasi-steady expectations are due to flow inertia, which tends to filter out fluctuations but which is absent from the quasi-steady framework.  

\section{Discussion and conclusions}

This work presents systematic experimental characterizations of Nikola Tesla's fluidic valve or diode across a wide range of both steady and unsteady flow conditions. Our findings support two main conclusions: 1) For steady forcing, an abrupt turn-on of diodic behavior is linked to the transition to turbulent flow, which is triggered at anomalously low Re; and 2) For unsteady forcing, the diodic performance is enhanced several-fold and especially so for high-frequency pulsatile flow. In more detail, our steady pressure/flow-rate investigation reveals a transitional Reynolds number $\textrm{Re} \approx 200$ marking the appearance of significant differences in the reverse versus forward hydraulic resistances. This turn-on of diodicity is accompanied by the onset of nonlinear scaling of pressure drop with flow rate, departure from the laminar-flow friction law, and flow instabilities that induce unsteadiness and strong mixing. These steady-forcing results serve as a baseline for understanding unsteady performance, which we assess using an AC-to-DC converter that transforms forced oscillations into one-way flows. The system acts as a pump or rectifier whose output rate varies nonlinearly with input driving parameters. While significant pumping is observed only when the diodic response is activated, a quasi-steady model informed by steady-flow characteristics fails to account for the observed pump rates. The model also does not account for the smooth operation of the pump, as quantified by the low pulsatility of its output DC flow, nor the counter-intuitive decrease in DC pulsatility for stronger AC driving.

The case of steady pressure/flow-rate considered in previous studies on Tesla-like channels provides a point of comparison to our results \cite{bardell2000diodicity,thompson2014numerical,truong2003simulation,zhang2007performance,mohammadzadeh2013numerical,nobakht2013numerical,gamboa2005improvements,thompson2013transitional,zhang2007performance,forster2007design,forster2002parametric,anagnostopoulos2005numerical,de2017design,losapio2018}. Experiments and simulations have reported weaker diodicity for corresponding conditions \cite{forster1995design,bardell2000diodicity,forster2002parametric,schluter2002micro,gamboa2005improvements,kord2012hybrid,losapio2018}, \textit{e.g.} $\textrm{Di} = 1.0-1.3$ at $\textrm{Re} = 500$ for which we measure $\textrm{Di} = 2.0$. This may be due to differences in geometry, both in that the diodes are simplifications of Tesla's original design and the measurement systems may not isolate the channel, thereby introducing additional pressure drops that dilute diodicity. The more encouraging results reported here are in better alignment with values $\textrm{Di} = 2-4$ reported for various Tesla-inspired geometries and at select Re values between $10^3$ and $10^6$ \cite{paul1969fluid,forster2002parametric,anagnostopoulos2005numerical,kord2012hybrid,kim2015hydrodynamic,losapio2018}. This regime of very high $\textrm{Re} > 2000$ would benefit from a systematic characterization of diodicity as presented here for $\textrm{Re} < 2000$. Future research might also include shape optimization towards improving performance
\cite{paul1969fluid,bendib2001analytical,schluter2002micro,forster2002parametric,forster2002parametric,gamboa2005improvements,anagnostopoulos2005numerical,kord2012hybrid,mohammadzadeh2013numerical,de2017design,losapio2018}, though it must be said that the diodicity value of 200 stated in Tesla's patent \cite{tesla1920valvular} seems well beyond reach.

Previous experimental and computational visualizations for Tesla-inspired channels have confirmed the reversible and irreversible flows expected at low \cite{stone2017fundamentals} and high Re \cite{bardell2000diodicity,thompson2014numerical,nobakht2013numerical,ansari2018flow}, respectively. Studies of the latter validate Tesla’s picture of the reverse flows taking a circuitous path around the periodic internal structures \cite{tesla1920valvular}. Our work bridges flow regimes by focusing on the newly-identified transitional $\textrm{Re} = 200$ and thereby revealing the destabilization mechanism for the reverse mode: Repeated interactions of the central flow with these structures lead to amplified lateral deflections that reroute streamlines and eventually induce complete mixing. These flow observations together with our hydraulic resistance characterizations point to the conclusion that Tesla’s conduit induces a turbulent-like flow state at unusually low Re. Namely, the signatures of pipe flow turbulence that first appear in smooth and rough-walled pipes at $\textrm{Re} \approx 2000$ are triggered in Tesla’s conduit at $\textrm{Re} \approx 200$. Signs of such early turbulence, though with less pronounced changes in the transitional Re, have been observed for textured channels and corrugated pipes \cite{kandlikar2005characterization,loh2011stability,cotrell2008instability}. Our measurements establish a link between early turbulence and the turn-on of diodicity, a connection that should be tested in other asymmetric conduits \cite{thiria2015ratcheting,groisman2004microfluidic,hawa2000viscous,fadl2007experimental} and might be exploited in the design of fluidic rectifiers and their applications.

The case of unsteady or pulsatile flow, while less studied, has been considered in previous experiments \cite{forster1995design,bardell2000diodicity,gamboa2005improvements,morris2003low,schluter2002micro,kord2012hybrid} and simulations \cite{anagnostopoulos2005numerical,mohammadzadeh2013numerical} aimed at small-scale fluidic devices. These works focus more on demonstrating pumping using Tesla-inspired geometries but less on the quantification needed for detailed comparison. Our findings offer guidelines that might be generally useful for such applications: 1) Reynolds numbers should be kept in the hundreds or higher to fully activate diodic response; 2) The output pump rate can be expected to increase super-linearly in both the amplitude and frequency of driving pulsations; 3) Stronger AC driving and high frequencies in particular can be used to optimally exploit unsteady effects and boost pump rate; and 4) Stronger driving and high amplitudes in particular optimally exploit inertial effects for low-pulsatility output flow. The second and third properties seem to bear out Tesla's vision that diodicity is amplified for highly unsteady conditions \cite{tesla1920valvular}. Future work might assess these findings in simulations, test whether they are general features of asymmetric conduits \cite{thiria2015ratcheting,groisman2004microfluidic,hawa2000viscous,fadl2007experimental}, and optimize shape for pulsatile flow conditions.

While this work emphasizes fluid mechanical characterization rather than practicalities, many applications of Tesla-like valves have been proposed, investigated and implemented \cite{zhang2007performance,mohammadzadeh2013numerical,arunachala2019stability,chandavar2019stability,hong2004novel,hossain2010analysis,yang2013development,wang2014tesla,forster1995design,bardell2000diodicity,morris2003low,de2017design,gamboa2005improvements,forster2007design,anagnostopoulos2005numerical,forster2002parametric,schluter2002micro,kord2012hybrid,kim2015hydrodynamic,losapio2018,porwal2018}. Here we touch on two engineering contexts and their physiological analogs. First, the pumping applications discussed above speak to Tesla's original motivation of a no-moving-parts valve that is resistant to wear or failure. Our findings indicate that its valving action is best at high frequencies. Thus suggests the specific application in which the vibrations intrinsic to all forms of machinery are harnessed to pump coolants, fuels, lubricants or other fluids needed for proper operation \cite{thiria2015ratcheting}. Here the leakiness or lossiness of the diode is no major detriment as the kinetic energy of the vibrational source is ample, ever-present and otherwise unused. This use is reminiscent of the return flow of blood up leg veins and of transport in the lymphatic system, both of which exploit ambient muscle contractions that squeeze vessels and activate serial valvular structures \cite{nielsen1964animal,buxton2006computational,schmid1990microlymphatics}. Secondly, the device may be exploited for its direction-dependent flow states and especially the turbulent mixing available at unusually low Re. A fluid drawn in reversely can be readily mixed with other fluids, heated/cooled or otherwise conditioned or treated, after which it may be expelled for further processing or use. Biological analogs exist in respiratory systems, where inhaled air is heated and humidified by so-called turbinates in the nasal cavity \cite{nielsen1964animal,keyhani1995numerical} and where structures resembling Tesla valves have recently been discovered in the turtle lung \cite{farmer2019tesla}.





\chapter{Valveless pumping and rectification in loopy models of avian lung}  \label{ch4}
This chapter is adapted from the preprint version of Q. M. Nguyen, A. U. Oza, J. Abouezzi, G. Sun, S. Childress, C. Frederick, and L. Ristroph, \href{https://doi.org/10.1103/PhysRevLett.126.114501}{"Flow rectification in loopy network models of bird lungs"} submitted to  Physical Review Letters (July 2020) \cite{nguyen2021flow}. The computational fluid dynamics simulations in COMSOL were performed by Dr. A. U. Oza and Dr. C. Frederick at the New Jersey Institute of Technology. QMN contributed to setting up the simulation parameters, compiling, interpreting and visualizing the results.

\section*{Abstract}

We demonstrate flow rectification, valveless pumping or AC-to-DC conversion in macroscale fluidic networks with loops. Inspired by the unique anatomy of bird lungs and the phenomenon of directed airflow throughout the respiration cycle, we hypothesize, test and validate that multi-loop networks exhibit persistent circulation or DC flow when subject to oscillatory or AC forcing at high Reynolds numbers. Experiments and simulations reveal a nonlinear response in which disproportionately stronger circulation is generated for higher frequencies and amplitudes of the imposed oscillations. Visualizations show that flow separation and vortex shedding at network junctions serve the valving function of directing current with appropriate timing in the oscillation cycle. These findings suggest strategies for controlling inertial flows through network topology and junction connectivity.

\section{Introduction}
Oscillatory, random or otherwise undirected movements can be induced into progressive motion by the presence of asymmetries \cite{magnasco1993forced,marquet2002rectified}. Rectification of fluids is a form of pumping \cite{glezer2002synthetic,prakash2008surface,lagubeau2011leidenfrost,thiria2015ratcheting,mo2016passive}, which is conventionally achieved via valves whose opening/closing motions are biased to direct flows appropriately. For example, the circulatory system involves pulsations produced by the beating heart that are rectified by flap-like valves to drive directed flow of blood through vessels. It is interesting and potentially useful that fluidic AC-to-DC conversion can also be achieved without moving elements but instead with entirely static geometries. This is permissible due to fluid dynamical irreversibility at high Reynolds numbers, for which the dominance of inertia over viscosity leads to phenomena such as flow separation and vortex shedding that respond sensitively to geometry and thus directionality \cite{tritton2012physical,schlichting2016boundary}. Such effects are exploited in fluidic diodes, devices whose asymmetric internal geometries lead to direction-dependent hydraulic resistance. Examples include conduits with diverging/converging walls, sawtoothed corrugations, and more complex geometries such as that proposed by Nikola Tesla \cite{tesla1920valvular,groisman2004microfluidic,thiria2015ratcheting,tao2020microfluidic}.

\begin{figure*}
\centering
\includegraphics[width=16.5cm]{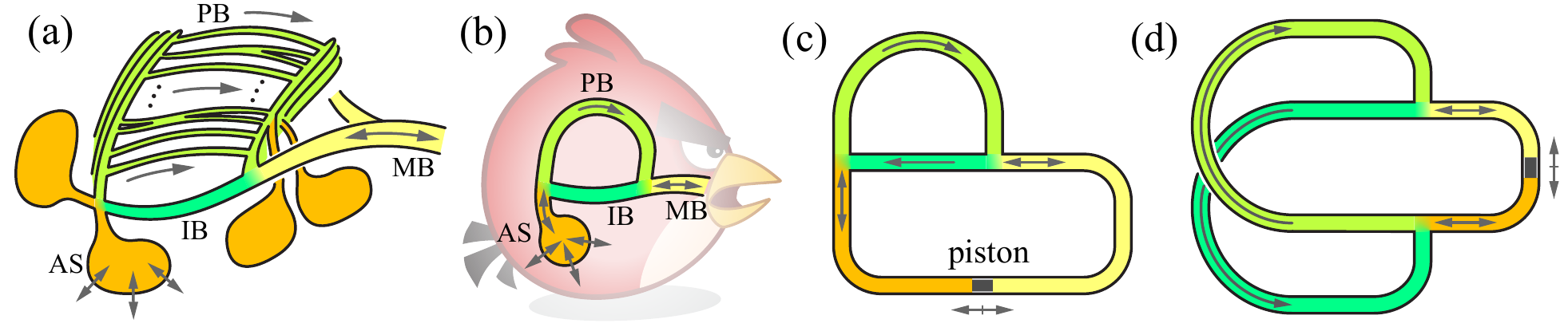}\vspace{-0.2cm}
\caption{Loopy network models of avian lungs. (a) Schematic of bird lung anatomy modified from \cite{scheid1972mechanisms}. Reciprocal expansion/contraction of air sacs (AS) drives oscillatory inhalation/exhalation flows in the main bronchus (MB). Directed flow is observed in the parabronchi (PB) that span the intrapulmonary bronchus (IB). (b) A `spherical bird' simplification with a single AS and PB. (c) A closed network formed by rerouting MB to AS, the latter replaced by a reciprocating piston. (d) A symmetrized network that modifies conduit lengths while preserving the connectivity of the T-junctions. }
\label{fig1}\vspace{-0.4cm}
\end{figure*}

Here we explore an alternative strategy for flow rectification in which the topology and connectivity of a network of symmetric pipes or channels leads to persistent circulation in response to oscillatory forcing. Our inspiration comes from a remarkable physiological observation: While air flow in the lungs of mammals oscillates with inhalation and exhalation, directional flows throughout the respiration cycle have been measured in birds \cite{hazelhoff1951structure,cohn1968respiration,bretz1971bird,brackenbury1971airflow,scheid1972mechanisms,powell1981airflow,maina2017pivotal,jones1981control}. This flow pattern may contribute to respiratory efficiency and has been viewed as an adaptation for the high metabolic demands of flight, although recent discoveries in reptiles suggest that the trait is ancestral \cite{farmer2010unidirectional,cieri2016unidirectional}. The avian lung differs fundamentally from that of mammals in that its conduits are not only branched but also \textit{reconnect} to form \textit{loops} \cite{hazelhoff1951structure,biggs1957new,duncker1974structure}. Indeed, the directed flows are observed along looped airways called parabronchi, as illustrated in the anatomical schematic of Fig. \ref{fig1}(a). The respiratory network is thus broadly analogous to a circulatory system, with alternating or AC forcing via air sac contraction/expansion driving directional or DC flows around loops \cite{hazelhoff1951structure,brackenbury1971airflow}. However, no valvular structures have been found in bird lungs \cite{duncker1974structure,king1958volumes}, spurring various hypotheses for rectification that invoke: static baffle-like structures that `guide' flows \cite{hazelhoff1951structure}; bronchial tube constrictions and divergences that induce separated flows and jets \cite{banzett1991pressure,wang1992aerodynamic,maina2000inspiratory,sakai2006numerical}; the curvature of tubes and their angles at junctions \cite{brackenbury1972physical, maina2009inspiratory}; and phased expansion/contraction of multiple air sacs and their compliance \cite{urushikubo2013effects,harvey2016robust}. Importantly, the Reynolds numbers $\textrm{Re} \sim 100-1000$ are high  \cite{bretz1971bird,butler1988inspiratory,kuethe1988fluid}, and studies that vary gas density and flow speed indicate the importance of inertial effects \cite{banzett1987inspiratory,kuethe1988fluid}.  

Here we propose, test and verify that loopy networks driven with oscillations or pulsations at high Reynolds numbers represent sufficient ingredients for AC-to-DC conversion. Towards isolating the factors essential for rectification, we begin with a series of geometrical idealizations of the avian respiratory network. In Fig. \ref{fig1}(b) we show a `spherical bird' approximation that isolates one parabronchial arc (PB), which connects to a single air sac (AS) on one end and to the main bronchus (MB) on the other, the latter idealized as a tube that opens to the outside at the mouth. A closed loop is formed by the PB and the intrapulmonary bronchus (IB). The air sac functions as a bellows that reciprocally expands and contracts to drive inspiration and expiration. The system is transformed into a closed circuit by replacing the air sac with a reciprocating piston and rerouting the main bronchus to attach as shown in Fig. \ref{fig1}(c). This closure is an experimental and computational convenience for studying flows internal to the network and without regard to exchange with the external environment.

\section{Experiments}
Laboratory demonstrations prove the rectification capability of the circuit of Fig. \ref{fig1}(c), as shown in the Supplemental Video 1. A system is constructed from clear rubber tubing and T-junction fittings, filled with water and flow-tracing particles, and driven via a reciprocating roller that serves the function of a piston. While the flow in the lower portion (yellow and orange) matches the driving motions and is purely AC, the upper segment (light green) displays directed or DC flows. Their sense is indicated by the arrow in Fig. \ref{fig1}(c) and matches that observed in the PB of bird lungs. Further, DC motions are apparent in the straight segment (dark green, IB) connecting the two junctions. In essence, circulation develops around the closed loop that includes both PB and IB, and the system acts as a circulating pump, fluidic rectifier or AC-to-DC converter.

The network of Fig. \ref{fig1}(c) contains asymmetries not critical to rectification, as demonstrated by considering further transformations that result in the symmetrized circuit of Fig. \ref{fig1}(d). The lengths and trajectories of the conduits are manipulated while preserving the network topology, junction geometry and connectivity. Specifically, the side branch of each T-junction feeds into the straight segment of the partner junction. This can be achieved in various 3D arrangements and in the quasi-2D layout of Fig. \ref{fig1}(d), which is planar except for a slight defect where the conduits cross. Demonstrations again show persistent DC circulation for AC driving. The network is mirror symmetric and so are the time-averaged DC flows, which emanate from the straight branches and return through the side branches of the junctions, as indicated in Fig. \ref{fig1}(d). These observations rule out an explanation of the observed rectification based on spontaneous symmetry breaking.

\begin{figure*}
\centering
\includegraphics[width=16.7cm]{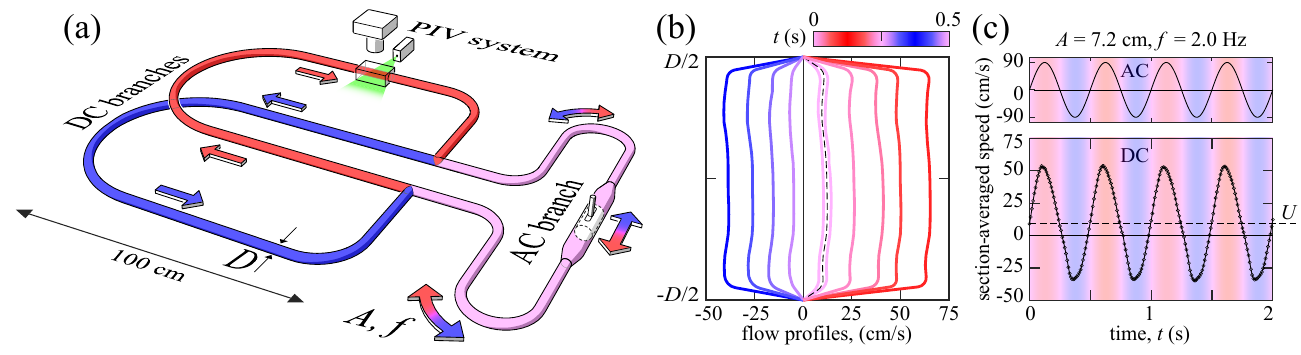}\vspace{-0.2cm}
\caption{Experimental system for characterizing flows in a forced fluidic network. (a) Rigid tubing of inner diameter $D = 1.6$ cm is formed and connected via T-junctions, and the network is filled with water seeded with microparticles. Oscillatory flow in the AC branch is driven by a motorized piston that oscillates with amplitude $A$ and frequency $f$, and the resulting flow field in a DC branch is measured via PIV. (b) Flow profiles in the DC branch measured at different times in the cycle for $A = 7.2$ cm and $f = 2.0$ Hz. Rectification is evident as the positive bias in the time-averaged profile (dashed curve). (c) Imposed oscillatory flow in the AC branch (top) and measured section-averaged flow in a DC branch (bottom); the time-average $U$ of the latter characterizes the pump rate.}
\label{fig2}\vspace{-0.1cm}
\end{figure*}

More careful experiments show that the rectified flows are reproducible for given forcing parameters and vary systematically with these parameters. As shown in the experimental schematic of Fig. \ref{fig2}(a), a version of the symmetric system is constructed from rigid tubing and custom-made junctions (all of uniform inner diameter $D = 1.6$ cm), filled with water, and connected to a motorized piston that oscillates sinusoidally with controllable amplitude $A \sim 0.3-8$ cm and frequency $f \sim 0.3-3$ Hz. This yields high Reynolds numbers that span those relevant to avian respiration \cite{bretz1971bird,butler1988inspiratory,kuethe1988fluid}: $\textrm{Re} = \rho A f D/ \mu \sim 10-5000$, where $\rho = 1.0~\textrm{g}/\textrm{cm}^{3}$ and $\mu = 8.9 \times 10^{-3}~\textrm{dyn} \cdot \textrm{s} / \textrm{cm}^{2}$ are the density and viscosity of water, respectively. Seeding with microparticles allows for flow measurement via particle-image velocimetry (PIV). A portion of the pipe in one of the DC branches is illuminated by a laser sheet and imaged with a high-speed camera, and a rectangular water jacket and thin-walled tubing ensure minimal optical distortion. The measured particle motions yield time- and space-resolved flow velocity fields. Profiles of the axially-averaged velocity, sampled throughout an oscillation period, are shown in Fig. \ref{fig2}(b) for selected values of $A$ and $f$. Rectification manifests as a bias towards positive velocities. Assuming that the flow is axisymmetric, the radial average of the flow velocity yields the section-averaged speed shown in Fig. \ref{fig2}(c). Its time average $U$ quantifies the emergent DC flow speed or, equivalently, the time- and section-averaged volumetric flux (volume flow rate per unit area).

\begin{figure*}
\centering
\includegraphics[width=16.5cm]{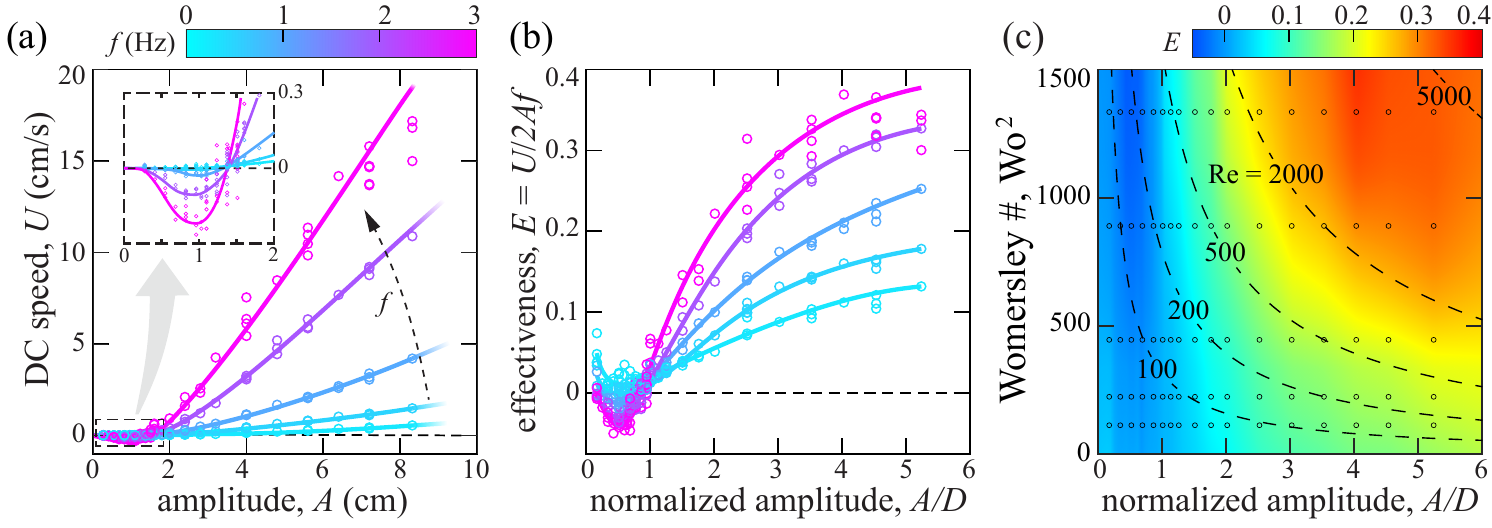}\vspace{-0.2cm}
\caption{Experimental characterization of AC-to-DC conversion in the symmetric circuit of Fig. 2. (a) DC flow speed $U$ across different values of the AC driving parameters $A$ and $f$. Inset: Magnified view of low-$A$ data. Multiple data points represent independent trials, and the curves are guides to the eye. (b) Effectiveness $E = U/2Af$ as a dimensionless measure of AC-to-DC conversion. (c) Map of $E$ across normalized amplitude $A/D$ and frequency $\textrm{Wo}^{2} \propto f$. Markers indicate where measurements are performed, and the map is interpolated and extrapolated elsewhere.}
\label{fig3}\vspace{-0.4cm}
\end{figure*}

The performance as an AC-to-DC converter can be characterized by measuring the output $U$ across different values for the input parameters $A$ and $f$. The plots of Fig. \ref{fig3}(a) show systematic trends in these data, and repeated measurements at each $A$ and $f$ indicate reproducibility of the phenomenon and reliability of the measurements. The pump rate increases with $A$ and $f$ for all but the smallest values of these parameters, for which $U$ is small or even negative (inset).

A dimensionless measure of pumping effectiveness is $E=U/2Af$, which compares the output DC speed to the characteristic input AC speed. More quantitatively, $E = 1$ represents perfect rectification in the following sense. A stroke of the piston in one direction, say the downward motion marked in red in Fig. \ref{fig2}(a), displaces fluid in the AC branch by an amount $2A$ in the duration $1/2f$. The entire flux $4Af$ is injected straight past the lower T-junction and into the DC branch colored red, and it returns in whole to the other side of the AC branch by turning at the upper junction. The side branch of the lower junction behaves as if sealed shut by a valve, deactivating the other DC branch (blue). The DC branches and T-junctions swap roles in the next stroke. This ideal behavior is thus half-wave rectification with cycle-averaged flux $U=2Af$ in each DC branch, hence $E=1$. The plots of Fig. \ref{fig3}(b) show the measured effectiveness $E(A,f)$ to be as high as 0.4, and the trends indicate increasing $E$ for greater $A$ and $f$.

The normalization $E = U/2Af$ fails to collapse these data, highlighting the nonlinear response of the system. A fully dimensionless characterization can be formed in terms of $A/D$, which represents the oscillatory amplitude relative to conduit diameter, and the square of the Womersley number $\textrm{Wo}^{2} =  \pi \rho f D^2 / 2\mu$, which is proportional to $f$ and assesses the unsteadiness of pulsatile flow \cite{womersley1955method}. Figure \ref{fig3}(c) displays $E(A/D,\textrm{Wo}^{2})$, showing increasingly effective pumping in both driving parameters. Also shown are the hyperbolic contours of the Reynolds number $\textrm{Re} = (2/\pi) (A/D) \textrm{Wo}^{2}$.

\begin{figure*}
\centering
\includegraphics[width=16.5cm]{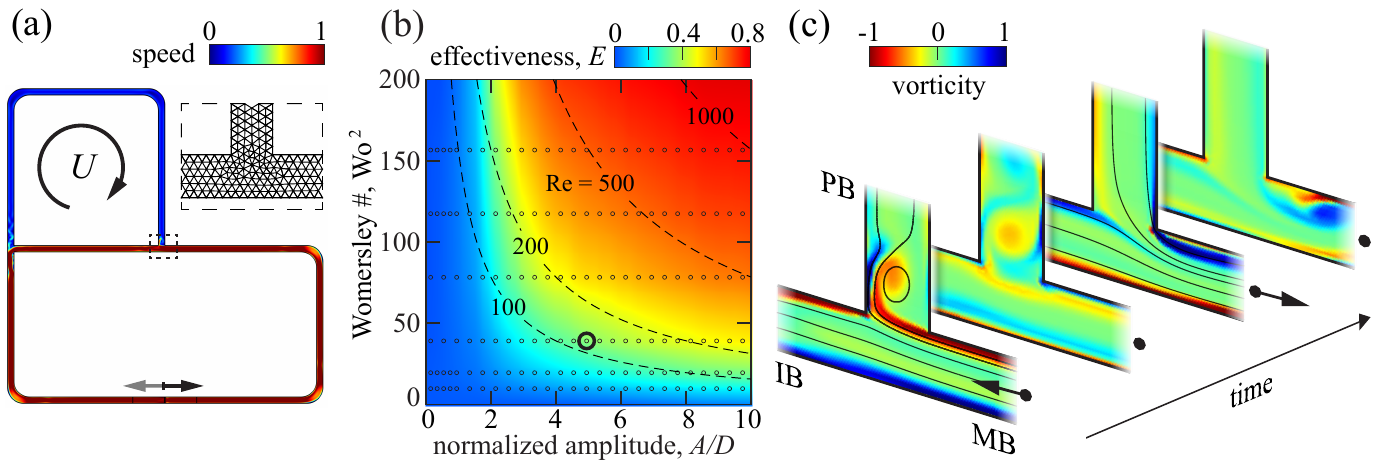}\vspace{-0.2cm}
\caption{Simulation of the asymmetric circuit. (a) The domain and spatial grid near a T-junction (inset). Color indicates flow speed normalized by $2 \pi fA $ and at the moment the piston is moving rightward at peak velocity. Here $D = 0.5$ cm, $A = 2.5$ cm and $f = 1$ Hz. (b) Effectiveness $E$ across $A/D$ and $\textrm{Wo}^2 \propto f$. (c) Streamlines and vorticity near one junction at four equally spaced phases in the cycle and for the parameters circled in (b). Arrows indicate the phase of the imposed flow from the MB. Vorticity is normalized by $ \pi^2f(A/D)^2 $, a scale appropriate for vortex shedding from a sharp edge in oscillatory flow \cite{agre2016linear}.}
\label{fig4}\vspace{-0.4cm}
\end{figure*}

\section{Simulations}
To further validate these findings, we implement computational fluid dynamics simulations relevant to the asymmetric circuit of Fig. \ref{fig1}(c). The two-dimensional Navier-Stokes equation for incompressible flow is numerically solved using the finite element method in the domain defined in Fig. \ref{fig4}(a), where no-slip boundary conditions are imposed on the static walls as well as the reciprocating piston. The simulations are carried out using the COMSOL Multiphysics software package with laminar flow settings \cite{comsol}. A detail of the spatial grid near one T-junction is shown in the inset. The system parameters span the ranges $A/D \sim 0.1-10$, $\textrm{Wo}^{2} \sim 10-200$ and $\textrm{Re} \sim 10-1000$. Compared to our experiments, the simulations are restricted to lower $\textrm{Wo}^{2}$ but permit larger $A/D$, leading to comparable Re. Additional implementation details are available in section \ref{ch4:supplementary}.

We assess the dependence of the terminal or long-time mean flow rate $U$ on the driving parameters $A$ and $f$. Our results reproduce all phenomenology and trends seen in the experiments, as shown by the $E(A/D,\textrm{Wo}^{2})$ map in Fig. \ref{fig4}(b) and the associated curves in the Supplemental Material. The simulations typically yield values of $E$ that are several times greater than those measured at corresponding conditions in experiments. This may be attributed to dimensionality (2D simulations versus 3D experiments), the use of asymmetric versus symmetric forms of the circuit, and the consequent differences in the lengths and curvatures of the conduits. Nonetheless, the features of emergent DC flows, their sense of direction, and qualitative trends for varying driving parameters are robust to these geometric differences.  

The simulations provide access to the flow fields, as shown in Supplemental Videos 2 and 3. In Fig. \ref{fig4}(c) we focus on streamlines and vorticity near the junction highlighted in (a) and at four instances in the oscillation cycle. At peak velocity during inhalation (first image), fluid is injected from the MB and predominantly goes straight past the junction down the IB, with little turning up the PB. This matches the intuition that inertia of the flow tends to maintain its straight course. A stagnation streamline impinges on the far (left) corner of the junction and divides the straight and turning flows, and a large vortex shed from the near (right) corner `plugs' the side branch. At peak velocity during exhalation (third image), fluid exits the junction via MB and draws more equally from IB and PB. The converging currents are divided by a separation streamline emanating from the far corner, and vorticity produced at the near corner does not detach but rather `hugs' the MB wall. Integrating in time over a cycle, the indicated junction thus has net flux incoming from the side branch (PB) and exiting via the straight branch (IB), which corresponds to net circulation in the clockwise sense around the PB-IB loop. (Consistent with mass conservation, the other junction also has net flux entering via its side branch and exiting via the straight branch, these associated with IB and PB, respectively.) In essence, flow separation and vortex shedding serve the valving function of closing and opening the side branch with appropriate timing in the cycle.
\section{Discussion and conclusion}
These events are ultimately rooted in flow irreversibility at high Reynolds numbers \cite{tritton2012physical,schlichting2016boundary}, which plays an analogous role in contexts such as expulsion versus suction from an orifice and reverse versus forward flow through an asymmetric conduit. The relevant asymmetry in our system is more subtle: Each T-junction has an axis of symmetry, but its anisotropic shape has distinct straight and side branches with differing end conditions due to their connectivity to the partner junction. Rectified flows of the type observed here are impossible for $\textrm{Re}=0$ or viscous-dominated conditions, for which the governing Stokes equation is reversible and the Scallop Theorem ensures flows induced in one stroke are retraced reversely in the return stroke \cite{tritton2012physical,purcell1977life}. Rectification is also precluded in the other extreme of inviscid and irrotational flow due to Kelvin's Circulation Theorem \cite{tritton2012physical}. Hence, the valving mechanism described here may operate at any finite Re, though it is expected to be exceedingly weak for $\textrm{Re} < 10$. The turbulent flow regime of $\textrm{Re} > 2000$, while less relevant to avian lungs, may be applicable to other problems and awaits future studies.

The phenomena reported here may arise more generally in the looped topologies common across many biological and physiological flow networks, which are often subject to unsteady forcing and pulsatile flows \cite{schmidt1997animal,fung2013biomechanics-circulation}. Loops play the essential role of providing routes around which circulation can be established. Our systems are minimal in that they comprise two interconnected loops, one whose AC flow is prescribed and the other free to display an emergent DC flow. Loops necessitate junctions, and the mechanism described above suggests that their geometrical anisotropy is critical. We posit that any anisotropy, appropriately mirrored for partner junctions, will in general induce rectification. In this sense, the T-shape used here is not critical, and the complexities of junction flows invite optimization over parameters such as branching angles \cite{aultandstone2016vortex,takagi1985study,haselton1982flow,mauroy2003interplay}. Future studies might vary not only the junction and conduit geometries but also the driving waveform and global network topology and connectivity, which could further inform on avian respiration and flow transport in complex networks generally. The value of the simpler circuits studied here is in identifying loops, junction anisotropy and inertial flows as sufficient ingredients for rectification.

\section{Supplementary Information on Numerical Simulations}\label{ch4:supplementary}
The numerical simulations presented above are performed using COMSOL Multiphysics. The incompressible Navier-Stokes equations are solved numerically in the two-dimensional domain shown in Fig. 1:
\begin{equation}
    \rho \left(\frac{\partial u}{ \partial t} + \boldsymbol{u} \cdot \nabla \boldsymbol{u} \right   ) = - \nabla p + \mu \Delta \boldsymbol{u}, \;\;
    \nabla \cdot \boldsymbol{u} =0
\end{equation}

where $\boldsymbol{u} = \boldsymbol{u}(\boldsymbol{x},t) \in \mathbb{R}^2 $ is the fluid velocity field, $p = p(\boldsymbol{x}, t) \in \mathbb{R}$ is the pressure, and $\rho = 1.0$ $\mathrm{ g}/\mathrm{cm}^{-3}$ and $\mu = 8.9 \times 10^{-3}$ $\mathrm{ g}/(\mathrm{cm} \cdot s)$ are, respectively, the density and dynamic viscosity of water. We assume the fluid to start at rest and thus impose the initial conditions $\boldsymbol{u}(\boldsymbol{x},0)=0$ and $p(\boldsymbol{x},0)=0$. The oscillatory motion of a piston is given by $x_p(t) = A \mathrm{cos}(2 \pi f t) $, where $A$ and $f$ are the forcing amplitude and frequency, respectively. No-slip conditions are imposed on all boundaries; specifically, $\boldsymbol{u}$ = 0 on the walls, and $\boldsymbol{u}= ( \dot{x_p}, 0)$ on the boundary of the piston. The piston has width $l = D/10$, where $D$ = 0.5 cm is the pipe diameter. The corners in the domain, with the exception of the two T-junctions, are rounded with a radius of curvature $L/ 20$.

\begin{figure*}
\centering
\includegraphics[width=15cm]{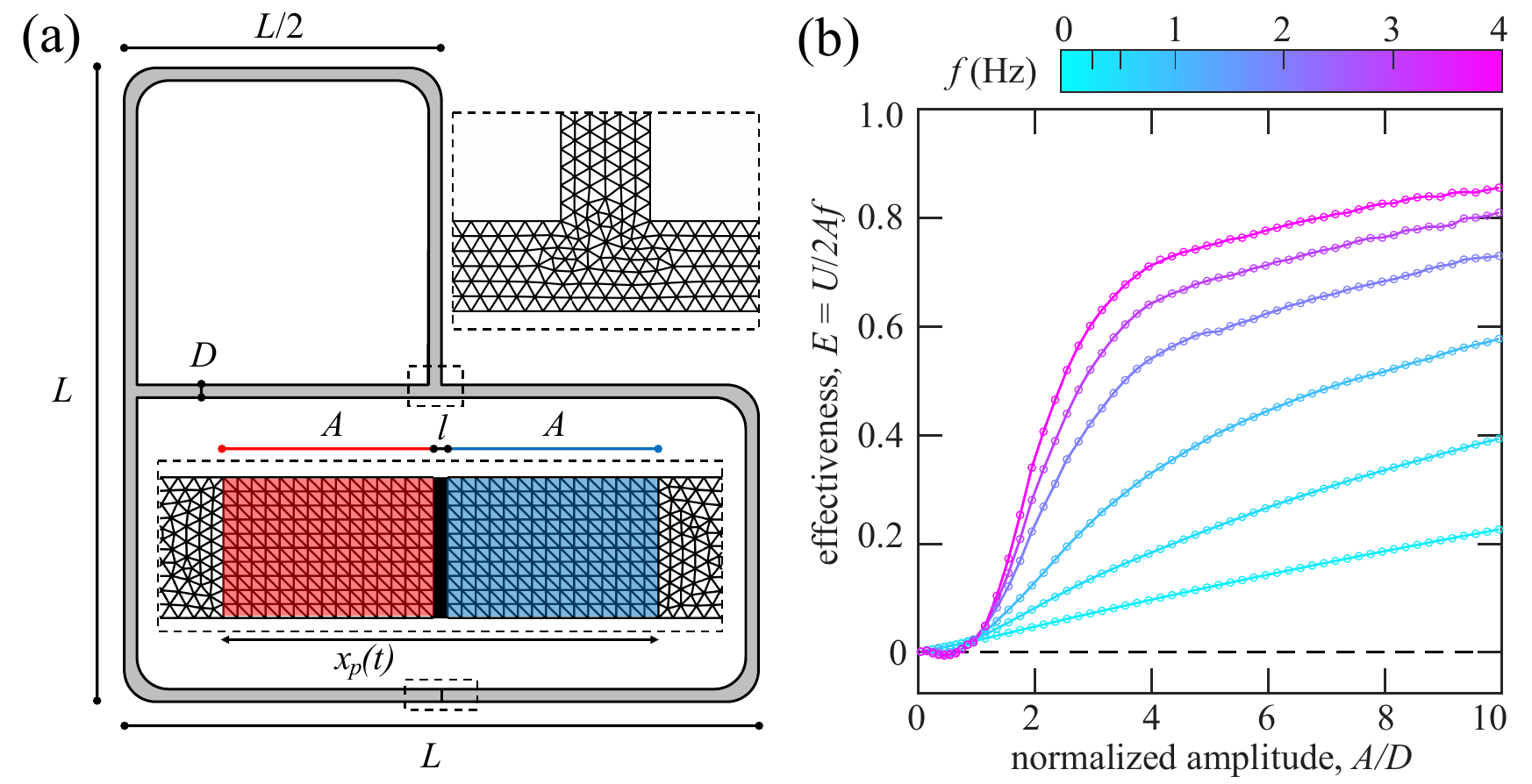}\vspace{-0.2cm}
\caption{(a) The 2D Navier-Stokes equations are solved in the gray region. The top inset shows the mesh near a junction, and the bottom inset shows the mesh around the piston (black region). (b) Simulation data showing the dependence of the effectiveness $E$ on the forcing frequency $f$ and dimensionless forcing amplitude $A/D$. The data are identical to that reported in Fig. 4(b) of the Main Text.}
\label{fig:ch5:sup}\vspace{-0.4cm}
\end{figure*}

The equations are solved numerically using the finite element method. The fluid velocity $\boldsymbol{u}$ is approximated by piecewise quadratic functions (P2), and the pressure by piecewise linear functions (P1), so as to satisfy the Lady\v{z}enskaja-Babu\v{s}ka-Brezzi (LBB) condition. A free triangular mesh is used throughout most of the domain (Fig. \ref{fig:ch5:sup}(a), top inset), and the maximum element size is $D/5$. In a neighborhood of the moving piston, we use an Arbitrary Lagrangian-Eulerian (ALE) formulation \cite{hirt1974arbitrary} to solve the PDE on time-dependent mesh points. The mesh is fixed in terms of the ALE coordinates $(X, Y )$, which identify points in a fixed reference domain corresponding to the colored regions in Fig. \ref{fig:ch5:sup}(a) (lower inset). These regions are discretized using a mapped mesh, in which the elements are split by inserting diagonal edges. In the domains to the left (red, -) and right (blue, +) of the piston, the ALE coordinates are related to the physical coordinates $(x_{\pm}(t); y)$ by the formula $x_{\pm}(t) = X +x_p(t)(A+l \mp X)/(A+l/2)$. Note that
$y = Y$ because the piston does not move vertically.

The simulations are performed for a maximum time $t_{\mathrm{max}}$ = 20T, where $T=1/f$ is the forcing period. The effectiveness $E = U/2Af$ is computed by averaging the flux in the DC loop over the last forcing cycle:
\begin{equation}
U = \frac{1}{DT } \int_{t_{\mathrm{max}-T}}^{t_\mathrm{max}} \mathrm{d} t \int_D    \mathrm{d} \boldsymbol{x} \; \boldsymbol{u} (\boldsymbol{x},t) 
\end{equation}
where $\int_D$ denotes an integral over the pipe diameter
within the DC loop. Both integrals are computed using
trapezoidal rule. Computing the flux along different pipe sections in the DC loop gives consistent results.
For example, for the simulations shown in Fig. 4(c) of
the Main Text, the values of $E$ computed along each of the four sides of the DC loop differ by less than $10^{-5}$.

The dependence of the effectiveness $E$ on both the forcing frequency $f$ and amplitude $A$ is shown in Fig. \ref{fig:ch5:sup}(b). The simulation results are largely unchanged upon refinement of the mesh. Specifically, all of the data points were recomputed with a more refined mesh, with a minimum element size of $D/10$. The maximum difference in the
effectiveness between the coarse and refined simulations is $\Delta E$ = 0.1, and the average difference is $\Delta E$ = 0.02.

\chapter{Quasi-steady theory for flows in network of tubes}  \label{ch5}
\section{Introduction}
We develop at framework for modeling flow in general complex networks consisting of branches and junctions (nodes) and apply it to the two loop network model of avian lung. Steve Childress at the Courant Institute of Mathematical Sciences conceived the idea of treating the avian lung problem as a one dimensional network of tubes (unpublished), and his analytical works result in a single ordinary differential equations for the flow in the DC branch, which correctly predicts the direction of rectification. We generalized Childress's works further to a framework applicable to any network topology while using fewer assumptions. Upon application to the avian model, our results show promising signs of qualitative agreements with the experiments and 2D simulations, and demonstrate that rectification is a manifestation of anisotropy. The results presented in this chapter find a fertile playground for further ideas on modeling fluid transport networks, however they should be considered preliminary.

\section{Navier-Stokes equations in 1D branches}
\begin{figure}[hbt!]
\centering
\includegraphics[width=0.5\textwidth]{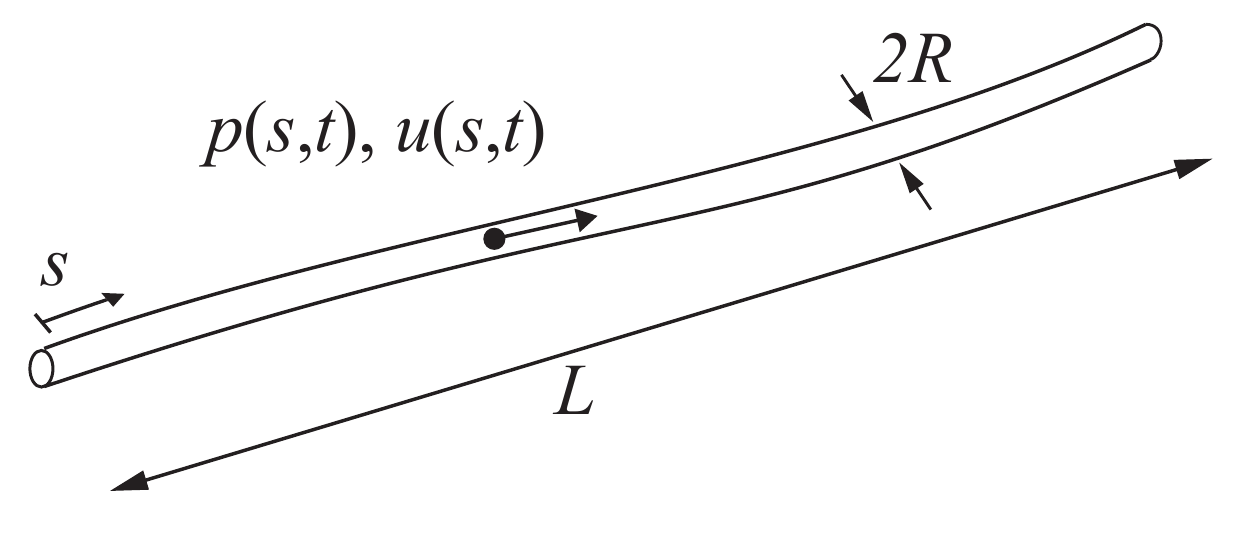}
\caption{The branches of the network are long and thin tubes, with radius $R$, characteristic length $L \gg R$, and large radius of curvature (not shown). At each point along the length of the branch which parametrized by axial coordinate $s$, the time-dependent pressure $p(s,t)$ and velocity $u(s,t)$ are cross-section averaged values.}
\label{circuitch6}
\end{figure}
We model the network in the limit $R/L \ll 1 $, where $R$ is the tube radius and $L$ is the  characteristic length of a branch. We use $s$ to indicate axial coordinate. Velocity $u(s,t)$ and pressure $p(s,t)$ are averaged values over the cross-section $\sigma$. The one-dimensional continuity and momentum equation for incompressible flows in long thin tubes are written as:

  \begin{equation} \label{cont}
   \frac{\partial \sigma}{\partial t} + \frac{\partial ( \sigma u(s,t))}{\partial s}  =0
     \end{equation}{}
     
       \begin{equation} \label{moment}
 \rho \left[ \frac{\partial u(s,t)}{ \partial t} + u \frac{\partial u(s,t)}{ \partial s}\right] +  \frac{\partial p(s,t)}{\partial s}  =  \frac{2 \pi R}{ \sigma}  \tau_{\mathrm{w}}
        \end{equation}{}
where $\rho$ is fluid density and $\tau_{\mathrm{w}}$ is boundary shear stress (or wall friction). Equation (\ref{moment}) may be derived as an  asymptotic limit of the full 3D Navier-Stokes equation assuming the diameter of the tube is small compare to its length and its radius of curvature, and the flow is axisymmetric. A derivation is given by Ottesen \cite{ottesen2003valveless}. We consider rigid cylindrical tubes with constant  cross-section $ \sigma= \pi R^2$, thus the one-dimensional Navier-Stokes equations become:
 \begin{equation} \label{cont2}
      \frac{\partial u }{ \partial s}  =0  
 \end{equation}{}
       \begin{equation} \label{eq:eomt}
 \rho \frac{ \mathrm{ d}u(t)}{ \mathrm{\mathrm{d}} t} +  \frac{\partial p(s,t)}{\partial s}  = \frac{2 }{R}  \tau_{\mathrm{w}}
        \end{equation}{}
The velocity $u$ in the branch has become only a function of time, while the pressure depends on both time and axial coordinates. We now address the wall shear stress $\tau_{\mathrm{w}}$.          
 \section{Resistance of branches to pulsatile flows}
  We make a step back to 3 dimensions momentarily to consider the wall shear stresss $\tau_{\mathrm{w}}$ which by definition,
 \begin{equation}
    \tau_{\mathrm{w}} = \mu \frac{\partial u_s (r,s,t)}{ \partial r} \bigg\rvert  _{r=R}
\end{equation}
where $ u_s (r,s,t)$ is the axisymmetric velocity with radial dependence retained (we define  $ u(s,t) = \\ 2 \pi \int_0^{R} u_s (r,s,t) r \mathrm{d}r$) and $\mu$ the dynamic viscosity of the fluid. Knowledge of $\partial u_s / \partial r$ or the shape of the flow profile is required to proceed. For steady flow in tubes (Poiseuille's flow), analytical solutions of $u_s$ are paraboloids \cite{tritton2012physical} and the wall shear stress is in phase with the the average flow rate given by $\tau_{\mathrm{w}} = (4 \mu/R)u $. For axisymmetric pulsatile flow (with both a steady component and oscillatory component) in tubes, solutions of which have been worked out by Womersley \cite{womersley1955method}, the shape of $u_s$ deviates from paraboloids, and velocity is out of phase with the stress. The exact solution for $\tau_{\mathrm{w}}$ is a complicated expression involving Bessel functions of Womersley number ($\mathrm{Wo} = R \sqrt{\rho 2 \pi f /\mu} $ where $ f$ is the frequency of the oscillation) and the amplitude of pressure gradient. While the exact stress could be useful for more elaborate models we can  approximate $\tau_{\mathrm{w}}$ for the case being considered here.

For developed axisymmetric pulsatile flows, the velocity profile is on dependent on time and the radial coordinate $r$. Following a common choice in blood flow modeling \cite{Olufsen1999structured}, we assume further that the velocity is  flat radially, except within a thin boundary layer of thickness $\delta$ near the wall, where it linearly drops to zero to satisfy the non-slip boundary condition.
\begin{equation}
u_s(r,t)= \begin{cases} 
     \;\; u(t), & 0 \le r < R-\delta \\
     \;\; \ u(t)(R-r)/\delta,  &  R -\delta \le r \le R 
      \end{cases}
\end{equation}
Using the approximated thickness of oscillating boundary layer (also called Stokes layer), $\delta \approx \sqrt{\mu/ \rho 2 \pi f}$ \cite{lighthill1975mathematical}, we arrive at an approximation for the wall shear stress $ \tau_{\mathrm{w}} \approx - \mu u(t)/\delta  =  - \sqrt{ \rho  \mu 2 \pi f}  u(t)$. The governing equation, Eq. \ref{eq:eomt} now reads
\begin{equation}
   \rho \frac{ \mathrm{ d}u(t)}{ \mathrm{d} t} +  \frac{\partial p(s,t)}{\partial s} =- k  u(t)
\end{equation}
where $k=2/R \sqrt{ \rho  \mu 2 \pi f}$ plays the role of the instantaneous friction coefficient. We now have the governing equation for flow inside a branch. The general network consists of many branches which are connected at nodes which obey conservation of mass.


\section{Connecting branches: Dynamics at junctions}
We are interested in a flow network that consists of branches and 90-degree T junctions.  To work with complex networks, we need a consistent scheme of labelling velocity and pressure. Each junction is associated with 3 branch points: point 1, 2 and 3 where point 1 and 2 are on the straight branches, and 3 is on the right angle branch. These hypothetical points are chosen such that they are close to the junction, but far enough that the values of velocity and pressure are distinct.
\begin{figure}[hbt!]
\centering
\includegraphics[width=0.5\textwidth]{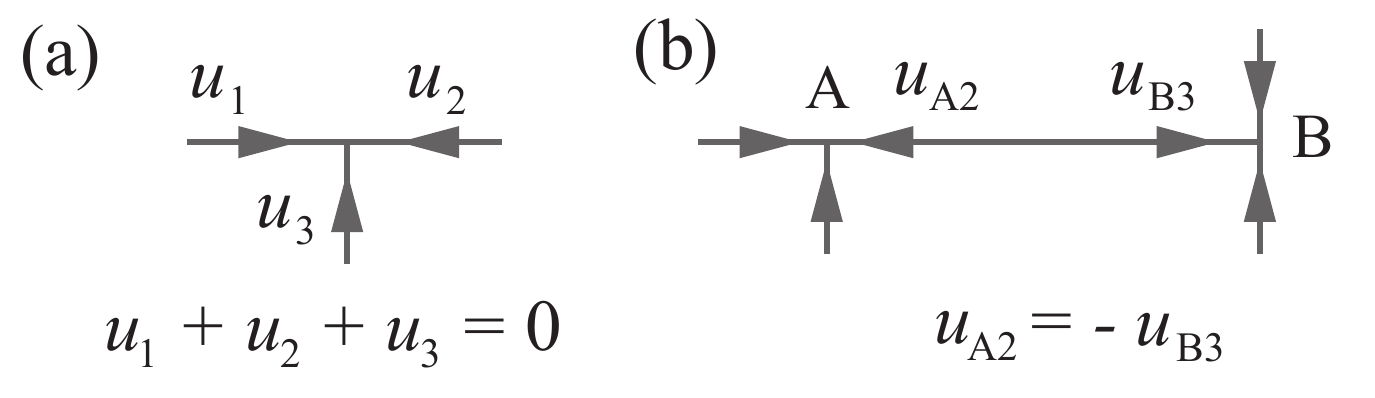}
\caption{Velocity labeling convention. (a) Each junction has 3 associated signed velocities. A velocity has positive value if the fluid flows towards the junction. (b) Each branch is associated with two equal and opposite velocity.}
\label{fig:convention}
\end{figure}
First we define the signed velocities at each junction as in Fig. \ref{fig:convention}(a). To keep track of velocities, we label them after each junction instead of branch, as in Fig. \ref{fig:convention}(b). A negative value means the velocity is leaving the junction, and vice versa. Thus conservation of mass means the sum of all velocity variables at each junction is zero. The velocity is constant in each branch but each branch connects two junctions and is assigned two equal and opposite velocities. This redundancy allows a system of equations to be written generically after naming the nodes. 
\subsection*{Modeling of flows at nodes}
At each node, we make two quasi-static  assumptions:
\begin{itemize}
    \item  Bernoulli's principle always applies  between point 1 and 2, i.e. $ p_1 + 1/2 \rho u_1^2 = p_2 +1/2 \rho u_2^2 $ 
    \item Bernoulli's principle between point 1 and 3 applies in some cases, and not other cases depend on the configuration of flow directions at the junction. 
    \end{itemize}

\begin{figure}[hbt!]
\centering
\includegraphics[width=1\textwidth]{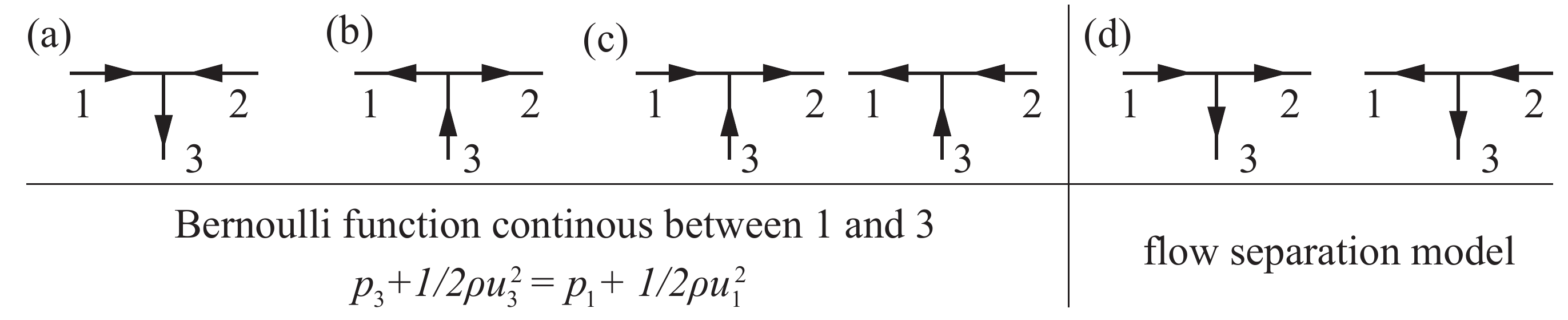}
\caption{Distinct states of actual flow directions at a T-junction, and modeling choices for each state. (a), (b), (c) show states without flow separation and Bernoulli's principle applies between point 1 and 3. (d) shows two flow states when flow in the straight branch separates/detaches at the sharp corner, and Bernoulli's principle doesn't apply.}
\label{fig:4states}
\end{figure}

In general, there are $2^3 = 8 $ combinations of velocity directions among 3 branches. Conservation of mass forbids two of them, reducing the number of states to 6 as shown in Fig. \ref{fig:4states}. Taking the symmetry between point 1 and 2 into account, there are only 4 distinct states. We assume that the Bernoulli function is continuous between point 1 and 3 in the states shown in Fig. \ref{fig:4states}(a), (b) and (c), but not in the  states shown in Fig. \ref{fig:4states}(d) when the flow in a straight branch splits at the junction. These choices are motivated by the vortices and streamlines in the simulation. When the flows combine at the junction, as shown in Fig. \ref{fig:4states}(c), we observe that streamlines converge smoothly. We also extend this to cases in Fig. \ref{fig:4states}(a) and (b). When flow is ejected from a straight branch and splits at the junction as in the two cases shown in Fig. \ref{fig:4states}(d), the flow streamlines tend to detach at the sharp corner, releasing a vortex that partially blocks the side branch. Because of this striking difference, these two cases are treated differently. 

\begin{figure}[hbt!]
\centering
\includegraphics[width=0.5\textwidth]{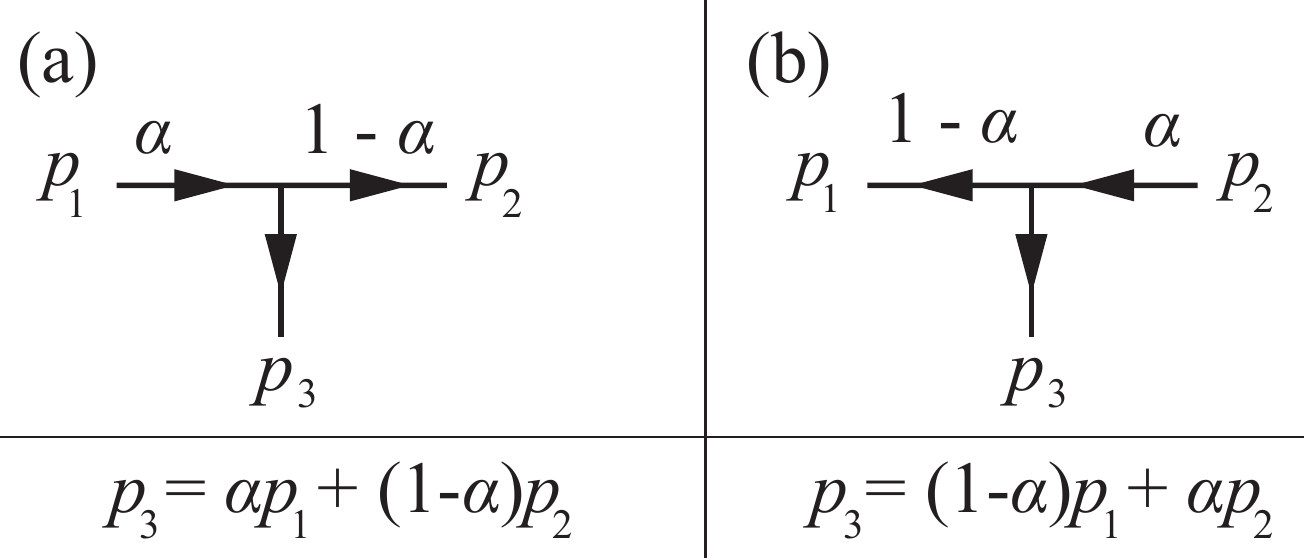}
\caption{Two states of flows at junction with separation. The pressure $p_3$ is modelled as a local static pressure, a weighted average of $p_1$ and $p_2$ with weight $\alpha $ and $1-\alpha$ given to the upstream and downstream velocity respectively.}
\label{fig:separationstates}
\end{figure}

In particular, when flow separation happens, we model $p_3$ as a local static pressure at the junction, and its value is equal to a weighted average of the pressure at upstream point and downstream point as shown in Fig. \ref{fig:separationstates}. The free parameter, $ \alpha $ is between 0 and 1. The 6 different flow states can be distinguished by the signs of the velocities. Let $F$ be a function that maps the values (to be precise, the signs) of $u_1$, $u_2$ and $u_3$ to a difference in pressure $p_3-p_1$:
\begin{equation}
p_3 - p_1 = F(u_1,u_2,u_3) =  \begin{cases} 
     \;\; \frac{1}{2} \rho u_1^2 - \frac{1}{2} \rho u_3^2 , & \textrm{Bernoulli's principle applies (4 states)} \\
     \;\;  (1-\alpha)(p_2-p_1),   &  \textrm{flow separation and  $u_1 > 0$ (1 state)}\\
       \;\;  \alpha(p_2-p_1),   &  \textrm{flow separation and $u_1 < 0$ (1 state)}
      \end{cases}
\end{equation}
To summarize, we have a closed system of algebraic and differential equations including: equations of motion relating velocity and pressure in branches, conservation of mass relating velocities at nodes, and Bernoulli's equation relating pressure between straight branches and $F$ relating pressures at the side branch and the straight branches. We can now test the model on the loopy network model of avian lung.

\section{Two loops network model of avian lung}

In general, for a network of $N$ nodes and $M$ branches and no free ends, there are  $3N$ pressures and $2M$  velocities, adding up to $3N+2M$ variables. The number of equation includes: $M$ opposite velocities relations, $M$ momentum equations, $N-1$ node conservation of mass equations (since the network is closed, the equation for the Nth node is not independent), N Bernoulli's equations, N pressure modelling equations, and one equation providing a known value either a pressure or a velocity. Thus the number of equations is $M+M+N-1+N+N+1=3N+2M$, equal to the number of variables. For complex networks, analytical solutions are not always attainable but the system could be solved numerically.
\begin{figure}[hbt!]
\centering
\includegraphics[width=0.4\textwidth]{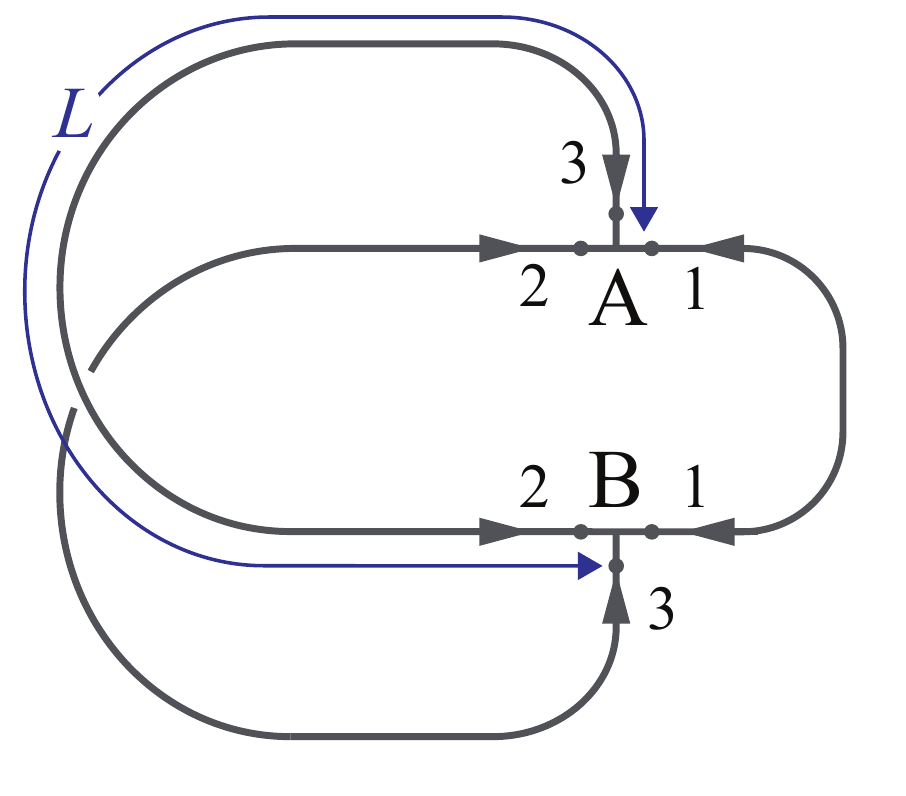}
\caption{The two loops network model of avian lung in reduced dimension. The network has 2 nodes A and B, and 3 branches. The gray arrow shows the labelled velocities. Branch A1-B1 is the so-called AC branch where the an oscillatory flow $  u_{\mathrm{A1}} = 2 \pi A f \mathrm{sin} (2 \pi f t)$ is specified. A3-B2 and A2-B3 (each has length $L$ as indicated by the blue curved double-headed arrow) are the so-called DC branches where flows are expected to have non-zero means. In particular, if the model reflects the experiment, there is a positive circulation in the close loop forming by the DC branches, where positive is defined as going in the order A2-B3-B2-A3.}
\label{fig:1dnetowrk}
\end{figure}
The two loops network model of avian lung has 2 T-junction A and B and 3 branches as labelled in Fig. \ref{fig:1dnetowrk}. For each node, there are 3 pressures, and 3 velocities. The total number of velocity and pressure variables in the network is 12. The system is completely described by 12 differential and algebraic equations. Suppose there is an oscillatory piston between $\mathrm{A}_1$ and $\mathrm{B}_1$ (AC branch), then
\begin{equation}
    u_{\mathrm{A1}} = 2 \pi A f \mathrm{sin} (2 \pi f t).
\end{equation}{}
The equal and opposite velocity relations for each branch are
\begin{equation}
    u_{\mathrm{A1}} = -u_{\mathrm{B1}}
\end{equation}{}
\begin{equation}
    u_{\mathrm{A2}} = -u_{\mathrm{B3}}
\end{equation}{}
\begin{equation}
    u_{\mathrm{A3}} = - u_{\mathrm{B2}} .
\end{equation}{}
Mass conversation for one node, say node A, is
\begin{equation}
     u_{\mathrm{A1}} + u_{\mathrm{A2}} +  u_{\mathrm{A3}} = 0.
\end{equation}{}
The momentum equation for the AC branch is not the 1D Navier-Stokes momentum equation. Assuming the piston is rigid, then 
\begin{equation}
    p_{\mathrm{A1}} = - p_{\mathrm{B1}}. 
\end{equation}{}
For the two DC branches, the momentum equation \ref{eq:eomt} is integrated over the branch length, giving
\begin{equation}
     \frac{\mathrm{d}u_{\mathrm{A2}}}{\mathrm{d}t}   + \frac{  p_{\mathrm{A2}} - p_{\mathrm{B3}} }{L} =- k u_{\mathrm{A2}}
\end{equation}{}
\begin{equation}
         \frac{\mathrm{d}u_{\mathrm{A3}}}{\mathrm{d}t} +  \frac{  p_{\mathrm{A3}}-p_{\mathrm{B2}} }{L} = - k u_{\mathrm{A3}} ,
\end{equation}{}
where  $k=2/R \sqrt{ \rho  \mu 2 \pi f}$ and $L$ is the length of the two DC branches. Pressure differences through the straight branch follow Bernoulli's equations:
\begin{equation}
     p_{\mathrm{A2}}- p_{\mathrm{A1}}    =\frac{1}{2}\rho u_{\mathrm{A1}}^2 -\frac{1}{2} \rho u_{\mathrm{A2}}^2  
\end{equation}{}
\begin{equation}
       p_{\mathrm{B2}} - p_{\mathrm{B1}} =    \frac{1}{2} \rho  u_{\mathrm{B1}}^2 - \frac{1}{2} \rho u_{\mathrm{B2}}^2.
\end{equation}{}
Pressure difference through the turn, one for each junction
\begin{equation}
p_{\mathrm{A3}} - p_{\mathrm{A1}} = F(u_{\mathrm{A1}},u_{\mathrm{A2}},u_{\mathrm{A3}})
\end{equation}
\begin{equation}
p_{\mathrm{B3}} - p_{\mathrm{B1}}  = F(u_{\mathrm{B1}},u_{\mathrm{B2}},u_{\mathrm{B3}}).
\end{equation}

There are many ways to implement $F$ numerically, one such example is
\begin{equation} \label{eq:ch6:F}
\begin{split}
    & F(u_1,u_2,u_3) = \\
    & \frac{1}{8} \Big \langle   \left( p_{2} - p_{1} \right)\left<1-\alpha,\alpha \right> \boldsymbol{\cdot} \left< 1+\mathrm{sign}(u_{1}), 1-\mathrm{sign}(u_{1}) \right> ,  \rho \left(u_{1}^2-u_{3}^2  \right), \rho \left(u_{1}^2-u_{3}^2  \right) \Big  \rangle \boldsymbol{\cdot} \\ 
    & \left<
     G(G-2)/2,  (2-G)(G+2),  G(G+2)/2 \right>
\end{split}
\end{equation}{}
where $G = \mathrm{sign}(u_1 u_2) + \mathrm{sign}(u_3)$, the angle brackets indicate vectors, and the dots indicate inner products. In practice, sign($u$) could be replaced by a smooth function such as tanh($cu$) where $c$ can be as large as necessary for a given accuracy. The only free parameter in the model $\alpha$ is set to 0.5. The system of differential algebraic equations (DAE) was solved in MATLAB using the solver \textit{ode23s} over a domain of $A$ and $f$. The results are presented in the next section

\section{Preliminary Results}
We first solve a simplified model when Bernoulli's principle applies in all routes, i.e. both between point 1 and 3, and point 1 and 2 of a junction. This was done by replacing $F$ in eq. (\ref{eq:ch6:F}) by
\begin{equation}
    F(u_1,u_2,u_3) =  \frac{1}{2} \rho u_{\mathrm{1}}^2 - \frac{1}{2} \rho u_{\mathrm{3}}^2.
\end{equation}
The velocity solutions are: $ u_{\mathrm{A2}} =  u_{\mathrm{A3}} = -0.5 u_{\mathrm{A1}}$ at junction A, for all values of $A$, $f$ and $t$ ( velocities at junction B follow mass conservation). The flow in AC branch always splits equally at the junctions, the flow in the DC branches are also purely oscillatory and no rectification emerges. This is expected from the fact that the simplified model is complete symmetric: the side and the straight branch obey the same law, on both half periods. 

\begin{figure}[hbt!]
\centering
\includegraphics[width=0.8\textwidth]{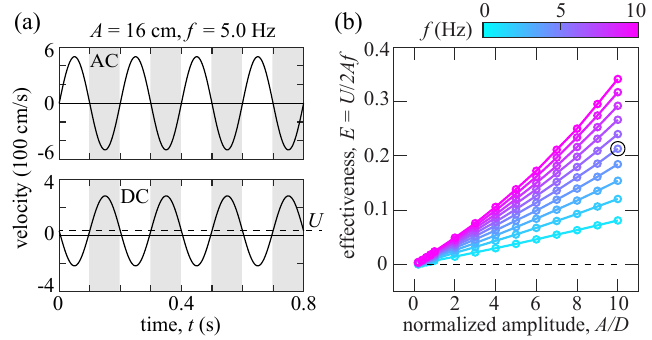}
\caption{Rectification in 1D model of the loopy network model of avian lung. (a) Imposed flow in the AC branch $ u_{\mathrm{A1}} = 2\pi Af$sin($2 \pi f t$) and the resulted flow in the DC branch $u_{\mathrm{A3}}$ whose time-average $U$ characterizes the pump rate. (b) Effectiveness $E=U/2Af$ as a dimensionless measure of AC-to-DC conversion. The circled point corresponds to the amplitude and frequency in (a).}
\label{fig:ch6:modeleff}
\end{figure}

In the full model, after a number of period $T= 1/f$ the solution of the DAE system reaches a terminal state in which the solution repeat itself every period. Fig. \ref{fig:ch6:modeleff}(a) shows a typical solution in a terminal state. The purely oscillatory flow in the AC branch is imposed through $ u_{\mathrm{A1}} = 2\pi Af$sin($2 \pi f t$) with $A$ = 16 cm and $f$ = 5 Hz, and the resulting flow in the DC branch ($u_{\mathrm{A3}}$ in particular) has both an oscillatory component with smaller amplitudes, and a  small drift $U$ which is the time average of $u_{\mathrm{A3}}$ in the terminal state. Thus rectification happens and the positive sign of $U$ agrees with experimental observation for large driving amplitude.  

As with the experiment, the effectiveness of the 1D network as an AC-to-DC converter is characterized by the non-dimensional quantity, $E = U/2Af$. Fig. \ref{fig:ch6:modeleff} shows $E$ for varying frequency $f$ and amplitude normalized by the tube's diameter, $A/D$. There is some quantitative agreement with experiments: the rectification is super-linear and $E$ increases with both $A$ and $f$. However, there are differing features: the model underestimates the experimental effectiveness, and lacks the flattening out of $E$ at high amplitudes or the $E<0$ region at low amplitude. 

\section{Discussion}
Though preliminary, the model confirms the idea that rectification in the two loops network emerges from the subtle anisotropy. The network has a mirror symmetry, and each junction also has an axis of symmetry but with distinct straight and side branches with differing end conditions due to their connectivity to the junction. When the difference in end conditions is removed, such as by enforcing the same conservation rule for both side and straight branch as we did above with Bernoulli's function, the system is completely symmetric and rectification does not manifest. In reality, complete symmetry would be realized in two ways. In a viscous dominated flow, Re $=0$ where the governing equations are reversible, the Scallop Theorem \cite{purcell1977life} ensures flows induced in one stroke are retraced reversely in the return stroke. In the other extreme of  inviscid flow, Re $=\infty $ Kelvin's Circulation Theorem \cite{tritton2012physical} ensures that if the circulation in the DC loop is initially zero ( true if the system starts from a state of rest), it will remain so for all time. 

While the model is able to demonstrate qualitatively the phenomenon of rectification with a single free parameter $\alpha$, it fails to capture the magnitude of effectiveness and its behavior at very low driving amplitude (where the circulation reverses in 3D experiment) and at high driving amplitude (where effectiveness seems to plateau in both experiment and 2D simulation). Perhaps these differences don't come as a surprise: One dimensional model versus three dimensional experiment and two dimensional simulation, and more importantly, the ad-hoc nature of the flow separation model with a parameter $\alpha$ independent of time, the driving amplitude, and frequency. Future studies of the detailed flow field and pressure field at a junction under pulsatile conditions through simulation and experiment would inform a better model. 

Although awaiting a systematic study, we conjecture that the friction coefficient $k$ plays a secondary role in achieving circulation: $k$ is not critical to whether rectification happens or not, but to the magnitude of rectification when it does happen. However, the one-dimensional modeling of $k$ or wall shear stress could be improved by including its time-dependence \cite{manopoulos2006one} based on Wormesley's theory for pulsatile flow. Another improvement is modifying the quasi-steady assumption. Our usage of the steady Bernoulli's equation might be justified at low frequency only. For higher frequency, the time-dependent Bernoulli's function $H= p + 1/2 \rho |\nabla \phi |^2 + \partial  \phi/\partial t $ where $\nabla \phi = \boldsymbol{u}$ should be used instead.

\chapter{Conclusion} \label{ch6}
Flows in macrofluidics are prevalent in many natural and artificial settings, and yet many of its aspects are not well understood. The relevant physics is that of flows at high Reynolds number, where irreversibility, instability and turbulence are relevant. On top of that, the non-linear nature of the governing Navier-Stokes equations presents theoretical challenges. Thus there are few general results regarding the mapping between network geometry and flow outcomes. Problems are solved on case-by-case basis and designs are based on trial and error and rely heavily on computational fluid dynamics simulations. This dissertation aims to shed light on some properties of flows in macrofluidic networks and their underlying physics, most notably the emergence of directed flows as a consequence of irreversibility, unsteady effects, and network geometry. 

Rectification is of interest in many physical systems, and is usually achieved by breaking symmetry. For example, flow rectification in macrofluidic networks can be achieved by imposing  geometrical asymmetry. We identified two kinds of asymmetry: asymmetry of the internal geometry of the branches, exemplified by Tesla's valve which has recently found to have a natural counterpart \cite{farmer2019tesla}; and asymmetry in the connectivity of the network, the minimal case of which is the two-loop network model of the avian lung. 

Regarding the asymmetry of channel geometry, we conducted the first characterization of Tesla's valve across flow regimes, under steady and unsteady conditions, accompanied by flow visualizations. We discovered non-trivial results: The turning-on of  diodicity is linked to the early transition to turbulence, and irreversibility (quantified by rectification effectiveness) increases with non-steadiness and favors low-amplitude, high-frequency forcing. Also counter-intuitive is the decrease of pulsatility at high amplitude. Our visualization highlights the destabilization mechanism in the reverse direction. The literature on Tesla's valve, although large, has been mostly focused on practical microdevice applications such as micropumps and micromixers. Our research contributes a fluid mechanical characterization of the device and suggests new understanding of the physics involved. 

What asymmetric geometry produces the most diodicity? While our research suggests that diodicity turns on at the transition to turbulence, an objective for optimization is not clear. To that end, we designed and tested a channel, the so-called "pocket" array (Fig. \ref{fig:twodiodes}), that outperforms Tesla's valve under steady flows. Left for future studies is to see why this is the case, and if the trend persists under unsteady flow, where the Tesla's valve receives a significant performance boost. 

Our data suggest that diodicity will keep increasing with Reynolds number, at least up to Re $\sim 10^4$. Is there an upper limit of diodicity, and if so, at what Reynolds number? The answers would benefit from further studies on the diodic behavior of asymmetric channels at higher Reynolds number. Naively, one would expect that at a high enough Reynolds number, the flow would be completely turbulent in both directions and it would make little difference which direction the flow is going, and the diodicity should decrease. Given that many facets of turbulence are not understood, experiments would be the best approach. If indeed diodicity is not monotonic and decreases to unity after some high Re, that would mean the flow is reversible much like the Re $ \rightarrow 0$ limit. This would be of fundamental interest. 


 Our idealized models of avian lung demonstrate fluid dynamical valving, and suggest that the minimal ingredients for AC-to-DC conversion are the presence of loops with appropriate asymmetric junction connectivity and irreversibility. However, more work needs to be done on real birds to corroborate our findings. The geometry of the complex airways in bird lungs has not been well mapped out. A 3D reconstruction of the airways would facilitate identifying the geometry suggested by our research. Many more questions arise: What is the effectiveness of real respiratory systems? And how is it optimized over the geometry and other parameters?
 
On the modeling front, a lot more could be done both with regards to the specific system of bird lung, and more general macrofluidic networks. Immediate progress could be made on modeling the fluid dynamics at a junction. We have used an ad-hoc model of flow separation to demonstrate that any anisotropy that breaks the inviscid flow assumption (enforced by Bernoulli's principles) would produce rectification. While there is a physical basis to our flow separation model, it could be fine-tuned further. 2D simulations, which faithfully reproduce the qualitative phenomena, provide the basic picture of the valving mechanism at the T-junction: Only in one half period, vortex shedding happens and blocks the side branch. In the other half period, the boundary layer remains thin and flow combines smoothly. Studying  dependence of the size of the vortex on amplitude and frequency, and how it modifies the pressure on the side branch would inform a more accurate and versatile 1D model. Further improvement could be made by incorporating time-dependent Bernoulli's equation, and time-dependent hydraulic resistance of branch under pulsatile flow. These building blocks could be used to study more complex fluidic networks.

\begin{figure}[hbt!]
\centering
\includegraphics[width=12cm]{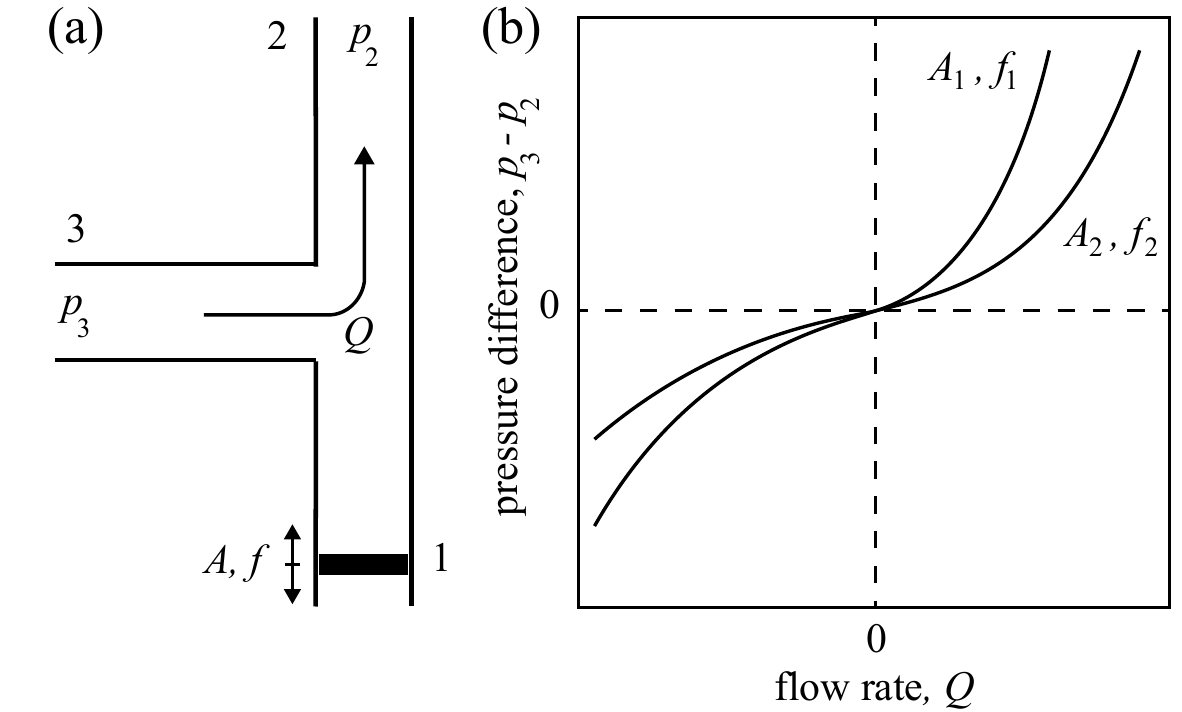}
\caption{Proposed experiment of pulsatile flow at T-junction. (a) An oscillatory flow rate is imposed at 1 by a sinusoidally moving piston at amplitude $A$ and frequency $f$, a flow rate $Q$ between 2 and 3 is produced, depends on the pressure difference (b) Hypothetical characteristic curves for difference set of $A$ and $f$. The asymmetry shows a bias in 3-to-2 direction.}
\label{fig:ch7:singleT}
\end{figure}

Experimentally, the analysis of macrofluidic networks would benefit from studies of pulsatile flows in a T-junction.
We propose that a junction could be studied in isolation as shown in Fig. \ref{fig:ch7:singleT}(a). The junction is filled with fluid and an oscillatory flow is imposed on one of the branches, say, branch 1 by a moving piston with amplitude $A$ and frequency $f$. A biased pressure difference is imposed between branch 2 and 3 (say, by connecting them each to large liquid tanks with different heights). As found by our experiments and simulations, under oscillatory conditions and no biased pressure imposed,
 there is on average a net flow from branch 3 to branch 2, which depends on $A$ and $f$. Adding a pressure difference would either increase or decrease the net flow. Thus we expect asymmetric characteristic curves as shown in \ref{fig:ch7:singleT}(b). Our experiments and simulations are limited to $\Delta p =p_3 - p_2 \sim 0$ and $Q$ is not measured directly. Done systematically, the proposed experiment could map out the behavior of a junction over the parameter space ($A,f, \Delta p$) and could be valuable for analysing pulsatile flows in networks. 

We observed reversal of rectification, corresponding to a net flow from branch 2 to branch 3 through the junction in \ref{fig:ch7:singleT}(b) at the the smallest driving amplitudes in our experiments, and to a lesser extent, in the 2D simulation. We suggest potential approaches for further studies. The turn at the  junction could be viewed as a curve with small radius of curvature. The flow in a curved pipe is known to have secondary flow in the axial direction due to centripetal force, also known as Dean's flow \cite{dean1927xvi}. The secondary flow is more complicated in unsteady conditions, and demonstrates counter-intuitive phenomenon \cite{lyne1971unsteady}. Potentially related is steady-streaming \cite{riley2001steady} and reversal of flow in the oscillatory boundary layer \cite{haddad2010pulsating}.

\begin{figure}[hbt!]
\centering
\includegraphics[width=1\textwidth]{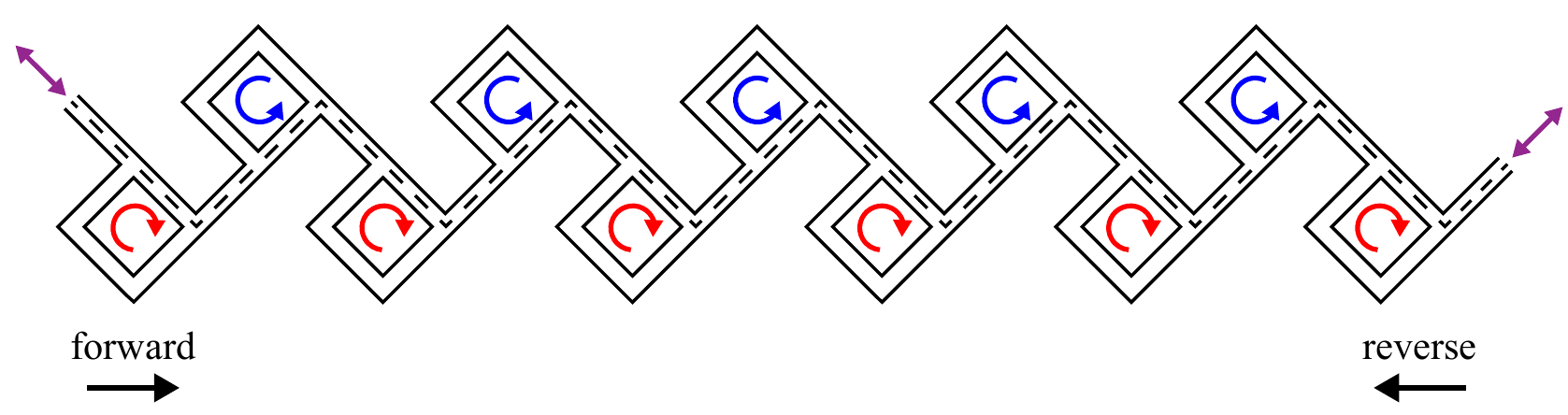}
\caption{The equivalent loopy network of Tesla's valve, with straight-angle junctions. Under oscillatory forcing (purple), in an alternative manner, the loops in the bottom row would sustain negative circulation (red) while the loops in the top row positive circulation (blue). The combine effect is a net drift in forward direction in the central corridor (dash line).}
\label{fig:ch7:teslaloopy}
\end{figure}

In bringing the two kinds of asymmetric geometries together, we observe that the internal geometry of Tesla's valve could be viewed as network of multiple loops in series, with acute junction angles. For clarity, the equivalent loopy network with right angle junctions is shown in Fig. \ref{fig:ch7:teslaloopy}. Our experiment, simulations and modeling of flow in loopy networks show that under oscillatory forcing, net circulation emerges in the loops with appropriate connectivity. In this case, if a sinusoidal pressure (indicated by purple arrows) is applied across the loops in this network, each loop would sustain a circulation. Specifically, each loop in the bottom row manifests a negative circulation (red) while each in the top a positive circulation (blue). They do so in alternating manner (coincidentally resembling \href{https://www.aps.org/units/dfd/pressroom/gallery/2009/kumar09.cfm}{ von Karman vortex street}) and a net drift in the forward direction emerges. Thus the loopy network when connected in series would behave as expected for Tesla's valve. Further, our findings show that the net circulation in each loop increases with the amplitude and frequency of driving. This suggests why Tesla's valve receives a boost in performance under pulsatile flows. We conjecture that the "pocket" array while having a higher diodicty than the Tesla's valve under steady flow, will not receive a strong boost of performance with pulsatile flows like the Tesla's valve because of the lack of a loopy topology. While the pattern of circulation shown here could be sustained in a channel without loop geometry \cite{thiria2015ratcheting,groisman2004microfluidic}, the presence of loops certainly make the the circulation more robust. 

Another important question left for further studies is whether the  effectiveness $E$, which in our data on Tesla's valve circuit and the loopy network circuit has a super-linear increase with driving amplitude and frequency, would keep increasing at higher $A$ and $f$. We have reasoned that $E=1$ for ideal valving action, and in the explored ranges of experiment parameters, we have only seen $E<1$. However, $E>1$ or more than ideal valving is not forbidden by any physical law. In a network, a circulation added to a closed loop would not violate conservation of mass. And $E>1$ would not violate conservation of energy, as energy could be added to the system through the source flow (say, due to a moving piston). Experimentally, this could be tested by ramping up unsteady effects though increasing the driving amplitude and frequency, increasing the size of the experiment or reducing the fluid viscosity. 

\label{chp-conclusion}








\cleardoublepage
\phantomsection

\addcontentsline{toc}{chapter}{Bibliography}

\printbibliography
\end{document}